\documentclass[sn-basic]{sn-jnl}% Basic Springer Nature Reference Style/Chemistry Reference Style
\usepackage{graphicx}%
\usepackage{multirow}%
\usepackage{amsmath,amssymb,amsfonts}%
\usepackage{amsthm}%
\usepackage{mathrsfs}%
\usepackage[title]{appendix}%
\usepackage[x11names]{xcolor}%
\usepackage{textcomp}%
\usepackage{manyfoot}%
\usepackage{booktabs}%
\usepackage{algorithm}%
\usepackage{algorithmicx}%
\usepackage{algpseudocode}%
\usepackage{listings}%
\usepackage{lineno}
\usepackage{lscape}

\usepackage{color}
\usepackage{rotating}
\usepackage{bigints}
\usepackage{enumerate}
\usepackage{hyperref}
\usepackage{cleveref}
\usepackage{float}
\usepackage{url}
\usepackage{tablefootnote}
\usepackage{soul}
\usepackage{array}

\newcommand{\beq}{\begin{equation}}
\newcommand{\eeq}{\end{equation}}
\newcommand{\bea}{\begin{eqnarray}}
\newcommand{\eea}{\end{eqnarray}}
\newcommand{\hmu}{\hat{\mu}}
\newcommand{\sym}{\mathrm{sym}}
\newcommand{\sat}{\mathrm{sat}}
\newcommand{\skin}{\mathrm{skin}}

\begin{document}
% \linenumbers

\title[]{Theoretical and experimental constraints for the equation of state of dense and hot matter}


\author[1]{Rajesh Kumar}
\author*[1]{Veronica Dexheimer}
\email{vdexheim@kent.edu}
\author[2]{Johannes Jahan}
\author[3]{Jorge Noronha}
\author[3]{Jacquelyn Noronha-Hostler}
\author[2]{Claudia Ratti}
\author*[3]{Nico Yunes}
\email{nyunes@illinois.edu}
\author[2]{Angel Rodrigo Nava Acuna}
\author[4]{Mark Alford}
\author[5]{Mahmudul Hasan Anik}
\author[15]{Debarati Chatterjee}
\author[6,7]{Katerina Chatziioannou}
\author[8,9]{Hsin-Yu Chen}
\author[1]{Alexander Clevinger}
\author[3]{Carlos Conde}
\author[3]{Nikolas Cruz-Camacho}
\author[10]{Travis Dore}
\author[11]{Christian Drischler}
\author[12]{Hannah Elfner}
\author[13]{Reed Essick}
\author[14]{David Friedenberg}
\author[15]{Suprovo Ghosh}
\author[2]{Joaquin Grefa}
\author[3]{Roland Haas}
\author[4]{Alexander Haber}
\author[16]{Jan Hammelmann}
\author[17]{Steven Harris}
\author[18,19]{Carl-Johan Haster}
\author[20]{Tetsuo Hatsuda}
\author[3]{Mauricio Hippert}
\author[16]{Renan Hirayama}
\author[14]{Jeremy W. Holt}
\author[2]{Micheal Kahangirwe}
\author[21]{Jamie Karthein}
\author[22]{Toru Kojo}
\author[23]{Philippe Landry}
\author[5]{Zidu Lin}
\author[24]{Matthew Luzum}
\author[3]{T. Andrew Manning}
\author[3]{Jordi Salinas San Martin}
\author[25]{Cole Miller}
\author[26,27,28]{Elias Roland Most}
\author[3]{Debora Mroczek}
\author[29]{Azwinndini Muronga}
\author[3]{Nicolas Patino}
\author[1]{Jeffrey Peterson}
\author[30]{Christopher Plumberg}
\author[2]{Damien Price}
\author[31]{Constanca Providencia}
\author[32]{Romulo Rougemont}
\author[5]{Satyajit Roy}
\author[2]{Hitansh Shah}
\author[3]{Stuart Shapiro}
\author[5,33]{Andrew W. Steiner}
\author[1]{Michael Strickland}
\author[3]{Hung Tan}
\author[22]{Hajime Togashi}
\author[2]{Israel Portillo Vazquez}
\author[14]{Pengsheng Wen}
\author[4]{Ziyuan Zhang}

\affil*[1]{Department of Physics, Kent State University, Kent, OH 44243 USA}

\affil[2]{Department of Physics, University of Houston, Houston, TX 77204, USA}

\affil*[3]{University of Illinois at Urbana-Champaign, Urbana, IL 61801, USA}

\affil[4]{Department of Physics, Washington University, St. Louis, MO 63130, USA}

\affil[5]{Department of Physics and Astronomy and University of Tennessee, Knoxville, Knoxville, TN 37996, USA}

\affil[6]{Department of Physics, California Institute of Technology, Pasadena, California 91125, USA}

\affil[7]{LIGO Laboratory, California Institute of Technology, Pasadena, California 91125, USA}

\affil[8]{Department of Physics, The University of Texas at Austin, Robert A Welch Hall, 105 E 24th st, Austin, TX 78712, USA}

\affil[9]{Department of Physics and Kavli Institute for Astrophysics and Space Research, Massachusetts Institute of Technology, 77 Massachusetts Ave, Cambridge, MA 02139, USA}

\affil[10]{Fakultat fur Physik, Universitat at Bielefeld, D-33615 Bielefeld, Germany}

\affil[11]{Department of Physics and Astronomy and Institute of Nuclear and Particle Physics, Ohio University, Athens, OH 45701, USA}

\affil[12]{GSI Helmholtz Centre for Heavy-ion Research, Planckstr. 1, 64291 Darmstadt, Germany}

\affil[13]{Perimeter Institute for Theoretical Physics, 31 Caroline Street North, Waterloo, ON N2L 2Y5}

\affil[14]{Department of Physics and Astronomy and Cyclotron Institute, Texas A\&M University, College Station, TX 77843, USA}

\affil[15]{Inter-University Centre for Astronomy and Astrophysics, Pune University Campus, Pune 411007, India}

\affil[16]{Frankfurt Institute for Advanced Studies, Ruth-Moufang-Strasse 1, 60438 Frankfurt am Main, Germany}

\affil[17]{Institute for Nuclear Theory, University of Washington, Seattle, WA 98195, USA}

\affil[18]{Department of Physics and Astronomy, University of Nevada, Las Vegas, 4505 South Maryland Parkway, Las Vegas, NV 89154, USA}

\affil[19]{Nevada Center for Astrophysics, University of Nevada, Las Vegas, NV 89154, USA}

\affil[20]{Interdisciplinary Theoretical and Mathematical Sciences Program (iTHEMS), RIKEN, Wako 351-0198, Japan}

\affil[21]{Center for Theoretical Physics, Massachusetts Institute of Technology, Cambridge, MA, 02139, USA}

\affil[22]{Department of Physics, Tohoku University, Sendai 980-8578, Japan}

\affil[23]{Canadian Institute for Theoretical Astrophysics, University of Toronto, Toronto, Ontario M5S 3H8, Canada}

\affil[24]{Instituto de Física, Universidade de São Paulo, Rua do Matão, 1371, Butantã, 05508-090, São Paulo, Brazil.}

\affil[25]{Department of Astronomy and Joint Space-Science Institute, University of Maryland, College Park, MD 20742}

\affil[26]{Princeton Center for Theoretical Science, Jadwin Hall, Princeton University, Princeton, NJ 08544, USA}

\affil[27]{Princeton Gravity Initiative, Jadwin Hall, Princeton University, Princeton, NJ 08544, USA}

\affil[28]{School of Natural Sciences, Institute for Advanced Study, 1 Einstein Drive, Princeton, NJ 08540, USA}

\affil[29]{Faculty of Science, Nelson Mandela University, Gqeberha, South Africa}

\affil[30]{Natural Science Division, Pepperdine University, Malibu, CA 90263, USA}

\affil[31]{CFisUC, Department of Physics, University of Coimbra, P-3004 - 516 Coimbra, Portugal}

\affil[32]{Instituto de Física, Universidade Federal de Goiás, Av. Esperança - Campus Samambaia, CEP 74690-900, Goiânia, Goiás, Brazil}

\affil[33]{Physics Division, Oak Ridge National Laboratory, Oak Ridge, TN 37830, USA}

\affil{(MUSES Collaboration)}

% \author*[1,2]{\fnm{First} \sur{Author}}

% \author[2,3]{\fnm{Second} \sur{Author}}

% \author[1,2]{\fnm{Third} \sur{Author}}


% \affil*[1]{Department of Physics, Kent State University USA 44240}

% \affil[2]{\orgdiv{Department}, \orgname{Organization}, \orgaddress{\street{Street}, \city{City}, \postcode{10587}, \state{State}, \country{Country}}}

% \affil[3]{\orgdiv{Department}, \orgname{Organization}, \orgaddress{\street{Street}, \city{City}, \postcode{610101}, \state{State}, \country{Country}}}


\abstract{This review aims at providing an extensive discussion of modern constraints relevant for dense and hot  strongly interacting matter. It includes theoretical first-principle results from lattice and perturbative QCD, as well as chiral effective field theory results. From the experimental side, it includes heavy-ion collision and low-energy nuclear physics results, as well as observations from neutron stars and their mergers. The validity of different constraints, concerning specific conditions and ranges of applicability, is also provided. }




\maketitle
\tableofcontents

\section{Introduction}

Depending on conditions (thermodynamic variables), such as temperature and density, matter can appear in many forms (phases).
Typical phases include solid, liquid, and gas; but many others can exist, such as plasmas, condensates, and superconducting phases (just to name a few).  How matter transitions from one phase to another can also take many forms.  A first-order phase transition is how water typically changes from solid to liquid or liquid to gas wherein the phase transition happens at a fixed temperature, free energy, and pressure, which leads to dramatic changes in certain thermodynamic properties (e.g., a jump in the density).  At extremely large temperatures and pressures, for water, a crossover phase transition is reached between the liquid and gas phases: depending on what thermodynamic observable one looks at, the substance could look more like a liquid or a gas.  In other words, the phase transition no longer takes place at a fixed temperature, free energy, and pressure, but rather across a range of them.  Finally, bordering these two regimes, there exists a critical point that separates a crossover phase transition from a first-order one.  
To describe these different phases of matter, one requires an equation of state (EoS) that depends on the thermodynamic variables of the system.  One should note, however, that the EoS is an equilibrium property, and, of course, out-of-equilibrium effects can also be quite relevant.  For instance, imagine a body of water that is flowing and being cooled at the same time.  In such a dynamical system, one also requires information about the transport coefficients in order to properly describe its behavior as it freezes. 

In this work, we will concern ourselves with phases of matter that appear at high energy, relevant when studying the strong force.  This is the force that binds together the nucleus, and leads to the generation of 99\% of the visible matter in the universe.  The theory that governs the strong force is quantum chromodynamics (QCD, \citealt{Gross:1973id,Politzer:1973fx}).  QCD describes the interactions of the smallest building blocks of matter (quarks and gluons).  Quarks and gluons are normally not free (or ``deconfined'') in Nature, but rather confined within hadrons. The latter comprise mesons (quark anti-quark pairs $q\bar{q}$), baryons (three-quark states $qqq$), or anti-baryons (three anti-quark states $\bar{q}\bar{q}\bar{q}$)\footnote{ 
Several pentaquark ($qqqq\bar{q}$) and tetraquark ($qq\bar{q}\bar{q}$) states have also been observed in the past two decades but are not directly relevant to this work; we refer the reader to \cite{Chen:2022asf} for an extended review of this topic.}. The quark content and their corresponding quantum numbers (see Table~\ref{tab:quarks}) yield the quantum numbers of the hadron itself.  One can calculate the thermodynamic properties of strongly interacting matter using either lattice QCD in the non-perturbative regime, or perturbative QCD (pQCD) where the coupling is small (high temperatures and/or extremely high densities).

\begin{table}[h!]
    \caption{Summary table of quarks including flavor, mass, and quantum numbers \citep{Workman:2022ynf}.}
    \centering
    \begin{tabular}{ccccccc}
    \hline \hline
      Flavor & Mass & Charge & Baryon  &  Spin &  Isopsin  & Strangeness\\
      &(MeV)&(e)&number&&(z-projection)\\
      \hline \hline
      Up~($u$) & $2.16^{+0.49}_{-0.26}  $  & $+\frac{2}{3}$ & $\frac{1}{3}$ & $\frac{1}{2}$ & $\frac{1}{2}$ ($+\frac{1}{2}$) & 0\\
      Down~($d$) & $4.67^{+0.48}_{-0.17}  $  & $-\frac{1}{3}$ & $\frac{1}{3}$ & $\frac{1}{2}$ & $\frac{1}{2}$ ($-\frac{1}{2}$)& 0\\
      Strange~($s$) & $93.4^{+8.6}_{-3.4}  $  & $-\frac{1}{3}$ & $\frac{1}{3}$ & $\frac{1}{2}$ & 0& -1\\
      Charm~($c$) & $1270 \pm 20  $  & $+\frac{2}{3}$ & $\frac{1}{3}$ & $\frac{1}{2}$ & 0& 0\\
      Bottom~($b$) & $4180^{+30}_{-20}  $  & $-\frac{1}{3}$ & $\frac{1}{3}$ & $\frac{1}{2}$ & 0& 0\\
      Top~($t$) & $1.73\times10^5 \pm 300  $  & $+\frac{2}{3}$ & $\frac{1}{3}$& $\frac{1}{2}$ & 0& 0\\
      \hline \hline
    \end{tabular}
    \label{tab:quarks}
\end{table}

Protons ($uud$ quark state), neutrons ($udd$ quark state), and, in rare cases, hyperons (baryons with strange quark content) form nuclei, the properties of which depend on the number of nucleons $A$, as well as the number of protons $Z$ and the number of neutrons $A-Z$ within the nucleus\footnote{In the rare case of hypernuclei, one must also consider the number of hyperons $H$ such that the total number of neutrons within the nucleus is $A-Z-H$.}. In principle, QCD also drives the properties of nuclei.  However, in the vast majority of cases, it would not be convenient to calculate the properties of nuclei or nuclear matter (beyond densities and temperatures at which nuclei dissolve into a soup of hadrons) directly from the Lagrangian of QCD, both because of the numerical challenges but also because it would not be the most effective way (it would be akin to calculating the properties of a lake from the microscopic interactions of H$_2$O molecules).  The objective of this review article is to put together the constraints derived from fundamental theories (that are gauge invariant and renormalizable) and observations. For this reason, we do not discuss relativistic mean-filed models. We also opted not to incorporate the different approaches used to describe nuclear matter, and refer instead to an excellent review on this subject \citep{Oertel:2016bki}. Two different approaches are generally used to obtain the equation of state of nuclear matter from these constraints: ab-initio many-body methods using realistic interactions (these include Green Function methods, variational and  Monte Carlo methods, (Dirac)--Brueckner--Hartree--Fock calculations and an example is the well known Akmal, Pandharipande, and Ravenhall EoS \citep{Akmal:1998cf}) or phenomenological approaches based in density functional theories applying effective interactions, including relativistic mean-field models with meson exchange forces and non-relativistic Skyrme and Gogny forces, see \cite{Oertel:2016bki} for a  review.
Using these methods, one can calculate thermodynamic quantities at low temperatures (on the MeV scale) and around nuclear saturation density, $n_{\sat}$,  which represents the point on the saturation curve where the binding energy per nucleon in a nucleus is at its lowest,  indicating a balance between the attractive and repulsive nuclear forces, therefore, maximal stability within the nuclear system \citep{Haensel:1981p}.

How can we solve QCD and study nuclear matter theoretically?
How can we probe QCD and nuclear matter experimentally? What systems in Nature and in the laboratory are sensitive to quarks and gluons, hadrons, or nuclei?  At large temperatures and vanishing net baryon densities $n_B=0$ (i.e., the same amount of baryons/quarks and anti-baryons/anti-quarks), the conditions are the same as those of the early universe and can be reproduced in the laboratory, at the Large Hadron Collider (LHC, \citealt{Citron:2018lsq}) and at the  Relativistic Heavy Ion Collider (RHIC, \citealt{STARnote,Cebra:2014sxa}) for top center-of-mass beam energies $\sqrt{s_{NN}}=200$ GeV. In equilibrium, lattice QCD can be used to calculate the EoS, which can be extended to finite $n_B$ using expansion schemes up to baryon chemical potentials (over temperature) of about $\mu_B/T\sim 3.5$.  Medium to low energy RHIC collisions explored in the Beam Energy Scan (BES) phase I and II ($\sqrt{s_{NN}}=7.7-200$ GeV in collider mode), as well as existing and future fixed target experiments at RHIC \citep{STARnote,Cebra:2014sxa}, SPS \citep{Pianese:2018xib}, HADES \citep{Galatyuk:2014vha,Galatyuk:2020lvg}, and FAIR \citep{Friese:2006dj,Tahir:2005zz,PANDA:2009yku,Durante:2019hzd} can reach temperatures in the range $T\sim 50-350$ MeV and baryon chemical potentials $\mu_B\sim 20 -800$ MeV, using a range of center of mass beam energies $\sqrt{s_{NN}}\sim 2-11$ GeV. Therefore, these low-energy experiments provide a significant amount of information that can also be used to infer the EoS \citep{Dexheimer:2020zzs,Lovato:2022vgq,Sorensen:2023zkk}.  However, these systems are probed dynamically and may be far from equilibrium, so one must not consider the EoS extracted from heavy-ions as data in the typical sense, but rather as a posterior model that is sensitive to priors and systematic uncertainties that may exist in that model. In the high temperature and/or chemical potential limit, systematic methods such as perturbative resummations can be used to calculate the EoS analytically directly form the QCD Lagrangian.

Low-energy nuclear experiments provide methods to extract key properties of nuclei. Most stable nuclei are composed of ``isospin-symmetric nuclear matter'', i.e $Z=0.5\ A$, such that the number of protons and neutrons are equal in the nucleus. For simplicity, one defines the charge fraction $Y_Q=Z/A$, which can also be related to the charge density $n_Q$ (assuming a system of only hadrons, no leptons) over the baryon density $n_B$ such that $Y_Q=n_Q/n_B$ as well. Then, for symmetric nuclear matter $Y_Q=0.5$ and this is where most nuclear experiments provide information.  However, heavy nuclei do become more neutron rich, such that $Y_Q\sim 0.4$. Note that, for the highest energies, heavy-ion experiments only probe $Y_Q=0.5$ as the nuclei basically pass through each other, and the fireball left behind cannot create net isospin ($Y_Q\ne0.5$) or strangeness ($Y_S=S/A=n_S/n_B\ne0$) during the very brief time of the collision (on the order of $\sim 10$ fm/c or $10^{-23}$ s). 

All thermodynamic properties change as $Y_Q$ varies. This can be measured experimentally in low-energy nuclear experiments around $n_{\sat}$ through the determination of the symmetry energy $E_{\sym}$, 
 which can be approximated as the difference between the energy per nucleon of $Y_Q=0$ (pure neutron matter $E_{\rm{PNM}}$) and $Y_Q=0.5$ matter (symmetric nuclear matter $E_{\rm{SNM}}$)\footnote{
The general definition of the symmetry energy is 
$E_{\sym}\equiv \frac{1}{2} {\partial^2 (E / N_B)}/{\partial \beta^2}$, where $E/N_B$ is the energy per baryon and  $\beta\equiv (n_n-n_p)/(n_n+n_p)$ in terms of neutron stars or $\beta = 1 - 2\,Y_Q$ \citep{Bombaci:1991zz,Haensel:1977,Muller-Kirsten:1998ijv}. 
Eq.~\eqref{eq:EsymSNMPNM}, commonly found in the literature, only shows terms up to second order in the expansion.}
\begin{equation}\label{eq:EsymSNMPNM}
    E_{\sym}\equiv \frac{E_{\rm{PNM}}-E_{\rm{SNM}}}{N_B}\ .
\end{equation}
The baryon number $N_B$ is more comprehensive than $A$, as it also includes quarks, with $N_B=1/3$. At saturation density, many other quantities can be determined such as the binding energy per nucleon, or the (in)compressibility of matter, in addition to $n_{\sat}$ itself.
At small $Y_Q$, matter in neutron stars provides information about both nuclear and QCD matter at low temperatures and medium-to-high densities. 
 Matter in this case is necessarily charge-neutral, as $Y_Q=Y_{lep}$, the charge fraction of leptons (electrons and muons). On the other hand, weak(-force) equilibrium ensures $\mu_Q=-\mu_e=-\mu_\mu$, meaning that the charge chemical potential, the difference between the chemical potential of protons and neutrons (in the absence of hyperons), or up and down quarks, equals the ones of electrons and muons.
 
 At saturation densities, a neutron star's internal composition is primarily made up of nucleons and leptons. However, as the density increases, other baryonic species may appear due to the rapid rise in baryon chemical potential associated with a higher density and reduce the ground state energy of the dense nuclear matter phase by opening new Fermi channels. 
Due to the long time-scales involved (when compared to weak interactions), matter in neutron stars can also include particles with net strangeness, hyperons. Here on Earth, hyperons can be produced but are unstable and quickly decay in $\sim 10^{-8}$ seconds via weak interactions into protons and neutrons. In the high density regime in the core of neutron stars, hyperons cannot decay back to nucleons due to Pauli blocking, meaning that producing additional nucleons would increase the energy of the system \citep{Joglekar:2020liw,Blaschke:2020qrs}. However, the appearance of hyperons softens the EoS of dense matter and lowers the maximum mass $M_{\max}$ of neutron stars predicted by a given theory, which is incompatible with the observations of massive stars, see Sect.~\ref{sec:astro}. This mismatch between   experimental observations  and theoretical calculations is referred to as the hyperon puzzle \citep{Bednarek:2011gd,Buballa:2014jta}. To make them compatible, additional repulsion is needed in the theory so that the EoS becomes stiffer. This additional effect can be introduced through the following known mechanisms, (i) hyperon-hyperon interaction via exchange of short-range vector mesons \citep{Rijken:2016uon}, (ii) three body repulsive hyperonic force \citep{Lonardoni:2014bwa,Gerstung:2020ktv,Logoteta:2019utx}, (iii) higher-order vector interactions \citep{Bodmer:1991hz,Dexheimer:2020rlp}, (iv) excluded volume for hadrons \citep{Hagedorn:1982qh,Dexheimer:2012eu}, and (iii) a phase transition to quark matter at a density less than or around the hyperon threshold \citep{Vidana:2005mx}. 

On the other hand, the generation of heavier non-strange baryons (resonances) in the core of neutron stars is still an open question \citep{Weissenborn:2011ut}. Initially \citep{Glendenning:1984jr}, it was argued that resonances appear at much higher densities beyond the density of a neutron star core and, thus, they are not relevant for nuclear astrophysics. Nevertheless, an early appearance of $\Delta$-baryons at 2--3 $n_{\sat}$ was obtained in several works \citep{Schurhoff:2010ph,Drago:2014oja,Li:2018qaw,Marquez:2022fzh}. It was shown that, due to the isospin rearrangement that takes place when the $\Delta$'s appear, they do not produce an effect analogous to the hyperon puzzle and are able to replace baryons without clashing with $M_{\max}$ constraints, producing smaller stars in better agreement with observations \citep{Dexheimer:2021sxs}.

\section{Executive summary}

In this work, we discuss theoretical and experimental constraints for dense and hot matter, including astrophysical observations.  For theoretical constraints, we restrict ourselves to those that are derived directly from first principles in particular regimes, where lattice QCD or pQCD calculations are possible, as well as from $\chi$EFT also in a particular regime, where it can be considered as the low-energy theory of QCD.  For experimental constraints, we focus on measurements and, whenever possible, avoid mentioning quantities inferred from data.  For example, by using yields of identified particles in heavy-ion collisions, it is possible to infer the temperature and baryon chemical potential at the point of chemical freeze-out.\footnote{Due to the rapid expansion and cooling of the quark-gluon plasma produced in heavy-ion collisions, at a point (chemical freeze-out) following the (pseudo)phase transition where quark and gluons have combined into hadrons, the particles become so far apart that chemical reactions are not longer possible. A second point (kinetic freeze-out) at even lower temperatures occurs (later in the reaction), where the particles become more dilute and kinetic reactions are no longer possible.  It is generally believed that chemical freeze-out occurs near the quark deconfinement transition and can be used as a (close but not precise) proxy for the phase transition line.}  However, the extracted $\left\{T,\mu_B\right\}$ at fixed $\sqrt{s_{NN}}$ and centrality are dependent on a number of assumptions such as the particle list, decay channels, decay widths, how interactions are described (if at all), etc.  Thus, we only provide the hadron yields measured directly from experiments and not the thermodynamic quantities inferred from them, which are model dependent. 

In the case of experimental low-energy nuclear results, the use of quantities inferred from data is unavoidable. Due to the importance of those results, we discuss them, while highlighting relevant dependencies.
For astrophysical observations, posteriors are extracted from a combination of measured data and modeling where the systematic uncertainties are carefully taken into account. Nonetheless, there are certain caveats when one considers these posteriors that we would be remiss not to discuss.  This context is important for theorists to understand before making comparisons between tidal deformabilities posteriors extracted from gravitational waves, mass-radius posteriors from NICER X-ray observations, and mass and/or radius extractions from other types of X-ray observations.

\subsection{Theoretical constraints: lattice QCD}
At vanishing $n_B$ or, equivalently (at finite temperature), $\mu_B=0$, lattice QCD calculations reliably provide the EoS for $T\gtrsim 125$ MeV. They rely on solving QCD numerically on a very large grid of points in space and time. In this case, it has been determined that the change of phase between a hadron resonance gas (HRG) at low temperatures into a quark-gluon plasma at high temperatures is a smooth crossover. At finite $\mu_B$, the exponential of the QCD action becomes complex and cannot be used as a weight for the configurations generated in Monte Carlo simulations, which is known as the sign problem \citep{Troyer:2004ge,Dexheimer:2020zzs}. However, expansions around $\mu_B=0$ allow one to obtain the lattice QCD EoS up to a chemical potential dependent on temperature $\mu_B\sim 3.5\, T$ \citep{Borsanyi:2021sxv,Borsanyi:2022qlh}. Furthermore, lattice QCD results can also constrain the hadronic spectrum through partial pressures \citep{Alba:2017mqu} and provide insight into strangeness-baryon number interactions using cross-correlators \citep{Bellwied:2019pxh}. Despite these successes, the expanded lattice QCD EoS cannot reach temperatures and densities relevant to low-energy heavy-ion collisions and neutron stars. 

\subsection{Theoretical constraints: perturbative QCD}

Although quarks are never truly free, due to asymptotic freedom, the coupling strength of the strong force ($\alpha_s$) decreases logarithmically with energy and, more importantly, in a deconfined medium, Debye screening reduces the effective interaction between quarks and gluons.  As a result, as the temperature and/or chemical potential(s) involved become large, one finds that perturbation theory becomes applicable and analytic calculations of the perturbative QCD (pQCD) EoS become reliable \citep{Andersen:2009tc,Andersen:2010ct,Andersen:2010wu,Andersen:2011ug,Mogliacci:2013mca,Haque:2013sja,Haque:2014rua,Haque:2020eyj,Ghiglieri:2020dpq}. This occurs at $T \gtrsim 300$ MeV at $\mu_B=0$ \citep{Haque:2013sja,Haque:2014rua,Haque:2020eyj} and at $n_B\gtrsim 40~n_{\sat}$ at $T=0$ \citep{Andersen:2002jz}. 
In the latter case, note however that causality and stability bounds allow pQCD to be applied at lower densities \citep{Komoltsev:2021jzg}.  In practice, achieving agreement between perturbative QCD and lattice QCD requires resummations at all orders.  The two main methods for accomplishing such resummations are effective field theory methods \citep{Braaten:1995cm,Braaten:1995jr} and hard-thermal-loop perturbation theory \citep{Andersen:2002ey,Andersen:2003zk}.  Both resummation schemes have been extended to N2LO (next-to-next-to leading order) in their respective loop expansions at $\mu_B=0$.  At finite chemical potential, N2LO \citep{Freedman:1976xs,Freedman:1976dm,Freedman:1976ub} and partial N3LO (next-to-next-to-next-to leading order) results are available \citep{Gorda:2018gpy,Gorda:2021znl,Gorda:2021kme}.

\subsection{Theoretical constraints: chiral effective field theory}

Chiral effective field theory ($\chi$EFT)
offers a systematic, model-independent framework for investigating the characteristics of hadronic systems at the low energies relevant for nuclear physics where the particle momentum is similar to the pion mass ($p \sim m_\pi$) with quantified uncertainties \citep{Weinberg:1978kz,Epelbaum:2008ga,Machleidt:2011zz,Drischler:2021kxf}. $\chi$EFT starts from the most general Lagrangian that is consistent with the symmetries, in particular the spontaneously broken chiral symmetry, of low-energy QCD with nucleons and pions as degrees of freedom. The theory offers an order-by-order expansion for two-nucleon and multi-nucleon interactions whose long-range features are governed by pion-exchange contributions constrained by chiral symmetry and whose short-distance details are encoded in a set of contact interactions with strengths fitted to two- and few-body scattering and bound-state data. Theoretical uncertainties can be estimated by examining the order-by-order convergence of the $\chi$EFT expansion,
% and by varying the resolution scale at which nuclear dynamics are resolved, 
a feature that provides a crucial benefit over phenomenological approaches \citep{Drischler:2021kxf}. Significant advances in the application of Bayesian statistical methods
% efforts across multiple groups 
have led to robust uncertainty quantification in calculations of the EoS up to fourth order in the chiral expansion that are applicable in the range of $n_B\lesssim 2~n_{\sat}$  \citep{Hebeler:2010jx,Sammarruca:2014zia,Tews:2018kmu,Drischler:2020hwi}.

\subsection{QCD phase diagram}

\begin{figure}[h!]
\includegraphics[scale=0.4]{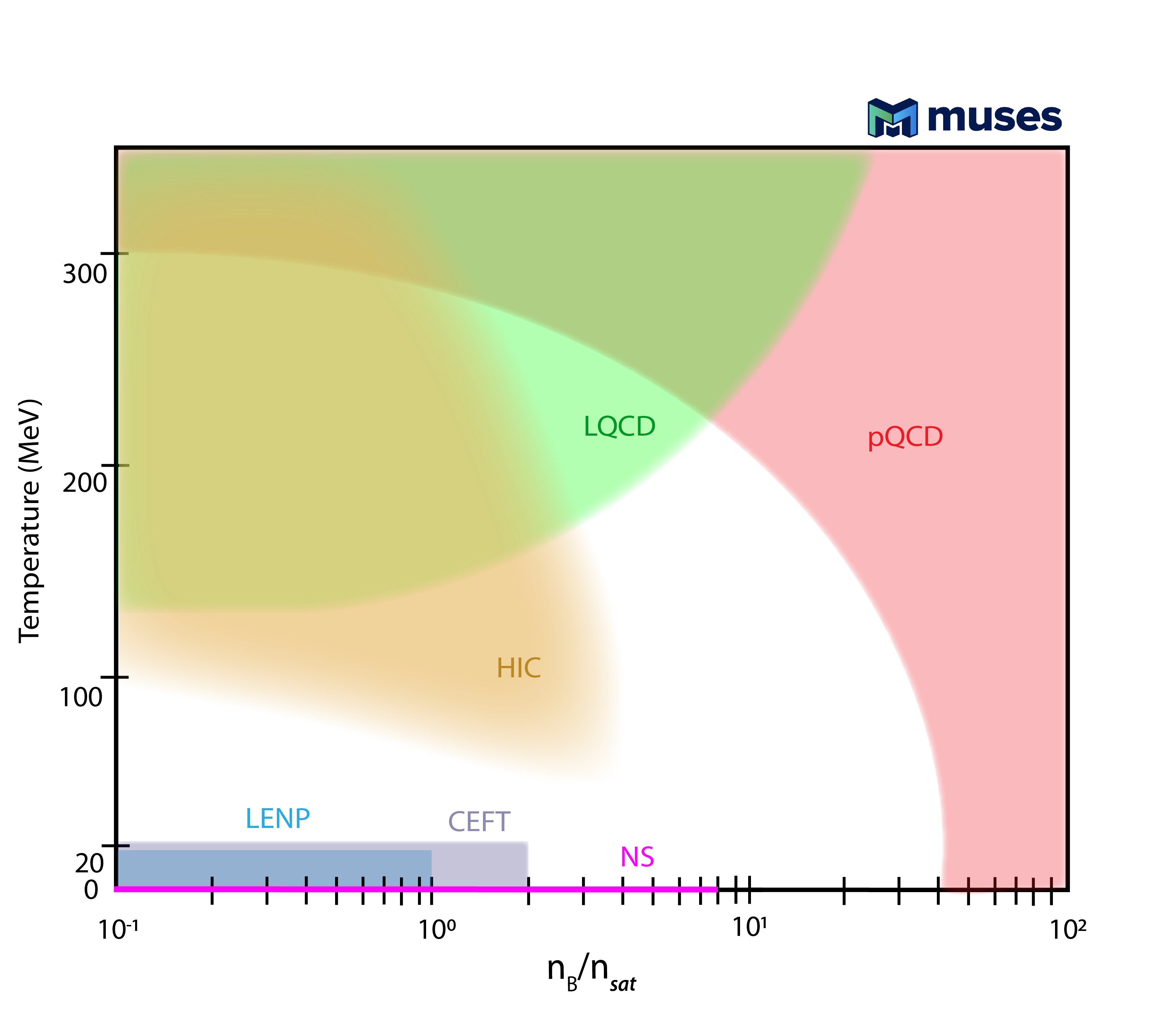}
\centering
\caption{Regions of the QCD phase diagram where constraints from heavy-ion collisions (HIC), lattice QCD (LQCD), perturbative QCD (pQCD), low-energy heavy-ion collisions (LENP), chiral effective field theory ($\chi$EFT), and astrophysics (neutron stars, NS) are available.}
\label{fig:phase diagram}
\end{figure}

From a combination of lattice QCD results (at $T\geq 135$ MeV and $\mu_B/T\leq 3.5$), pQCD calculations (limits include $T\gtrsim 300$ MeV at $\mu_B/T\lesssim 1$ and at $n_B\gtrsim 40~n_{\sat}$ at $T=0$), and $\chi$EFT bands ($T\lesssim20$ MeV and $n_B\lesssim 2~n_{\sat}$) we now have three theoretical points of reference (or rather regimes) in the QCD phase diagram, see Fig.~\ref{fig:phase diagram}. Effective models -- e.g., chiral models \citep{Nambu:1961tp,Nambu:1961fr,Hatsuda:1994pi,Dexheimer:2009hi,Motornenko:2019arp} and holography \citep{Rougemont:2017tlu,Critelli:2017oub,Grefa:2021qvt,Grefa:2022sav,Demircik:2021zll,Kovensky:2023mye} -- some describing the microscopic degrees of freedom and their interactions, are used  to connect these regimes in the phase diagram and even propose entirely new phases of dense and hot matter. These models are fixed to be in agreement with theoretical and experimental (low-energy nuclear physics, heavy-ion collisions, and astrophysics) results in the relevant regimes.

\subsection{Experimental constraints: heavy-ion collisions}
\label{sub:HIC_exec_summary}

In the laboratory, heavy-ion collisions probe finite temperatures in the range of $T\sim 50-650$ MeV, depending on the center of mass beam energy $\sqrt{s_{NN}}$, such that higher $\sqrt{s_{NN}}$ probe high temperatures and lower $\sqrt{s_{NN}}$ probe lower temperatures. The temperature and density of the system vary in space and time throughout the evolution, which is the hottest at early times. Depending on the choice of the experimental observables, one can obtain information at different temperatures and densities within the collisions. The final distribution of hadrons reflects the temperature and chemical potentials at chemical freeze-out (although certain momentum dependent observables are also sensitive to the kinetic freeze-out, see e.g. \citealt{STAR:2017sal} for a comparison between chemical and kinetic freeze-out).

When temperatures are high enough (i.e., high $\sqrt{s_{NN}}$) for a quark-gluon plasma to be produced, such that hydrodynamics is a good dynamical description, lowering $\sqrt{s_{NN}}$ corresponds to a lower initial temperature, a lower freeze-out temperature, and a larger $n_B$. However, for very low beam energies, below  $\sqrt{s_{NN}}\lesssim 4-7$ GeV, the hadron gas phase dominates, such that hadron transport models may be used. This then means  that higher $\sqrt{s_{NN}}$ reaches larger $n_B$ whereas lower $\sqrt{s_{NN}}$ reach a smaller range of $n_B$. The exact switching point from a quark-gluon plasma dominated- to hadronic-dominated dynamical description is unknown and still hotly debated within the community. The initial collision temperature $T_0$ is model-dependent, so we do not include estimates for it in this work.  The freeze-out temperature, however, can be more directly extracted from experimental data (with certain caveats that we will explain here) using particle yields and assuming thermal equilibrium at freeze-out. Additionally, the emission of photons and lepton pairs (dileptons), which are immune to strong interactions and can traverse the QGP, can be used to extract average temperatures at different points in the heavy-ion collision evolution, which can be used to pin down the temperature evolution \citep{Strickland:1994rf,Schenke:2006yp,Martinez:2008di,Dion:2011pp,Shen:2013vja,Gale:2014dfa,Bhattacharya:2015ada,Ryblewski:2015hea,Paquet:2015lta,Kasmaei:2018oag,Kasmaei:2019ofu}. On the other hand, the extraction of $n_B$ is more model dependent. If a QCD critical point exists, then susceptibilities of the pressure will diverge exactly at the critical point and may have non-trivial behavior in the surrounding critical region \citep{Stephanov:2008qz,Parotto:2018pwx,Mroczek:2020rpm}. In equilibrium, these would determine the cumulants of the distribution of protons, such as the kurtosis. Measurements of the kurtosis \citep{STAR:2020tga,STAR:2022qmt,HADES:2020wpc,ALICE:2019nbs}, 6th-order cumulants \citep{STAR:2021rls}, and fluctuations of light nuclei \citep{STAR:2022hbp} exist from BES-I across a variety of beam energies with large statistical error bars.  BES-II \citep{Tlusty:2018rif} will significantly improve the measurement precision. However, the data has not yet been released. 

Looking to the future, the Compressed Baryonic Matter (CBM) Experiment at FAIR (GSI, Germany)  will be an experimental facility that will be dedicated to explore low beam energies in fixed target mode with high luminosity, i.e., with a high collision rate \citep{Spies:2022sfg}. CBM \citep{PANDA:2009yku,Durante:2019hzd} will allow us to constrain the EoS at high $\mu_B$ and moderate temperatures.
Eventually, from the wealth of experimental data in heavy-ion collisions, it will be possible to extract an EoS using model-to-data comparisons. However, that will require more sophisticated dynamical models that do not yet exist \citep{Bluhm:2020mpc}.
It has already been identified that the azimuthal anisotropies of the momentum distribution of particles in collisions, otherwise known as flow harmonics, are sensitive to the EoS at these low beam energies \citep{Danielewicz:2002pu,Spieles:2020zaa}.  However, there are still significant questions remaining about the correct dynamical model and other free parameters such as transport coefficients. Depending on the model assumptions, one can obtain radically different posteriors of the EoS, or find different EoSs consistent with the data at these beam energies (a few examples include comparing the different results and conclusions from  \cite{Danielewicz:2002pu,Spieles:2020zaa,Schafer:2021csj,Shen:2022oyg,Oliinychenko:2022uvy}). Thus, in this work, we will only include a discussion on some of the key experimental measurements but cannot yet clarify the precise implications of the data. 

\subsection{Experimental constraints: low-energy nuclear physics}

At significantly lower beam energies (approaching the $T\rightarrow 0$ limit) there are a number of experiments that probe dense matter.  These experiments study the properties of nuclei at (or near) saturation density.  While most nuclei contain symmetric nuclear matter such that $Y_Q\sim 0.5$, heavy nuclei become more neutron-rich and may reach $Y_Q\sim 0.4$, while unstable nuclei close to the neutron drip line have much smaller values of $Y_Q$.  However, neutron stars are composed of primarily asymmetric nuclear matter, with $Y_Q\sim 0.001-0.2$.  Thus, one can use a Taylor series to expand between symmetric and asymmetric matter, known as the symmetry energy expansion. In this case, a few of its coefficients can be inferred from experimental measurements of, e.g., the neutron skin.
Symmetric-matter properties include binding energy per nucleon, compressibility, and saturation density, and can also be inferred, together with the EoS. However, in this case there is model dependence which can be investigated by using different kinds of models.

In addition to the compressibility and the binding energy per nucleon, the effective mass of nucleons at saturation was shown to be important to study the nuclear EoS of hot stars \citep{Raduta:2021coc}, the EoS of neutron stars with exotic particles at finite temperature \citep{Raduta:2022elz}, thermal effects in supernovae \citep{Constantinou:2015mna}, and neutron-star mergers (see \citealt{Raithel:2019gws} and references therein). Nevertheless, the experimental determination of this quantity still includes large uncertainties and, therefore, will not be discussed in this review. 

Beyond nucleons, properties of hyperons and $\Delta$-baryons can also be determined for symmetric matter. The most useful observable to constrain effective models is optical potentials at $n_{\sat}$, which provides the result of the balance between attractive and repulsive strong interactions. At finite temperature, there is also data concerning the critical point for the liquid-gas phase transition \citep{Elliott:2012nr}, where nuclei turn into bulk hadronic matter.  

\subsection{Experimental constraints: astrophysics}

\begin{figure}[h!]
\includegraphics[width=\textwidth]{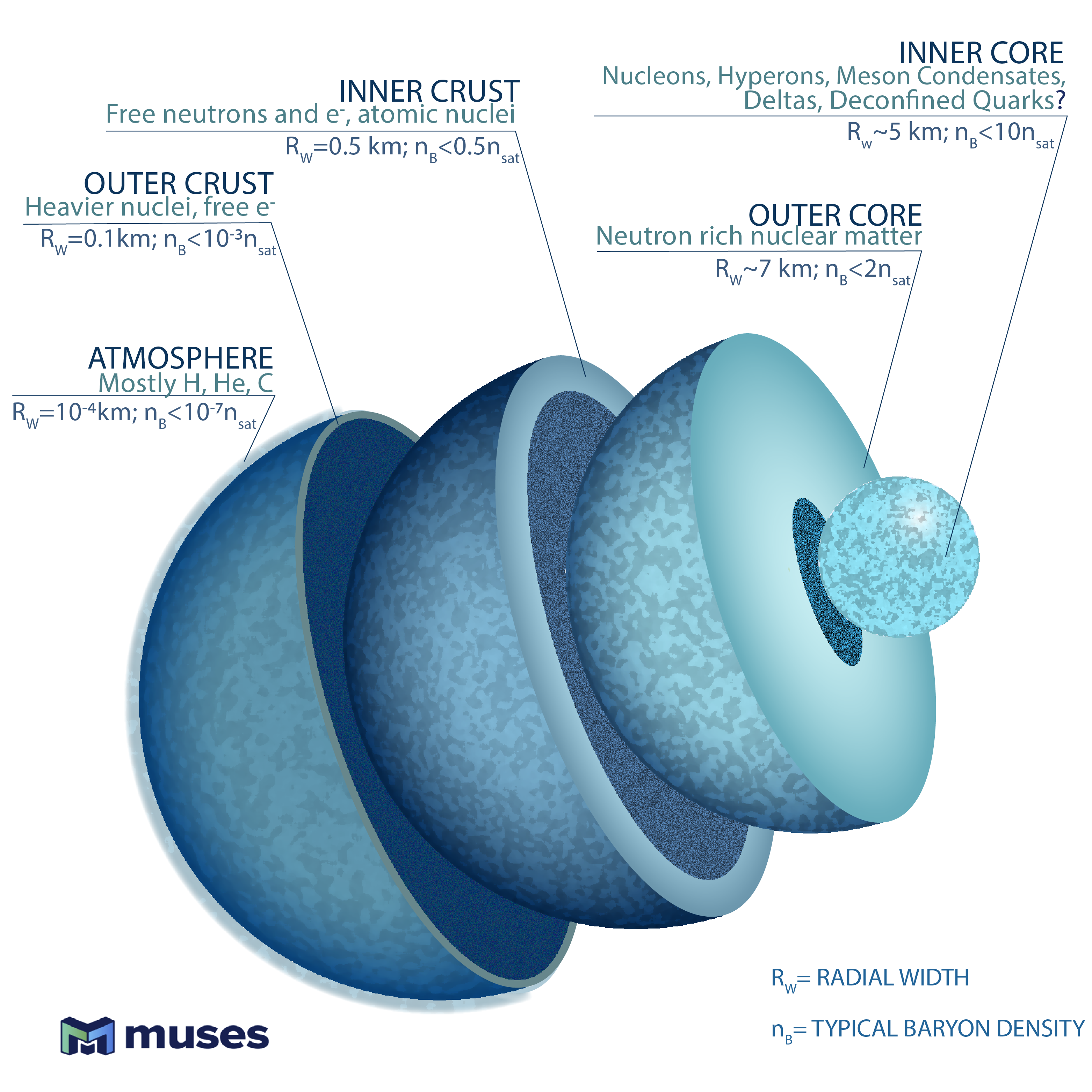}
\centering
\caption{Composition  of a  typical neutron star.}
\label{fig:neutronstar}
\end{figure}

The high baryon density inside neutron stars makes them a natural laboratory to understand strong interaction physics under conditions that are impossible to achieve in a laboratory setting.
Neutron stars are the end-life of massive stars, which run out of fuel for fusion and collapse gravitationally, violently exploding as supernovae.
As a result, the cores of the remnant neutron stars possess densities of the order of several times $n_{\sat}$. 
Neutron stars are stratified according to density, with different types of co-existing phases categorized according to the radial coordinate (see Fig.~\ref{fig:neutronstar}).
The outermost layer is the atmosphere, with a thickness of a few centimeters, which contains mostly hydrogen, helium, and carbon atoms.
 A little bit deeper, starts the
outer crust with a layer of $^{56}$Fe.  Electrons disassociate from specific nuclei and, moving towards the center of the star, nuclei become neutron richer and more massive, such that at the transition to the inner crust the nucleus $^{118}$Kr was determined as the  most stable within several theoretical models  \citep{Ruester:2005fm}. There are, however, some models that may predict slightly larger nuclei at the outer-inner crust transition.
In the inner crust, neutrons start to ``drip out'' of nuclei and, as a result, matter becomes a mixture of unbound electrons, unbound neutrons, and nuclei. 
At even higher densities in the outer core, above $\sim n_{\sat}$, matter turns into a neutron--rich ``soup" with no isolated nuclei. 
Going deeper into higher densities for the inner core, at about  $2~n_{\sat}$, hyperons, $\Delta$'s, and meson condensates may appear and, eventually, deconfined quark matter may form.
Regardless of the phase, fully evolved neutron stars fulfill chemical equilibrium and charge neutrality, either locally or globally, with mixtures of phases occurring in the latter case.

The EoS is related to microscopic equilibrium properties (pressure, energy density, etc.).
Therefore, the nuclear EoS is not directly comparable to astrophysical observations, but it serves as an important input in calibrated models to calculate experimental observables, such as mass-radius relationships of compact stars.
This is achieved by solving the Tolman--Oppenheimer--Volkoff (TOV) equations \citep{Tolman:1939jz,Oppenheimer:1939ne}, which are valid as long as rotational frequency ($\nu$) and magnetic field ($B$) effects are not significant.
These results can be compared with astrophysical observations from neutron-star electromagnetic emissions, usually radio and X-ray, and most recently, gravitational wave emission from neutron-star mergers.
In particular, observations from the National Radio Astronomy Observatory's Green Bank Telescope (GBT, \citealt{Demorest:2010bx}), NASA's Neutron Star Interior Composition Explorer (NICER, \citealt{Gendreau:2016kc,Baubock:2015ixa,Miller:2016kae,Ozel:2015ykl}), and NSF’s Laser Interferometer Gravitational-wave Observatory (LIGO) together with Virgo \citep{LIGOScientific:2017vwq,LIGOScientific:2017ync,Gendreau:2016kc} put strong constraints on the EoS \citep{Gendreau:2016kc,Annala:2017llu}.
Many EoS models have been updated since these observations were made, to be in agreement with observations \citep{Baym:2017whm}.

The most accurate neutron-star mass estimates come from the timing of radio pulsars in orbital systems with relativistic dynamical effects \citep{Antoniadis:2013pzd,NANOGrav:2019jur,Fonseca:2021wxt}; they inform the EoS insofar as they set a lower bound on the maximum mass it must be able to support against gravitational collapse.
The most reliable radius measurements stem from X-ray pulse profile modeling of rotating neutron stars \citep{Miller:2019cac,Riley:2019yda,Miller:2021qha,Riley:2021pdl}, and constrain the mass-radius relation predicted by the EoS.
Meanwhile, gravitational-wave observations of merging neutron stars constrain the EoS via their tidal deformability \citep{LIGOScientific:2017vwq,LIGOScientific:2018cki,LIGOScientific:2018hze,LIGOScientific:2020aai}.
If the gravitational waves are accompanied by a kilonova counterpart, as was the case for the binary neutron-star merger GW170817 \citep{LIGOScientific:2017ync}, the lightcurve and spectrum of the electromagnetic emission, as well as its implications for the fate of the merger remnant, also inform the EoS \citep{Bauswein:2017vtn,Margalit:2017dij,Radice:2017lry,Rezzolla:2017aly,Ruiz:2017due,Shibata:2017xdx}. Eventually, upgraded gravitational wave detectors will also be able to detect the post-merger signal \citep{Carson:2019rjx} (the post-merger starts at the point where the two neutron stars touch). This signal is also sensitive to finite temperature effects that may even reach temperatures and densities similar to heavy-ion collisions \citep{HADES:2019auv}, potential out-of-equilibrium effects due to the long-time scales associated with weak interactions \citep{Alford:2017rxf,Alford:2019kdw,Alford:2019qtm,Alford:2021lpp,Gavassino:2020kwo,Celora:2022nbp,Most:2022yhe}  as well as  deconfinement to quark matter \citep{Bauswein:2018bma,Most:2018eaw,Weih:2019xvw,Blacker:2020nlq,Tootle:2022pvd,Constantinou:2021hba}.

\subsection{Organization of the paper}

This review paper aims at compiling  up-to-date constraints from high-energy physics, nuclear physics, and astrophysics that relate to the EoS and are, therefore, fundamental to the understanding of current and future data from heavy-ion collisions to gravitational waves, making them relevant to a very broad community.  Additionally, precise knowledge of the dense and hot matter EoS can help physicists to look beyond the standard model either for dark matter, which may accumulate in or around neutron stars, or for modified theories of gravity. 
 
The paper is organized as follows: we first discuss the theoretical constraints of lattice (Sect.~\ref{sec:lattice}) and perturbative QCD  (Sect.~\ref{sec:pqcd}), followed by $\chi$EFT (Sect.~\ref{sec:eft}). 
Then, we discuss experimental constraints from heavy-ion collisions   (Sect.~\ref{sec:hic}), (isospin symmetric and asymmetric) low-energy nuclear physics  (Sect.~\ref{sec:nuclear}), and astrophysical observations (Sect.~\ref{sec:astro}). We provide a future outlook in Sect.~\ref{sec:outlook}, since a significant amount of new data is anticipated over the next decade.

%%%%%%%%%%%%%%%%%%%%%%%%%%%%%%%%%%%%%%%%%%%%%%%%%%%%%%%%%%%%%%%%%%%%%%%%%%%%%%%%%%%%%%%%%%%%%%%%%%%%%%%%%%%%%%%%%%%%%%%%%%%%%%%%%%%%%%%%%%%%%%%%%%%%%%%%%%%%%%%%%%%%%%%%%%%%%%%%%%%%%%%%%%%%%%%%%%%%%%%%

\section{Theoretical constraints: lattice QCD}
\label{sec:lattice}

Lattice QCD is the most suitable method to study strong interactions around and above the deconfinement phase transition region in the QCD phase diagram, due to its non-perturbative nature \citep{Drischler:2019xuo}. As discussed in the introduction, due to the sign problem, first-principles lattice QCD results for the EoS at finite $\mu_B$ are currently restricted. Since direct lattice simulations at $\mu_B=0$ and imaginary $\mu_B$ are feasible, observables can be extrapolated using techniques involving zero or imaginary chemical potential simulations, i.e., analytical continuation, Taylor series and other alternative expansions. In this section, we  present various constraints on the EoS, BSQ (baryon number, strangeness, and electric charge) susceptibilities, and partial pressures evaluated using lattice QCD.

\subsection{Equation of state}

\begin{figure}[h!]
\includegraphics[trim={0 -.5cm 0 0.3cm},clip,scale=0.49]{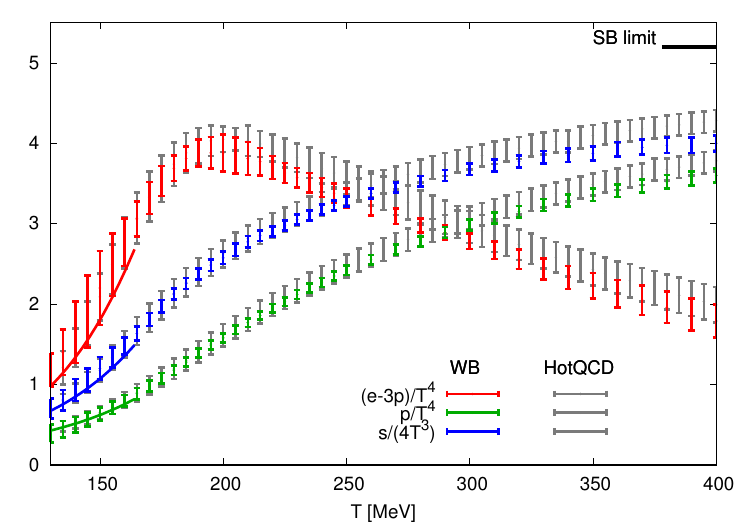}
\includegraphics[trim={0 0 0 0},clip,scale=0.87]{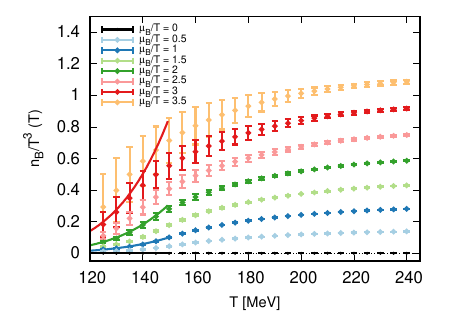}
\centering
\caption{Left: Comparison between the lattice QCD EoS at $\mu_B=0$ from the WB collaboration \cite{Borsanyi:2013bia} (colored points) and the HotQCD one \citep{HotQCD:2014kol} (gray points). Right: Baryonic density as a function of the temperature, for different values of $\mu_B/T$. At low temperature, full lines show results from the HRG model. Images reproduce with permission from [left] \cite{Ratti:2018ksb}, copyright by IOP, and [right] from \cite{Borsanyi:2021sxv}, copyright by the author(s). }
\label{figure:EoS_Lattice}
\end{figure}

In \cite{Borsanyi:2013bia} and \cite{HotQCD:2014kol}, the EoS was obtained in lattice QCD simulations at $\mu_B=0$. It was found that the rigorous continuum extrapolation results for 2+1 quark flavors are perfectly compatible with previous continuum estimates based on coarser lattices \citep{Aoki:2005vt,Borsanyi:2010cj}. The obtained pressure, entropy density, and interaction measure are displayed in the left panel of Fig. \ref{figure:EoS_Lattice} alongside the predictions of the hadron resonance gas (HRG) model  \citep{Venugopalan:1992hy} at low temperatures and the Stefan-Boltzmann (or conformal) limit of a non-interacting massless quarks gas at high $T$. They show full agreement with HRG results in the hadronic phase, and reach about 75\% of the Stefan--Boltzmann limit at $T\simeq 400$ MeV.

Furthermore, a Taylor series can be utilized to expand many observables to finite $\mu_B/T$
\begin{equation}
\frac{p(T,\mu_B,\mu_Q, \mu_S)}{T^4}=\sum_{i,j,k}\frac{1}{i!j!k!}\chi_{ijk}^{BQS}
\left(\frac{\mu_B}{T}\right)^i\left(\frac{\mu_Q}{T}\right)^j\left(\frac{\mu_S}{T}\right)^k\ ,
\end{equation}
where the susceptibilities $\chi_{ijk}^{BQS}$ are defined as follows 
\begin{equation}
\chi_{lmn}^{BSQ}=\frac{\partial^{l+m+n}(p/T^4)}{\partial(\mu_B/T)^l\partial(\mu_S/T)^m\partial(\mu_Q/T)^n}\ .
\label{eq:susceptibilities}
\end{equation}

They were obtained from lattice QCD calculations up to $\mathcal{O}(\mu_B/T)^4$ for the full series of coefficients and up to $\mathcal{O}(\mu_B/T)^8$ for some of the coefficients. 
The range of applicability of the Taylor expansion has recently been extended from $\mu_B/T\leq2$ \citep{Guenther:2017hnx,Bazavov:2017dus} to $\mu_B/T\leq2.5$ \citep{Bollweg:2022rps}. Isentropic trajectories in the $T-\mu_B$ plane have been extracted in \cite{Guenther:2017hnx}, for which the starting points are the freeze-out parameters at different collision energies at RHIC \citep{Alba:2014eba}. Strangeness neutrality and electric charge conservation were enforced by tuning the strange and electric charge chemical potentials, $\mu_S(\mu_B, T)$ and $\mu_Q(\mu_B,T)$, to reproduce the conditions $Y_S=0$ and $Y_Q=0.4$ \citep{Guenther:2017hnx}. 
     
A new expansion scheme for extending the EoS of QCD to unprecedentedly large baryonic chemical potential up to $\mu_B/T<3.5$ has been proposed recently \citep{Borsanyi:2021sxv}. The drawbacks of the conventional Taylor expansion approach, such as the challenges involved in carrying out such an expansion with a constrained number of coefficients and the low signal-to-noise ratio for the coefficients themselves, are significantly reduced in this new scheme \citep{Borsanyi:2021sxv}. In the hadronic phase, a good agreement is found for the thermodynamic variables with HRG model results. This scheme is based on the following identity
\begin{equation}
    \frac{\chi_1^B(T,\mu_B)}{\mu_B/T}=\chi_2^B(T',0)\ ,
\end{equation}
with
\begin{equation}
    T'(T,\mu_B)=T(1+\kappa_2^{BB}(T)\left(\frac{\mu_B}{T}\right)^2+\kappa_4^{BB}(T)\left(\frac{\mu_B}{T}\right)^4+\mathcal{O}((\mu_B/T)^6)\ .
\end{equation}

The baryonic density as a function of the temperature for different values of $\mu_B/T$ from \cite{Borsanyi:2021sxv} is shown in the right panel of Fig.~\ref{figure:EoS_Lattice}.
This extrapolation method was then generalized to include the strangeness neutrality condition \citep{Borsanyi:2022qlh}, which requires $\mu_S\ne0$. The extrapolation approach is devoid of the unphysical oscillations that afflict fixed order Taylor expansions at higher $\mu_B$, even in the strangeness neutral situation. Effects beyond strangeness neutrality are estimated by computing the baryon-strangeness correlator to strangeness susceptibility ratio  $\frac{\chi^{BS}_{11}}{\chi^S_2}$ (discussed in the following subsection) at finite real $\mu_B$ on the strangeness neutral line. This permits a leading order extrapolation in the ratio $R={\chi^S_1}/{\chi^B_1}$ \citep{Borsanyi:2022qlh}.

\subsection{BSQ susceptibilities}
\label{sub:susceptibilities}

Fluctuations of different conserved charges have been postulated as a signal of the deconfinement transition because they are sensitive probes of quantum numbers and related degrees of freedom. In heavy-ion collisions, one needs to relate fluctuations of net baryon number, strangeness, and electric charge with the event-by-event fluctuations of particle species. Non-diagonal correlators of conserved charges, like fluctuations, are useful for studying the chemical freeze-out in heavy-ion collisions. In thermal equilibrium, these correlators may be estimated using lattice simulations, and linked to moments of 
 event-by-event distributions of multiplicity (i.e., number of particles of a given species in some kinematic region, typically these are all charged particles) distributions.
They are defined as derivatives of the pressure with respect to the chemical potentials according to Eq.~\eqref{eq:susceptibilities}.
The quark number chemical potentials appear as parameters in the Grand Canonical partition function. The derivative of this function with respect to these chemical potentials yields the susceptibilities and the non-diagonal correlators of the quark flavors. Quark flavor chemical potentials can be related to the conserved charge ones through the following relationships:
$\mu_u=\frac13\mu_B+\frac23\mu_Q$, 
$\ \mu_d=\frac13\mu_B-\frac13\mu_Q$, and
$\ \mu_s=\frac13\mu_B-\frac13\mu_Q-\mu_S$.
 
\begin{figure}[h!]
\includegraphics[trim={0 .1cm .5 .4cm},clip,scale=0.54]{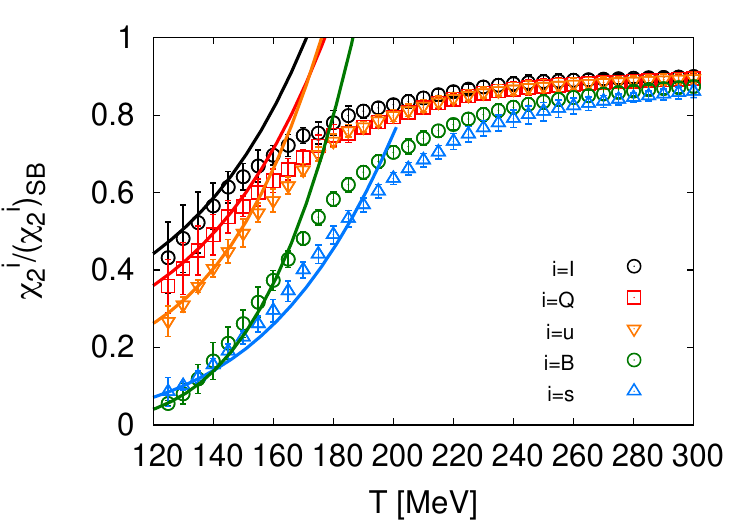}
\includegraphics[trim={0 =-.65cm .0 .0cm},clip,scale=0.481]{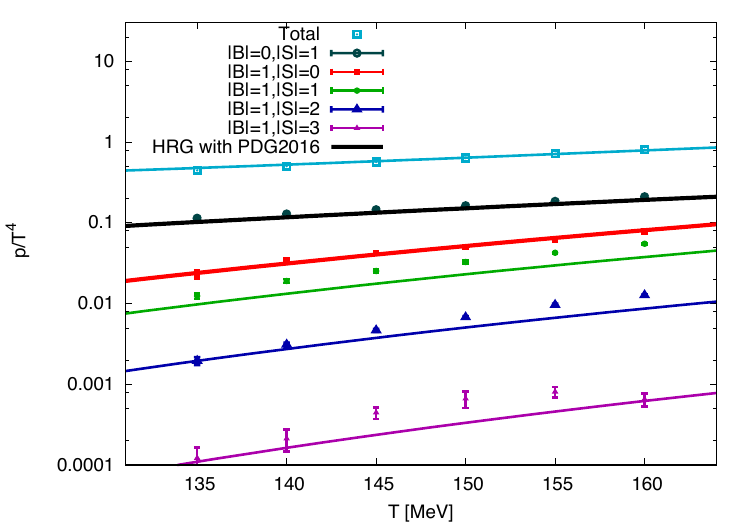}
\centering
\caption{Left: Baryon number, strange quark, electric charge, isospin number, and up quark susceptibilities as functions of the temperature at $\mu_B=0$. Right: Compilation of partial pressures for different hadron families as functions of the temperature. Images reproduced with permission from [left] \cite{Borsanyi:2011sw}, copyright by SISSA, and [right] from \cite{Alba:2017mqu}, copyright by APS.}
\label{figure:susc}
\end{figure}

For $T=125-400$ MeV and at $\mu_B=\mu_S=\mu_Q=0$, the Wuppertal-Budapest lattice QCD collaboration computed the non-diagonal (us) and  diagonal (B,s,Q,I,u) susceptibilities for a system of 2+1 staggered
quark flavors \citep{Borsanyi:2011sw}, where I stands for isospin. Selected susceptibilities are shown in the left panel of Fig.~\ref{figure:susc}. A Symanzik-improved gauge and a stout-link improved staggered fermion action were used in this analysis. The ratios of fluctuations were found, whose behavior may be recreated using hadronic observables, i.e. proxies, to compare either to lattice QCD findings or experimental observations \citep{Bellwied:2019pxh}.

Continuum extrapolated lattice QCD findings for $\chi_{2,2}^{u,s},~\chi_{2,2}^{u,d},~\chi_{1,1}^{u,d},~\chi_4^u,~\chi_4^B$ were presented in \cite{Bellwied:2015lba}. 
Second and fourth-order cumulants of conserved charges were constructed in a temperature range spanning from the QCD transition area to the region of resummed perturbation theory. It was found that, in the hadronic phase ($T \sim 130$ MeV), the HRG model predictions accurately reflect the lattice data, whereas in the deconfined region ($T \gtrsim 250$ MeV), a good agreement was found with three loop hard-thermal-loop (HTL) outcomes \citep{Bellwied:2015lba}.
Different diagonal  and non-diagonal fluctuations of conserved
charges are estimated up to sixth-order on a lattice size of 48$^3 \times$ 12 \citep{Borsanyi:2018grb}. Higher-order fluctuations at zero baryon/charge/strangeness chemical potential are calculated. The ratios of baryon-number cumulants as functions of $T$ and $\mu_B$ are derived from these correlations and fluctuations, which fulfill the experimental criteria of proton/baryon ratio  and strangeness neutrality and in turn, describe the observed cumulants as functions of collision energy from the STAR collaboration \citep{Borsanyi:2018grb}. Ratios of fourth-to-second order susceptibilities for light and strange quarks were presented in \cite{Bellwied:2013cta}. 

\subsection{Partial pressures}

Under the assumption that the hadronic phase can be treated as an ideal gas of resonances, and using lattice simulations, the partial pressures of hadrons were determined with various strangeness and baryon number contents. To explain the difference between the results of the HRG model and lattice QCD for some of them, the existence of missing strange resonances was proposed \citep{Bazavov:2013dta,Alba:2017mqu}. Note that partial pressures are only possible within the hadron resonance gas phase because i.) they require hadronic degrees-of-freedom and ii.) they are applicable under the assumption that the pressure can be written as separable components by the quantum number of the hadrons, i.e.,
\begin{eqnarray}
\label{eq:pressure}
P(\hmu_B,\hmu_S) &=& P^{BS}_{00}+P^{BS}_{10}\cosh(\hat{\mu}_B)+P^{BS}_{01} \cosh(\hat{\mu}_S) 
+ P^{BS}_{11} \cosh(\hat{\mu}_B-\hat{\mu}_S)
\nonumber \\
&+& P^{BS}_{12} \cosh(\hat{\mu}_B-2\hat{\mu}_S)+ P^{BS}_{13} \cosh(\hat{\mu}_B-3\hat{\mu}_S) \ ,
\end{eqnarray}
where the coefficients $P^{BS}_{ij}$ indicate the baryon number $i$ and strangeness $j$ of the family of hadrons for which the partial pressure is being isolated, and the dimensionless chemical potentials are written as $\hat{\mu}=\mu/T$.
The calculations were made feasible by taking imaginary values of strange chemical potential in the simulations. For strange mesons, more interaction channels should be incorporated into the HRG model, in order to explain the lattice data \citep{Alba:2017mqu}.
The right panel of Fig. \ref{figure:susc} shows a compilation of these partial pressures.

\subsection{Pseudo-phase transition line} 
\label{sub:pseudo-trans_line}

In a crossover, there is no sudden jump in the first derivatives of the pressure. Nevertheless, a ( chiral) pseudo-phase transition line can be calculated based on where the order parameters change more rapidly. 
The exact location of the QCD transition line is a hot topic of research in the field of strong interactions. The most recent results are contained in \cite{Borsanyi:2020fev}. The transition temperature, obtained from the chiral condensate and its susceptibility, as a function of the chemical potential can be parametrized as
\begin{equation}
\frac{T_c(\mu_B)}{ T_c(\mu_B=0)}=1
- \kappa_2 \left(\frac{\mu_B}{T_c(\mu_B)}\right)^2
- \kappa_4 \left(\frac{\mu_B}{T_c(\mu_B)}\right)^4+\dots\ .
\label{eq:kappa}
\end{equation}
The crossover or pseudo-critical temperature $T_c$ has been determined with extreme accuracy and extrapolated from imaginary up to real $\mu_B \approx 300$ MeV. Additionally, the width of the chiral transition and the peak value of the chiral susceptibility were calculated along the crossover line. Both of them are constant functions of $\mu_B$. This means that, up to $\mu_B=300$ MeV, no sign of criticality has been observed in lattice results. In fact, at the critical point the height of the peak of the chiral susceptibility would diverge and its width would shrink. The small error reflects the most precise determination of the $T-\mu_B$ phase transition line using lattice techniques. Besides $T_c=158 \pm 0.6$ MeV, the study provides updated results for the coefficients $\kappa_2=0.0153\pm0.0018$ and $\kappa_4=0.00032\pm0.00067$ \citep{Borsanyi:2020fev}.
Similar coefficients for the extrapolation of the transition temperature to finite strangeness, electric charge, and isospin chemical potentials were obtained in \cite{HotQCD:2018pds}, and are displayed in Table \ref{tab:kappa_SQI}. 

\begin{table}[h!]
    \caption{Continuum-extrapolated values of $\kappa_2^X$ and $\kappa_4^X$ (with  $\mu_Q=\mu_S=0$ for $X=B$, \\
    $\mu_B=\mu_Q=0$ for $X=S$ and $\mu_B=\mu_S=0$ for $X=Q,I$) from \cite{HotQCD:2018pds}.}
    \label{tab:kappa_SQI}
    \centering
    \begin{tabular}{c c|c c|c c|c c}
            \hline \hline
        $\kappa_2^B$ & $\kappa_4^B$ & $\kappa_2^S$ & $\kappa_4^S$ & $\kappa_2^Q$ & $\kappa_4^Q$ & $\kappa_2^I$ & $\kappa_4^I$ \\
        \hline \hline
        0.016(6) & 0.001(7) & 0.017(5) & 0.004(6) & 0.029(6) & 0.008(1) & 0.026(4) & 0.005(7)\\
                \hline \hline
    \end{tabular}
\end{table}
 
\subsection{Limits on the critical point location}

As mentioned in the previous subsection, in \cite{Borsanyi:2020fev}, by extrapolating the proxy for the transition width as well as the height of the chiral susceptibility peak from imaginary to real $\mu_B$, the strength of the phase transition was evaluated and no indication of criticality was found up to $\mu_B \approx$ 300 MeV. On the other hand, a phase transition temperature at $\mu_B=0$ of $T_c=132^{+3}_{-6}$ MeV was found in the chiral limit by the HotQCD collaboration using lattice QCD calculations  with ``rooted'' staggered fermions \citep{HotQCD:2019xnw}. This transition temperature is computed with two massless light quarks and a physical strange quark based on two unique estimators. Since the curvature of the phase diagram is negative, a critical point in the chiral limit would sit at a temperature smaller than this one. The expectation is that the temperature of the critical point at physical quark masses has to be smaller than the one of the critical point in the chiral limit, and therefore definitely smaller than $T_c=132^{+3}_{-6}$ MeV. 

%%%%%%%%%%%%%%%%%%%%%%%%%%%%%%%%%%%%%%%%%%%%%%%%%%%%%%%%%%%%%%%%%%%%%%%%%%%%%%%%%%%%%%%%%%%%%%%%%%%%%%%%%%%%%%%%%%%%%%%%%%%%%%%%%%%%%%%%%%%%%%%%%%%%%%%%%%%%%%%%%%%%%%%%%%%%%%%%%%%%%%%%%%%%%%%%%%%%%%%%
 
\section{Theoretical constraints: perturbative QCD}
\label{sec:pqcd}

It is possible to compute analytically the QCD EoS directly from the QCD Lagrangian using finite temperature/density perturbation theory.  
However, in thermal and chemical equilibrium, when $ T \gg \mu_i$ (with quark chemical potentials $\mu_i$), one finds that the naive loop expansion of physical quantities is ill-defined and diverges beyond a given loop order, which depends on the quantity under consideration.  In the calculation of QCD thermodynamics, this stems from uncanceled infrared (IR) divergences that enter the expansion of the partition function at three-loop order.  These IR divergences are due to long-distance interactions mediated by static gluon fields and result in contributions that are non-analytic in the strong coupling constant $\alpha_s = g^2/4\pi$, e.g., $\alpha_s^{3/2}$ and $\log(\alpha_s)$, unlike vacuum perturbation expansions which involve only powers of $\alpha_s$. 

It is possible to understand at which perturbative order terms that are non-analytic in $\alpha_s$ appear by considering the contribution of non-interacting static gluons to a given quantity.  For simplicity, we now discuss the case of $\mu_B=0$ for this argument, but the same holds true at finite chemical potential.  For the pressure of a gas of gluons one has $P_\text{gluons}\sim\int d^3p \,p\, f_B(E_p)$, where $f_B$ denotes a Bose-Einstein distribution function and $E_p$ is the energy of the in-medium gluons. The contributions from the momentum scales $\pi T$, $g T$ and $g^2T$ can be expressed as
\begin{eqnarray}
P_\text{gluons}^{p\sim \pi T}&\sim& T^4f_B(\pi T)\;\sim\; T^4+{\mathcal O}(g^2) \, ,  \label{eq:orders1} \\
P_\text{gluons}^{p\sim g T}&\sim& (gT)^4 f_B(g T)\;\sim\; g^3T^4+{\mathcal O}(g^4) \, ,  \label{eq:orders2}\\
P_\text{gluons}^{p\sim g^2 T}&\sim& (g^2T)^4 f_B(g^2 T)\;\sim\; g^6T^4 \ , \label{eq:orders3}
\end{eqnarray}
where we have used the fact that $f_B(E)\sim T/E$ if $E\ll T$.  This fact is of fundamental importance, since it implies that when the energy/momentum are \emph{soft}, corresponding to electrostatic contributions $p_\text{soft} \sim g T$, %that 
one receives an enhancement of $1/g$ compared to contributions from {\em hard} momenta, $p_\text{hard} \sim T$, due to the bosonic nature of the gluon.  For \emph{ultrasoft} (magnetostatic) momenta, $p_\text{ultrasoft} \sim g^2 T$, the contributions are enhanced by $1/g^2$ compared to the naive perturbative order.  As Eqs.~\eqref{eq:orders1}-\eqref{eq:orders3} demonstrate, it is possible to generate contributions of the order $g^3 \sim \alpha_s^{3/2}$ from soft momenta and, in the case of the pressure, although perturbatively enhanced, ultrasoft momenta only start to play a role at order $g^6 \sim \alpha_s^3$.

Due to the infrared enhancement of electrostatic contributions, there is a class of diagrams called hard-thermal-loop (HTL) graphs that have soft external momenta and hard internal momenta that need to be resummed to all orders in the strong coupling \citep{Braaten:1989mz,Braaten:1989kk,Braaten:1990it}.  There are now several schemes for carrying out such soft resummations \citep{Arnold:1994ps,Arnold:1994eb,Zhai:1995ac,Braaten:1995cm,Braaten:1995jr,Kajantie:1997tt,Andersen:1999fw,Andersen:1999sf,Andersen:1999va,Blaizot:1999ip,Blaizot:1999ap,Blaizot:2000fc,Blaizot:2001vr,Andersen:2002ey,Andersen:2003zk,Andersen:2010ct,Andersen:2011sf,Haque:2013sja,Haque:2014rua}.  We note however, that even with such resummations, if one casts the result as a strict power series in the strong coupling constant the convergence of the perturbative series for the QCD free energy is quite poor.  To address this issue, one must treat the soft sector non-perturbatively and re-sum contributions to all orders in the strong coupling constant.  This can be done using effective field theory methods \citep{Ghiglieri:2020dpq}, approximately self-consistent two-particle irreducible methods \citep{Blaizot:1999ip,Blaizot:1999ap,Blaizot:2000fc,Blaizot:2001vr}, or the hard-thermal-loop perturbation theory reorganization of thermal field theory \citep{Andersen:1999fw,Andersen:1999sf,Andersen:2002ey,Andersen:2003zk,Andersen:2010ct,Andersen:2011sf,Haque:2013sja,Haque:2014rua}.

Thus, the calculation of the QCD EoS  requires all-orders resummation, which can be accomplished in a variety of manners.  Despite the fact that different methods exist, they all rely fundamentally on the use of so-called hard-thermal- or hard-dense-loops, which self-consistently include the main physical effect of the generation of in-medium gluon and quark masses at the one-loop level.  By reorganizing the perturbative calculation of the QCD EoS around the high-temperature hard-loop limit of quantum field theory, the convergence of the perturbative series can be extended to phenomenologically relevant temperatures and densities.  Below we summarize the results that have been obtained and compared to lattice QCD calculations where available.

\subsection{The resummed perturbative QCD EoS}

\begin{figure}[h!]
\begin{center}
\includegraphics[width=0.47\linewidth]{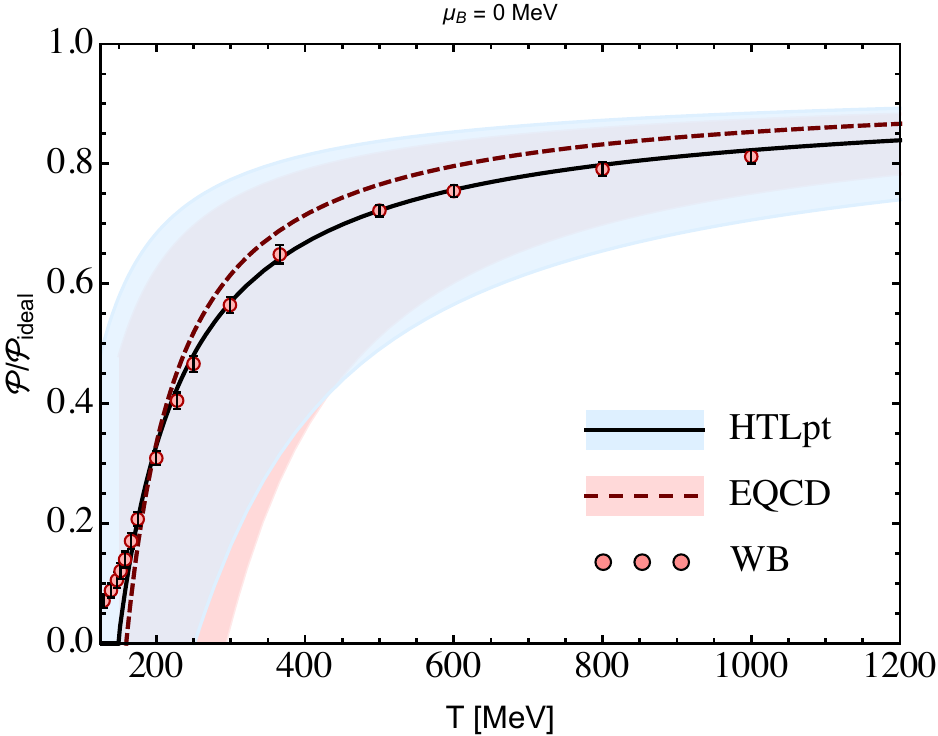}
\hspace{2mm}
\includegraphics[width=0.47\linewidth]{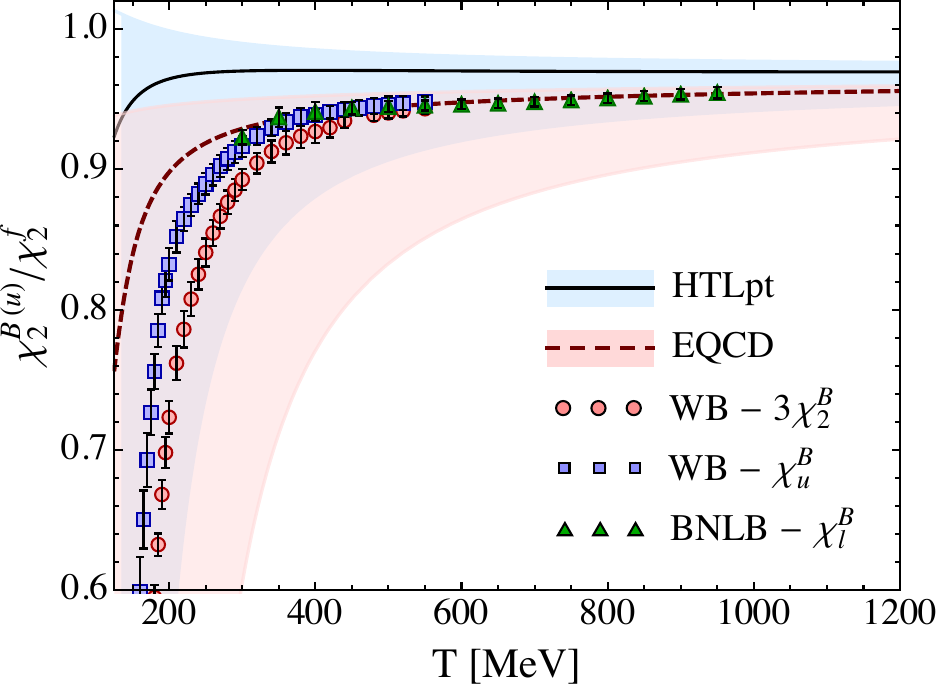}
\end{center}
% \vspace{-5mm}
\caption{Left: The resummed QCD pressure for $\mu_B=0$ obtained using the three-loop EQCD and HTL perturbation theory results
with lattice data from the Wuppertal-Budapest (WB) collaboration \citep{Borsanyi:2010cj}.
Right: The second-order light quark (and baryon) number susceptibilities. 
Lattice data are from the WB \citep{Borsanyi:2012rr,Borsanyi:2013hza} and BNLB  collaborations \citep{Bazavov:2013uja}.}
\label{fig:strickland-p0-xi2}
\end{figure}

\begin{figure}[h!]
\begin{center}
\includegraphics[width=0.47\linewidth]{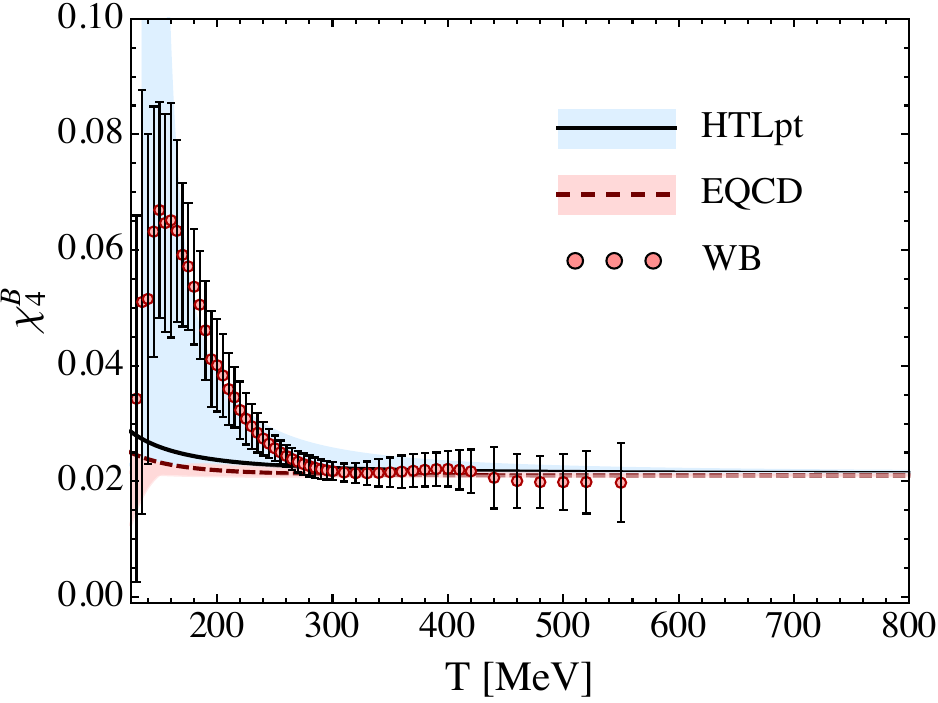}
\hspace{2mm}
\includegraphics[width=0.47\linewidth]{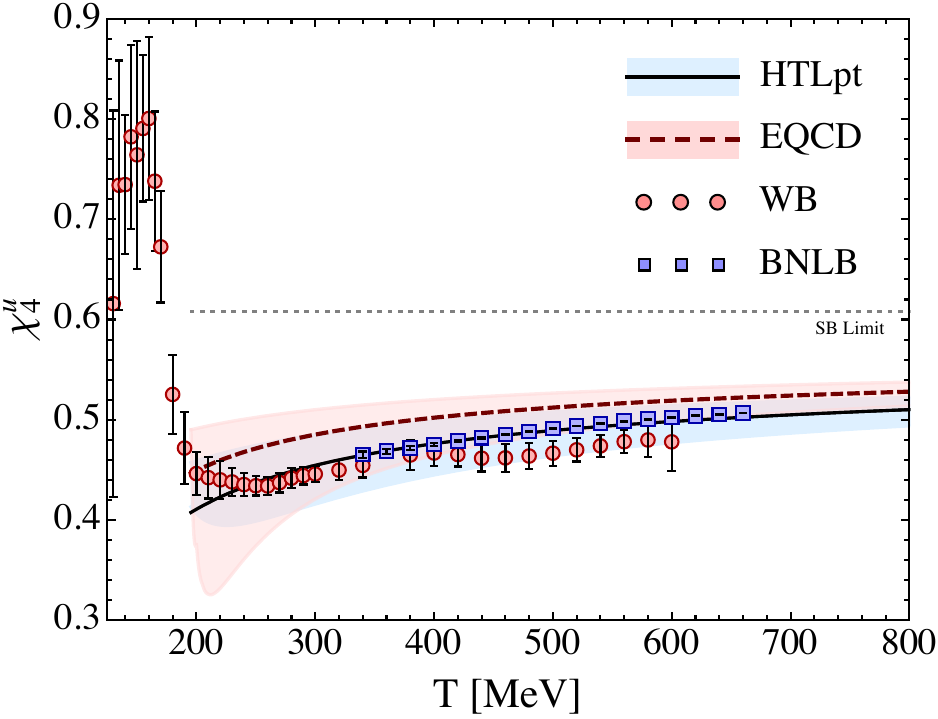}
\end{center}
\vspace{1mm}
\caption{Left: The 4th baryon number susceptibility.  Right: The 4th light quark number susceptibility.  Lattice data sources are the same as in Fig.~\ref{fig:strickland-p0-xi2}. }
\label{fig:strickland-xi4b-xi4u}
\end{figure}

The QCD EoS of deconfined quark matter at high chemical potential can be evaluated in terms of perturbative series in the  running coupling constant $\alpha_s$. As a result, the neutron-star EoS can be studied using the weak coupling expansion \citep{Kurkela:2014vha,Annala:2017llu,Shuryak:1977ut,Zhai:1995ac,Braaten:1995jr,Braaten:1995ju,Arnold:1994eb,Arnold:1994ps,Toimela:1984xy,Kapusta:1979fh,Annala:2019puf,Kurkela:2014vha,Kurkela:2009gj,Freedman:1976dm,Freedman:1976ub,Ecker:2022xxj,Altiparmak:2022bke,Ecker:2022dlg}. The EoS and trace anomaly of deconfined quark matter have been calculated to three-loop order using HTL perturbation theory framework at small $\mu_B$ and arbitrary $T$. Renormalization of the vacuum energy, the HTL mass parameters, and $\alpha_s$ eliminate all UV divergences. The  three-loop results for the thermodynamic functions  are observed to be in  agreement with lattice QCD data for $T \gtrsim 2-3 T_c$ after choosing a suitable mass parameter prescription \citep{Andersen:2011sf}. Furthermore, the QCD thermodynamic potential  at nonzero temperature and chemical potential(s) has been calculated using N2LO  at three-loop HTL perturbation theory which was used further to calculate the pressure, entropy density,  trace anomaly, energy density, and speed of sound, $c_s$, of the QGP \citep{Haque:2014rua}.  These findings were found to be in very good agreement with the data obtained from lattice QCD using the central values of the renormalization scales.  This is illustrated in Figs.~\ref{fig:strickland-p0-xi2} and \ref{fig:strickland-xi4b-xi4u}, which present comparisons of the resummed perturbative results with lattice data for the pressure and fourth-order baryonic and light-quark susceptibilities. In these figures, HTLpt corresponds to the N2LO hard thermal loop perturbation theory calculation of the EoS and EQCD corresponds to a resummed N2LO electric QCD effective field theory calculation of the same.  The shaded bands indicate the size of the uncertainty due to the choice of renormalization scale.

%%%%%%%%%%%%%%%%%%%%%%%%%%%%%%%%%%%%%%%%%%%%%%%%%%%%
%%%%%%%%%%%%%%%%%%%%%%%%%%%%%%%%%%%%%%%%%%%%%%%%%%%%

%%%%%%%%%%%%%%%%%%%%%%%%%%%%%%%%%%%%%%%%%%%%%%%%%%%%
%%%%%%%%%%%%%%%%%%%%%%%%%%%%%%%%%%%%%%%%%%%%%%%%%%%%

\subsection{The curvature of the QCD phase transition line}

\begin{figure}[h!]
	\centering
	\includegraphics[width=0.47\linewidth]{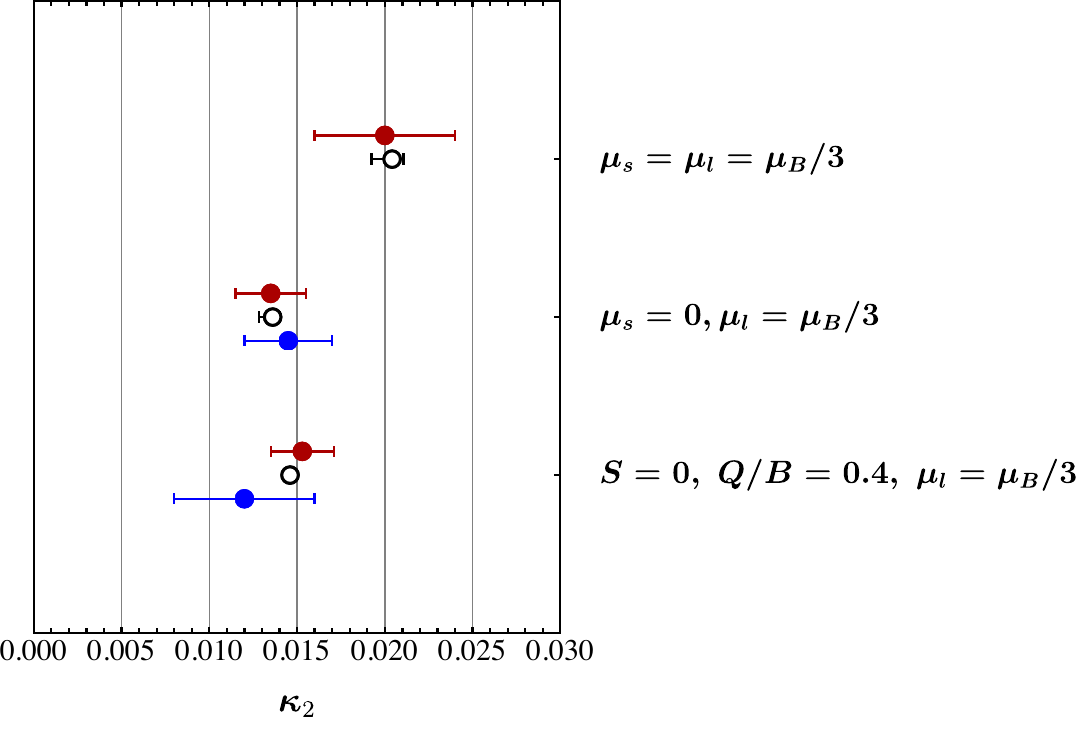}
        \hspace{2mm}
        \includegraphics[width=0.47\linewidth]{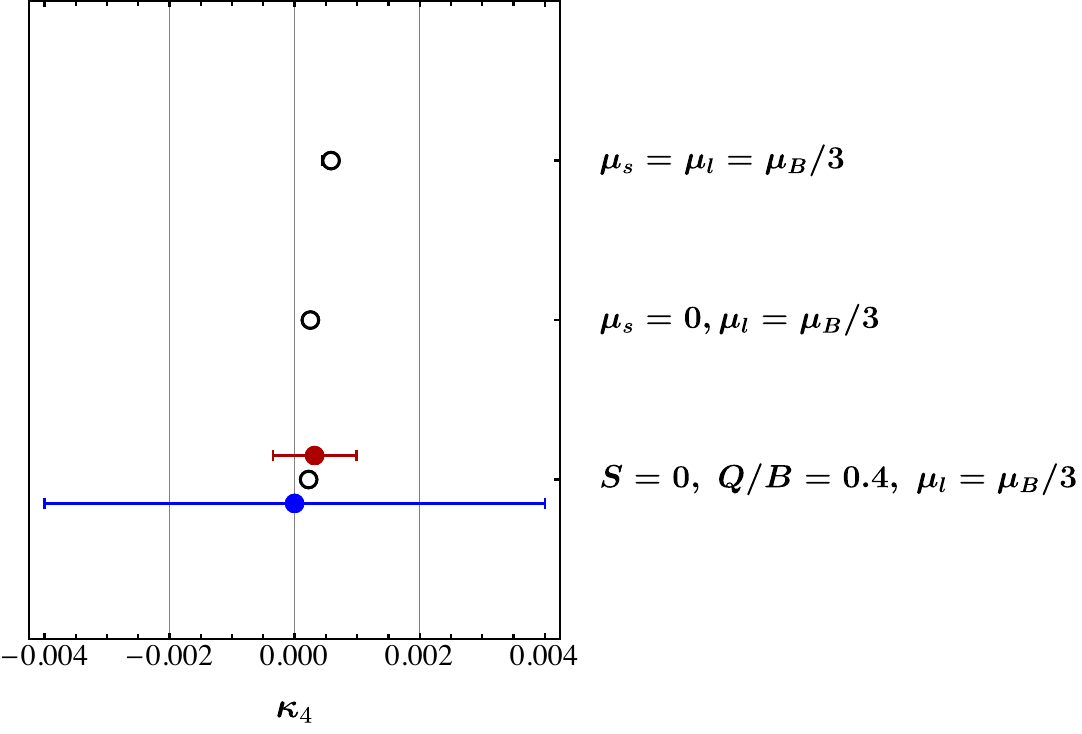}
	\caption{Left: Filled circles are lattice calculations of the quadratic curvature coefficient $\kappa_2$ \citep{Cea:2015cya,Bonati:2015bha,Bonati:2018nut,Borsanyi:2020fev,HotQCD:2018pds}, from top to bottom, respectively.  Red-filled circles are results obtained using the imaginary chemical potential method and blue-filled circles are results obtained using Taylor expansions around $\mu_B=0$.  Black open circles are the N2LO HTL perturbation theory predictions.  Right: Filled circles are lattice calculations of quartic coefficient $\kappa_4$ from \cite{Borsanyi:2020fev,HotQCD:2018pds}, from top to bottom, respectively.  The error bars associated with the HTL perturbation theory predictions result from variations of the assumed renormalization scale.}
	\label{fig:strickland-kappa2-kappa4}
\end{figure}

In another study,  for the second- and fourth-order curvatures of the QCD phase transition line, the N2LO HTL perturbation theory predictions were shown. In all three situations, (i) $\mu_{s}=$ $\mu_{l}=\mu_{B} / 3$, (ii) $\mu_{s}=0, \mu_{l}=\mu_{B} / 3$, and (iii) $S=0, Q / B=0.4, \mu_{l}=\mu_{B} / 3$, it was shown that N2LO HTL perturbation theory is compatible with the already available lattice computations of $\kappa_2$ and $\kappa_4$ as defined in Eq.~\eqref{eq:kappa} \citep{Haque:2020eyj}.  This is illustrated in Fig.~\ref{fig:strickland-kappa2-kappa4}, which presents comparisons of the resummed perturbative results with lattice data for the coefficients $\kappa_2$ and $\kappa_4$.

%%%%%%%%%%%%%%%%%%%%%%%%%%%%%%%%%%%%%%%%%%%%%%%%%%%%%%%%%%%%%%
%%%%%%%%%%%%%%%%%%%%%%%%%%%%%%%%%%%%%%%%%%%%%%%%%%%%%%%%%%%%%%

\subsection{Application at high density}

\begin{figure}[ht]
	\centering
	\includegraphics[width=\textwidth]{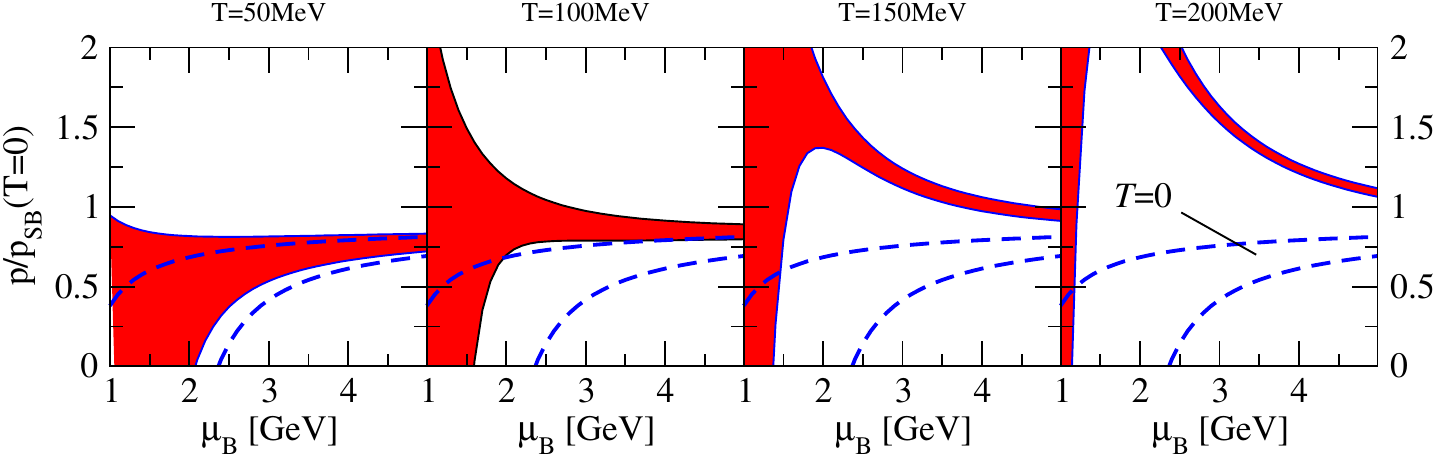}
	\caption{EoS of deconfined quark matter as a function of $\mu_B$ at  four different temperatures. The new result is shown by the red bands, with the widths resulting from a change in the renormalization scale $\tilde{\Lambda}$ \citep{Kurkela:2016was}. The corresponding $\mathcal{O}(g^4)$ result at absolute zero is shown by the dashed blue lines \citep{Freedman:1976ub,Baluni:1977ms,Vuorinen:2003fs}. Image reproduced with permission from \cite{Kurkela:2016was}, copyright by the author(s). }
	\label{fig:pQCDEoS}
\end{figure}

\cite{Gorda:2021znl}, at $T=0$, calculated the N3LO  contribution emerging from non-Abelian interactions among long-wavelength, dynamically screened gluonic fields using the weak-coupling expansion of the dense QCD EoS. In particular, they used the HTL effective theory to execute a comprehensive two-loop computation that is valid for long-wavelength, or soft, modes. In the plot of the EoS, unlike at high temperatures, the soft sector behaves well within cold quark matter, and the novel contribution reduces the renormalization-scale dependence of the EoS at high density \citep{Gorda:2021znl}. Working at exactly zero temperature is often a good approximation for fully evolved neutron stars but for the early stages of neutron-star evolution and neutron-star
mergers, it is  essential to incorporate temperature effects \citep{Shen:1998gq}.  However, the inclusion of finite temperature in high-$\mu_B$ quark matter gives rise to a technical difficulty for weak coupling expansions. It is no longer sufficient under this regime to handle simply the static sector of the theory nonperturbatively, but the $T=0$ limit's accompanying technical simplifications are also unavailable. In the EoS plots (see Fig.~\ref{fig:pQCDEoS}), the breakdown of the weak coupling expansion is observed by a rapid increase in the uncertainty of the result with an increase in temperature for tiny values of $\mu_B$  \citep{Kurkela:2016was}.
%
%%%%%%%%%%%%%%%%%%%%%%%%%%%%%%%%%%%%%%%%%%%%%%%%%%%%%%%%%%%%%%

\begin{figure}[h!]
	\centering
	\includegraphics[width=0.75\textwidth]{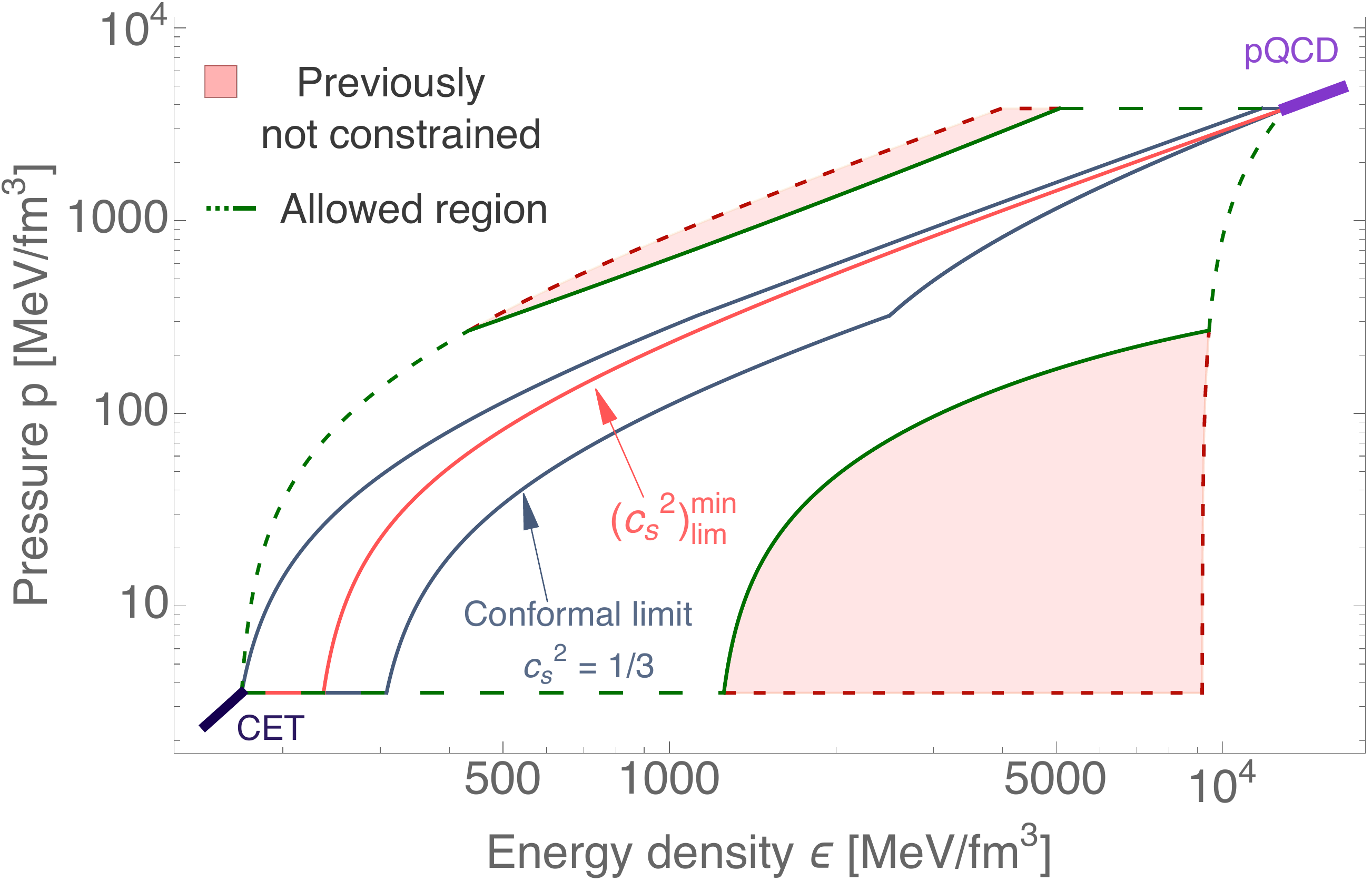}
	\caption{ The extracted EoS constraints from pQCD and $\chi$EFT(CET in the figure)  as a function of energy density  are shown (see Sect.~\ref{sec:eft} for details). The excluded regions are in pink. Image reproduced with permission from \cite{Komoltsev:2021jzg}, copyright by APS.}
	\label{fig:pQCD}
\end{figure}

The most up-to-date pQCD  results at $T=0$ and finite densities can be found in \cite{Gorda:2021kme}. The EoS derived in these calculations was applicable starting at $n_B\sim 40~n_{\sat}$ and above. However, there is an overall renormalization scale parameter, $X$, that is unknown. One can extrapolate down to lower densities assuming that the speed of sound squared should be bounded by causality and stability, i.e., $0\leq c_s^2 \leq 1$.  The results were shown in \cite{Komoltsev:2021jzg} where they varied $X$ in the range $1\leq X \leq 4$.  The results of the constrained regime can be seen in Fig.~\ref{fig:pQCD}. Various groups have then used these constraints in their neutron star EoS analyses \citep{Marczenko:2022jhl,Somasundaram:2022ztm}. 
%%%%%%%%%%%%%%%%%%%%%%%%%%%%%%%%%%%%%%%%%%%%%%%%%%%%%%%%%%%%%%

\subsection{Transport coefficients at finite $T$ and $\mu_B$}

\begin{figure}[h!]
	\centering
	\includegraphics[width=0.48\linewidth]{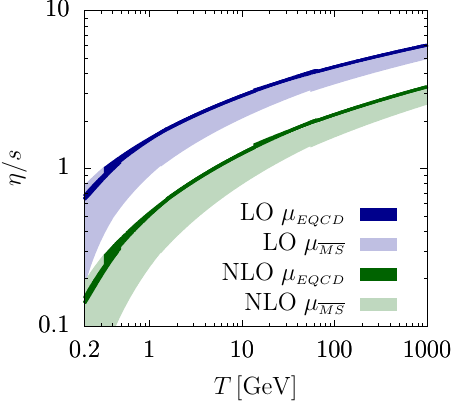}
        \hspace{2mm}
        \includegraphics[width=0.48\linewidth]{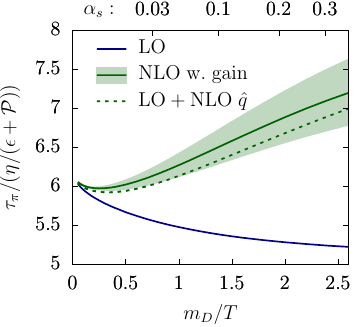}
	\caption{Left: Shear viscosity to entropy density ratio vs. temperature derived from pQCD at LO (leading order)  and NLO (next-to leading order) for two different choices of the running coupling for 3 flavors. Right: Coefficient of the relaxation time for shear viscosity at leading order and next-to-leading order as a function of the Debye mass over temperature  for QCD with 3 light flavors. Images reproduced with permission from [left] \cite{Ghiglieri:2018dib} and from [right] \cite{Ghiglieri:2018dgf}, copyright by the author(s).}
	\label{fig:etas}
\end{figure}

The quark-gluon plasma probed in heavy-ion collisions is not in equilibrium and viscous effects from shear and bulk viscosities are important for the evolution of the system.  At the moment it is not yet possible to reliably compute the shear and bulk viscosities using first principle calculations \citep{Meyer:2011gj}. However, it is possible to perform calculations of these coefficients in the weak-coupling limit of QCD.   The shear viscosity $\eta$ and relaxation time $\tau_\pi$ (the timescale within which the system relaxes towards its Navier--Stokes regime, \citealt{Denicol:2021}) are usually related through
\begin{equation}
    \tau_\pi=C \frac{\eta}{\varepsilon+p}\ ,
\end{equation}
where $C$ is a constant determined by the theory.
Calculations of $\eta/s$ in QCD have been completed up to NLO (next-to leading order, \citealt{Ghiglieri:2018dib}) and the constant $C$ of the relaxation time at NLO \citep{Ghiglieri:2018dgf} for $\mu_B=0$, as shown in Fig.~\ref{fig:etas}. 

\begin{figure}[h!]
	\centering
	\includegraphics[width=0.75\textwidth]{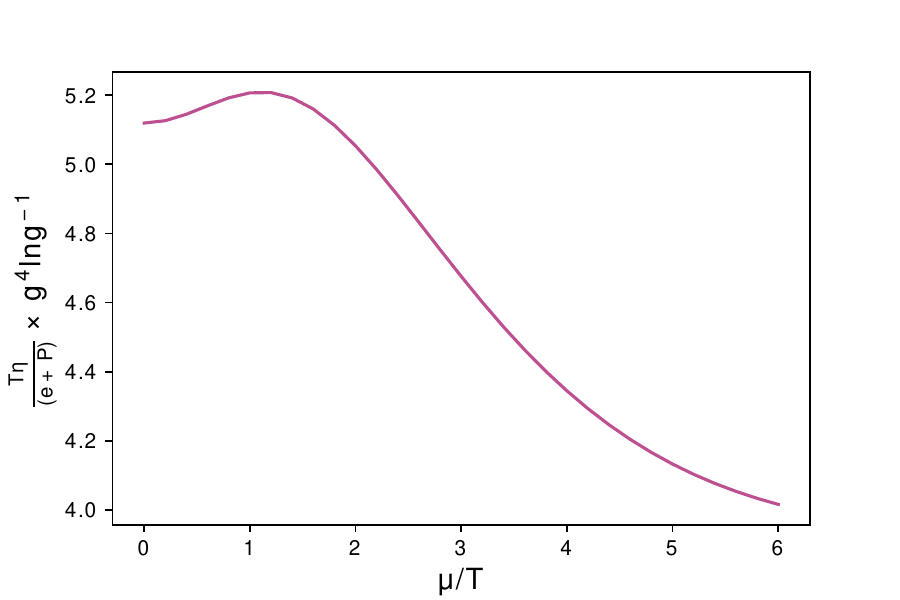}
	\caption{Shear viscosity times temperature divided by the enthalpy versus chemical potential over temperature in 3 flavor QCD. Image reproduced with permission from \cite{Danhoni:2022xmt}, copyright by the author(s). }
	\label{fig:etasMUB}
\end{figure}

Recently, the first calculations of shear viscosity at leading-log at finite $\mu_B$ were performed in QCD in \cite{Danhoni:2022xmt}, as shown in Fig.~\ref{fig:etasMUB}. 

Note, however, that at finite $\mu_B$ the most natural dimensionless quantity involves the enthalpy ($w=\varepsilon+p$), such that $\eta T/w$  is the relevant quantity (the factor of $T$ is to ensure that it remains dimensionless) to be used \citep{Liao:2009gb}.  In the limit of vanishing baryon chemical potential, then
\begin{equation}
 \lim_{\mu_B\rightarrow 0}  \frac{\eta T}{\varepsilon+p}= \frac{\eta}{s}\ ,
\end{equation}
such that these results should be smoothly connected regardless of $\mu_B$. The relaxation time has not yet been calculated in QCD at finite $\mu_B$. Finally we note that, in typical relativistic viscous hydrodynamics simulations performed in heavy-ion collisions, a number of other transport coefficients are also needed. For example, using perturbative QCD, the bulk viscosity has been computed in \cite{Arnold:2006fz}, conductivity and diffusion in \cite{Arnold:2000dr}, and some second-order transport coefficients can be found in \cite{York:2008rr}. In practice, these perturbatively-determined expressions are not the ones used in simulations, which often rely on simple formulas involving the transport coefficients determined from, for instance, kinetic theory models \citep{Denicol:2012cn,Denicol:2014vaa} or holography \citep{Kovtun:2004de,Finazzo:2014cna,Rougemont:2017tlu,Grefa:2022sav}, see \cite{JETSCAPE:2020mzn}. 

%%%%%%%%%%%%%%%%%%%%%%%%%%%%%%%%%%%%%%%%%%%%%%%%%%%%%%%%%%%%%%%%%%%%%%%%%%%%%%%%%%%%%%%%%%%%%%%%%%%%%%%%%%%%%%%%%%%%%%%%%%%%%%%%%%%%%%%%%%%%%%%%%%%%%%%%%%%%%%%%%%%%%%%%%%%%%%%%%%%%%%%%%%%%%%%%%%%

\section{Theoretical constraints: chiral effective field theory}
\label{sec:eft}

In the opposite regime of low density and temperature, Chiral Effective Field Theory ($\chi$EFT) is used to calculate the EoS relevant around $n_B\sim n_{\sat}$ of neutron stars. 
 For ab initio $\chi$EFT calculations, it is possible to study the EoS at arbitrary isospin asymmetry at both zero and nonzero temperature within a many-body perturbation theory\cite{Drischler:2013iza,Wellenhofer:2016lnl,Wen:2020nqs,Somasundaram:2020chb}, or a many-body Brueckner–-Hartree-–Fock approach \citep{Logoteta:2016hxh,Logoteta:2016nzc}. Recently, 
several
benchmark calculations  have been performed,  considering  the first and second generation
of $\chi$EFT Norfolk NN and 3N interactions,
to  assess the possible  error which is associated with the chosen  method when solving
the many-body Schr\"odinger equation \citep{Piarulli:2019pfq,Lovato:2022apd}.  The authors have obtained a good agreement among the many-body techniques tested up  to approximately the nuclear saturation density. 

However, in practice it is convenient to first compute the EoS for symmetric nuclear matter ($Y_Q=0.5$) and pure neutron matter ($Y_Q=0$) and then interpolate between the two using the quadratic approximation for the isospin-asymmetry dependence of the EoS. From the density-dependent symmetry energy, one can extract the coefficients $E_{\sym}$, $L$, $K_{\sym}$ from $\chi$EFT calculations
\begin{align}    
E_{\mathrm{NS}}=E_{Y_Q=0.5} + N_B\left[E_{\sym,\sat}+\frac{L_{\sat}}{3}\left(\frac{n_B}{n_{\sat}}-1\right)+\frac{K_{\sym,\sat}}{18}\left(\frac{n_B}{n_{\sat}}-1\right)^{2}\right]  \big(1-2Y_{Q}\big)^2,
\end{align}
where $E_{\mathrm{NS}}$ is the ground-state energy at a given density and isospin asymmetry, $E_{Y_Q=0.5}$ is the total energy for isospin-symmetric nuclear matter, $E_{\sym,\sat}=\left(\frac{E_{Y_Q=0}-E_{Y_Q=0.5}}{N_B}\right)_{n_{\sat}}$ is the symmetry energy at saturation, 
$L_{\sat}=3n_{\sat}\left(\frac{dE_{\sym}}{dn_B}\right)_{n_{\sat}}$ is the slope of the symmetry energy at saturation, and 
$K_{\sym,\sat}=9n_{\sat}^2\left(\frac{d^2E_{\sym}}{dn_B^2}\right)_{n_{\sat}}$ is the symmetry energy curvature at saturation. Using this expansion scheme, it is possible to obtain the neutron star outer core EoS with quantified uncertainties from $\chi$EFT. The properties of the low-density crust \citep{Lim:2017luh,Grams:2022lci} and the high-density inner core \citep{Hebeler:2010jx,Tews:2018kmu,Lim:2020zvx,Drischler:2020fvz,Brandes:2022nxa} require additional modeling assumptions.

The energy per baryon of symmetric nuclear matter, $E_{Y_Q=0.5}/N_B$, and pure neutron matter, $E_{Y_Q=0}/N_B$, has been computed from $\chi$EFT at different orders in the chiral expansion and different approximations in many-body perturbation theory. As a representative example, in \cite{Holt:2016pjb} the EoS was computed up to third order in many-body perturbation theory, including self-consistent second-order single-particle energies. Chiral nucleon-nucleon interactions were included up to N3LO, while three-body forces were included up to N2LO in the chiral expansion. Using the above approach, the authors gave error bands on the EoS (including $E_{\sym,\sat}$ and slope parameter $L_{\sat}$), taking into account uncertainties from the truncation of the chiral expansion and the choice of resolution scale in the nuclear interaction. The incorporation of third-order particle-hole ring diagrams (frequently overlooked in EoS computations) helped to reduce theoretical uncertainties in the neutron matter EoS at low densities, but beyond $n_B\gtrsim 2n_{\sat}$ the EoS error bars become large due to the breakdown in the chiral expansion. Recent advances in automated diagram and code generation have enabled studies at even higher orders in the many-body perturbation theory expansion \citep{Drischler:2021kxf,Drischler:2017wtt}. In the following we focus on selected results obtained in many-body perturbation theory and refer the reader to \cite{Lynn:2019rdt, Carlson:2014vla,Gandolfi:2020pbj, Tews:2020wrl,Rios:2020oad,Hagen:2013nca} for comprehensive review articles on many-body calculations in the frameworks of quantum Monte Carlo, self-consistent Green's functions method, and coupled cluster theory.

\begin{figure}[h!]
\raisebox{0.2cm}{\includegraphics[scale=0.275]{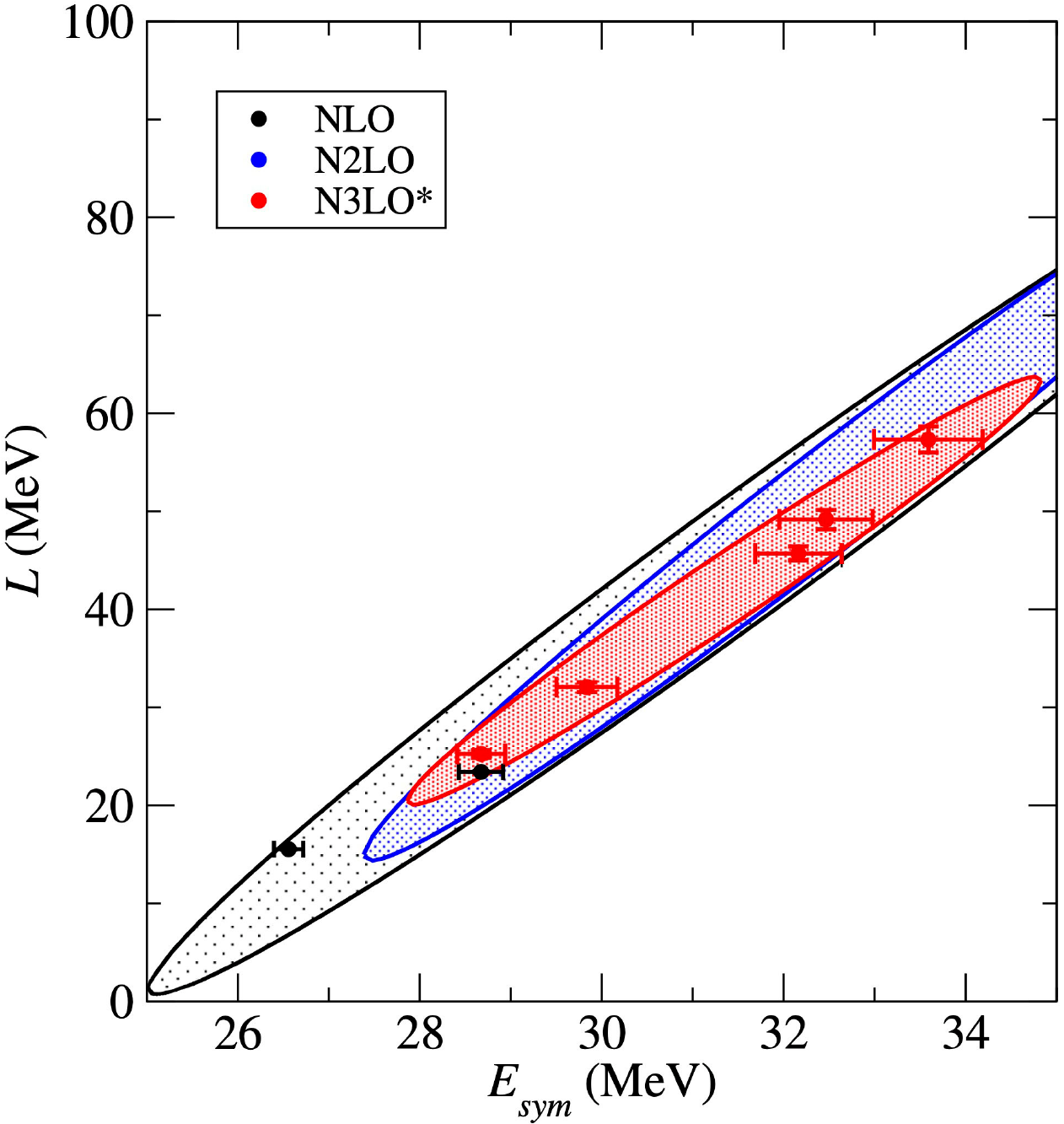}}\hspace{.1in}
\includegraphics[scale=.904]{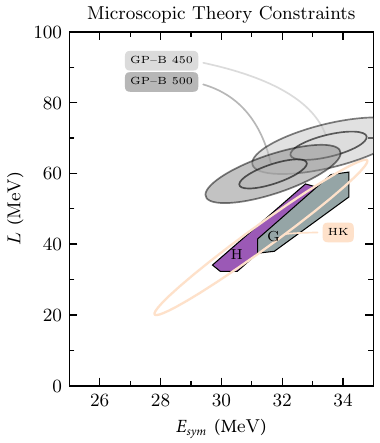}
\centering
\caption{Adapted from \cite{Holt:2016pjb,Drischler:2021kxf}. Left: Correlation between the symmetry energy $E_{\sym}$ and its slope, $L$ at saturation, from $\chi$EFT calculations at different orders in the chiral expansion \citep{Holt:2016pjb}. Right: Microscopic constraints on the $E_{\sym} - L $ correlation from \cite{Hebeler:2010jx,Gandolfi:2011xu,Holt:2016pjb,Drischler:2020hwi}.}
\label{CEFT_S_vs_L}
\end{figure}

The left panel of Fig.~\ref{CEFT_S_vs_L} shows the correlation between the symmetry energy $E_{\sym}$ and slope parameter $L$ at saturation for different choices of the high-momentum regulating scale (shown as different data points) and order in the chiral expansion (denoted by different colors) all calculated at 3rd order in many-body perturbation theory including self-consistent (SC) nucleon self-energies at second order \citep{Holt:2016pjb}.
The ellipses show the 95\% confidence level at orders NLO, N2LO, and the N3LO$^*$ (where the star denotes that the three-body force is included only at N2LO) \citep{Holt:2016pjb}. Interestingly, one finds that even EoS calculations performed at low order in the chiral expansion produce values of $E_{\sym}$ and $L$ at saturation that tend to lie on a well-defined correlation line. The N3LO* (red) ellipse illustrates the range of symmetry energy $28\, \mathrm{MeV}< E_{\sym,\sat}< 35\, \mathrm{MeV}$ and slope parameter $20\, \mathrm{MeV}<L_{\sat}< 65\, \mathrm{MeV}$, both of which are quite close to the findings of prior microscopic computations \citep{Hebeler:2010jx,Gandolfi:2011xu} that also used neutron matter calculations plus the empirical saturation properties of symmetric nuclear matter to deduce $E_{\sym}$ and $L$. All three sets of results are shown in the right panel of Fig.~\ref{CEFT_S_vs_L} and labeled `H' \citep{Hebeler:2010jx}, `G' \citep{Gandolfi:2011xu}, and `HK' \citep{Holt:2016pjb} respectively. In contrast, a recent work \citep{Drischler:2020hwi} that analyzed correlated $\chi$EFT truncation errors in the EoS for neutron matter and symmetric nuclear matter using Bayesian statistical methods found $E_{\sym} = 31.7\pm1.1$ MeV and $L = 59.8 \pm 4.1$ MeV at saturation, shown as `GP-B 500' in the right panel of Fig.~\ref{CEFT_S_vs_L}. The obtained value of  $E_{\sym}$ was similar to those from \cite{Gandolfi:2011xu,Hebeler:2010jx,Holt:2016pjb}, but $L$ was systematically larger. The results, however, are in good agreement with standard empirical constraints \citep{Drischler:2020hwi,Li:2019xxz} discussed in Sect.~\ref{sec:sym_E}.

Chiral effective field theory has also been employed to study the liquid-gas phase transition and thermodynamic EoS at low temperatures ($T<25$ MeV) in isospin-symmetric nuclear matter \citep{Wellenhofer:2014hya}. The EoS has been computed using $\chi$EFT nuclear potentials at resolution scales of 414, 450, and 500 MeV. The results from this study are tabulated in Table~\ref{table:lg1}. In particular, the values of the liquid-gas critical endpoint in temperature, pressure, and density agree well with the empirical multifragmentation and compound nuclear decay experiments discussed in Sect.~\ref{subsecLG}. In addition, at low densities and moderate temperatures, the pure neutron matter EoS is well described within the virial expansion in terms of neutron-neutron scattering phase shifts. The results from chiral effective field theory have been shown \citep{Wellenhofer:2015qba} to be in very good agreement with the model-independent virial EoS. Since finite-temperature effects are difficult to reliably extract empirically, chiral effective field theory calculations have been used to constrain the temperature dependence of the dense-matter EoS in recent tabulations 
\cite{Du:2018vyp,Du:2021rhq} for astrophysical simulations.

\begin{table}[h!]
\caption{The N3LO  contributions  for binding energy per nucleon at saturation, saturation density, compressibility at saturation, and liquid-gas phase-transition critical values of temperature, density, and pressure at different resolution scales. Table adapted from \cite{Wellenhofer:2014hya}.}.
\label{table:lg1}
\centering
\begin{tabular}{ccccccc}
\hline \hline   Resolution scale & $\frac{B}{A}$(MeV)  & $n_{\sat}\left(\mathrm{fm}^{-3}\right)$ & $K(\mathrm{MeV})$ & $T_{c}(\mathrm{MeV})$ & $n_{c}\left(\mathrm{fm}^{-3}\right)$ & $P_{c}\left(\mathrm{MeV}/\mathrm{fm}^{3}\right)$ \\
\hline \hline 
$414\left(M^{*} / M\right)$ & $-15.79$ & $0.171$ & 223 & $17.4$ & $0.066$ & $0.33$ \\
450 $\left(M^{*} / M\right)$ & $-15.50$ & $0.161$ & 244 & $17.2$ & $0.064$ & $0.32$ \\
500 $\left(\mathrm{no}~\mathrm{M}^{*} / M\right)$ & $-16.51$ & $0.174$ & 250 & $19.1$ & $0.072$ & $0.42$ \\
\hline  \hline 
\end{tabular}
\end{table}

%%%%%%%%%%%%%%%%%%%%%%%%%%%%%%%%%%%%%%%%%%%%%%%%%%%%%%%%%%%%%%%%%%%%%%%%%%%%%%%%%%%%%%%%%%%%%%%%%%%%%%%%%%%%%%%%%%%%%%%%%%%%%%%%%%%%%%%%%%%%%%%%%%%%%%%%%%%%%%%%%%%%%%%%%%%%%%%%%%%%%%%%%%%%%%%%%%%

\section{Experimental constraints: heavy-ion collisions}
\label{sec:hic}

Given that hot and dense matter can be created experimentally in heavy-ion collisions, constraints on its EoS can be extracted via experimental measurements obtained from such collisions. 
As explained previously in Sect.~\ref{sub:HIC_exec_summary},
we will especially focus in this paper on the data themselves, and avoid citing quantities inferred from the data, as the latter come with an associated model dependence. 
We will mention particle production yields and their ratios, as well as fluctuation observables of particle multiplicities. Then, we will review experimental results on flow harmonics, to end with Hanbury--Brown--Twiss (HBT) interferometry measurements, also referred to in the field as femtoscopy. 
% }

We remind the reader that, as a general rule of thumb, high center of mass energy collisions, $\sqrt{s_{NN}}\gtrsim 200$ GeV, are in the regime of the phase diagram where $\mu_B \ll T$, such that the numbers of particles and anti-particles are approximately equal (i.e., $n_B\sim 0$).  As one lowers $\sqrt{s_{NN}}$, baryons are stopped within the collision such that higher $n_B$ is reached.  At sufficiently low beam energies, $\sqrt{s_{NN}}\lesssim 4-7$ GeV\footnote{Note that the exact beam energy where this occurs is still under debate.}, matter is dominated by the hadron gas phase, such that lowering $\sqrt{s_{NN}}$ leads to lower temperatures and lower $n_B$. 

 \subsection{Particle yields}

Particle production spectra are part of the simplest experimental observables used in heavy-ion collisions, to access thermodynamic properties and characteristics of the hot and dense matter. 
Starting with the integrated production yields of identified hadrons, they can be measured to help determine properties from the evolution of the system, in particular at chemical and kinetic freeze-out. These steps designate the ending of inelastic collisions between formed hadrons (what fixes the chemistry of the system) and in turn the ceasing of all elastic collisions (after which particles stream freely to the detectors) in the evolution of a heavy-ion collision.
Statistical hadronization models \citep{Hagedorn:1965st,Dashen:1969ep,Becattini:2001fg,Wheaton:2011rw, Petran:2013dva, Andronic:2017pug,Andronic:2018qqt, Vovchenko:2019pjl} are fitted to these yields and ratios by varying over $T$ and $\mu_B$, in order to extract the respective chemical freeze-out values.
Despite their ability to reproduce particle yields successfully, those models are limited in scope since  they do not reproduce the dynamics of a collision and hinge on the assumption of thermal equilibrium, which is not necessarily achieved within the short time scales of heavy-ion collisions.
Similar information can also be inferred for kinetic freeze-out, using a so-called blast-wave model \citep{Schnedermann:1993ws}.
The data from low transverse momentum particles (i.e., $p_T \lesssim 2-3$ GeV/c) measured at mid-rapidity (approximately transverse to the beam line direction) is used for statistical hadronization fits, because these particles spend the longest time within the medium (so they are more likely to be thermalized) and they have low enough momentum to avoid contributions from jet physics.

\begin{figure}[h!]
    \centering
    \includegraphics[width=0.4\textwidth]{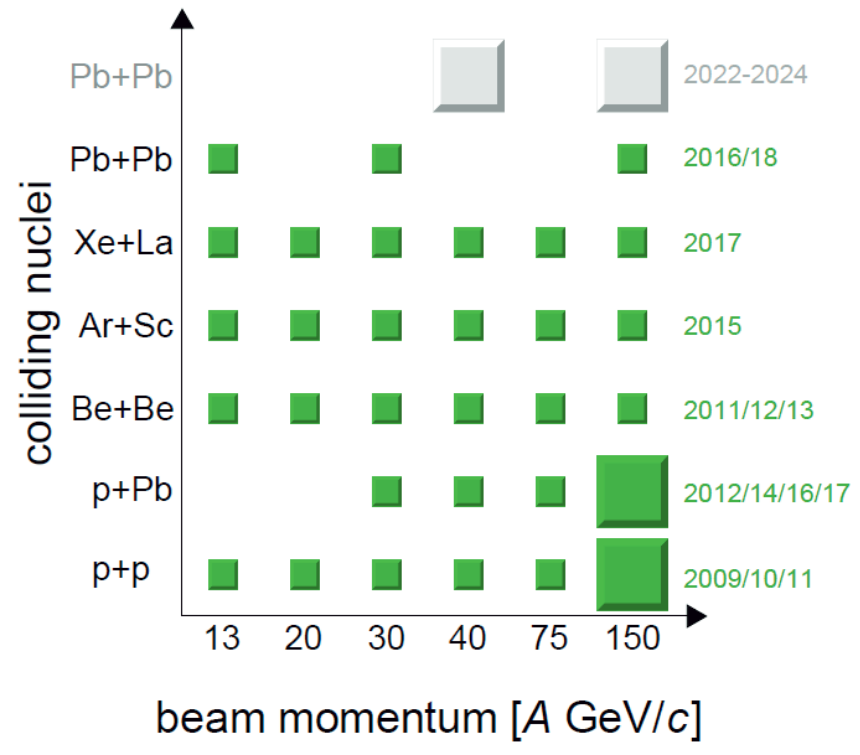}
    \includegraphics[width=0.59\textwidth]{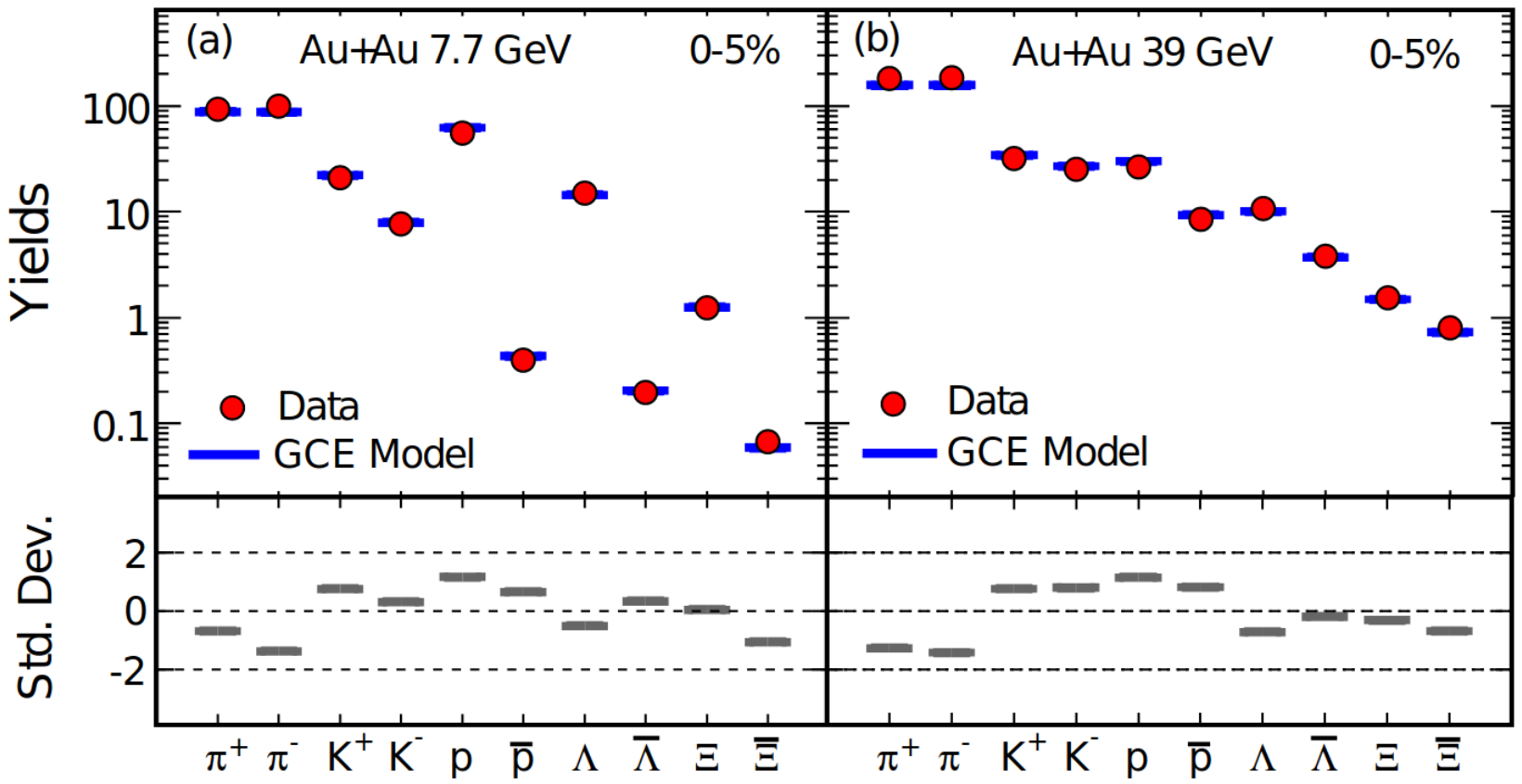}
    \caption{
    Left: diagram of different collided systems as a function of collision energy.
    Right: yields of hadrons measured by STAR in central Au+Au collisions at 7.7 and 39 GeV/A, compared with results from a grand canonical statistical hadronization model. Images reproduced with permission from [left] \cite{Kowalski:2022ugf}, copyright by the author(s), and [right] \cite{STAR:2017sal}, copyright by APS.
    }
    \label{fig:NA61_scan+ALICE_yields}
\end{figure}

Measurement of yields for the most common light hadron species (namely $\pi^\pm$, $K^\pm$, $p/\overline{p}$) and strange hadrons ($\Lambda/\overline{\Lambda}$, $\Xi^-/\overline{\Xi}^+$ and $\Omega^-/\overline{\Omega}^+$) was achieved by STAR at RHIC and ALICE at the LHC.
As part of the BES program, STAR  has measured these hadron yields in Au+Au collisions at center-of-mass energies of $\sqrt{s_{NN}}=$7.7, 11.5, 14.5, 19.6, 27, 39, 62.4 and 200 GeV %\jj{necessary to cite all energies?} 
\cite{STAR:2008med, STAR:2017sal, STAR:2019vcp, STAR:2019bjj} and in U+U collisions at $\sqrt{s_{NN}} = 193$ GeV  \citep{STAR:2022ypn}.
Motivated by results from the past SPS-experiment NA49 obtained from Pb+Pb collisions at  $\sqrt{s_{NN}} < 20$ GeV \citep{NA49:2007stj}, the NA61/SHINE experiment at CERN has conducted a scan in system-size and energy (in the same energy range as NA49). The diagram of all collided systems as a function of collision energy and nuclei is displayed in the left panel of Fig. \ref{fig:NA61_scan+ALICE_yields}. They published data for light hadrons in Ar+Sc \citep{NA61SHINE:2021nye} and Be+Be collisions \citep{NA61SHINE:2020czq} so far, while results from Xe+La and Pb+Pb collisions should become available in the next few years \citep{Kowalski:2022ugf}.
At LHC energies, the higher $\sqrt{s_{NN}}$ produces significantly more particles, allowing more precise measurements of the species. For this reason, the ALICE experiment has measured yields not only for light and (multi)strange hadrons, but also for light nuclei and hyper-nuclei, in Pb+Pb collisions at 2.76 TeV/A in particular \citep{ALICE:2013mez, ALICE:2013cdo, ALICE:2013xmt, ALICE:2015wav, ALICE:2017jmf, ALICE:2015oer}, and more recently in Pb+Pb collisions at 5.02 TeV/A too \citep{ALICE:2019hno, ALICE:2022boh}, as well as Xe+Xe collisions at 5.44 TeV/A \citep{ALICE:2021lsv}. 

% \\

\begin{figure}[ht]
    \centering
    \includegraphics[width=0.6\linewidth]{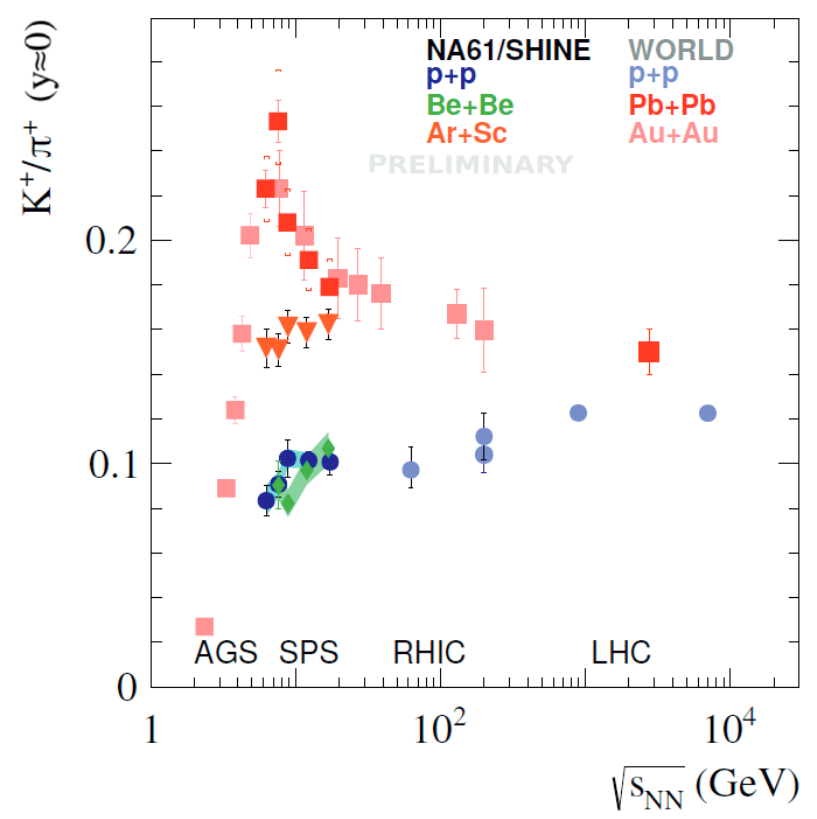}
    \caption{Preliminary results for the ratio of $K^+/\pi^+$ yields at mid-rapidity as a function of the collision energy, compared for different collision systems. Image reproduced with permission from \cite{Kowalski:2022ugf}, copyright by the author(s). 
    %\jj{OK even if it's preliminary ?}
    } 
    \label{fig:Horn_structure_K-pi_ratio}
\end{figure}

The yields of particle production can also be used as an indicator of the onset of deconfinement, notably thanks to the strangeness enhancement: strange quark-antiquark pairs are expected to be produced at a much higher rate in a hot and dense medium than in a hadron gas. Hence, one should expect in particular an increase of multi-strange baryons compared to light-quark-compound hadrons in collision systems where the QGP has been formed, which has been observed experimentally in heavy-ion collisions at several energies \citep{NA57:2006aux, STAR:2007cqw, ALICE:2013xmt}.
Moreover, the distinctive non-monotonic behavior of the $K^+/\pi^+$ ratio as a function of the collision energy can also be considered as a sign of the onset of deconfinement, according to some authors \citep{Gazdzicki:1998vd, Poberezhnyuk:2015wea}. This so-called ``horn'' in the $K^+/\pi^+$ ratio has been notably observed in Pb+Pb \citep{NA49:2002pzu, NA49:2007stj} and Au+Au collisions \citep{E802:1996owm, E866:1999ktz, STAR:2008med, STAR:2009sxc, STAR:2017sal}, but absent from p+p data \citep{NA61SHINE:2017fne, STAR:2009sxc, ALICE:2011gmo, ALICE:2014juv} and Be+Be collisions results \citep{NA61SHINE:2020czq}. Recent results from Ar+Sc collisions \citep{Kowalski:2022ugf} have, however, stirred up doubts regarding the interpretation of this observable, as the value of this ratio from such collision system is closer to the one measured in big systems, while no horn structure is seen, similar to small systems. All these results for different systems can be seen in Fig.~\ref{fig:Horn_structure_K-pi_ratio}. 

\subsection{Fluctuation observables}

In heavy-ion collisions, observables measuring fluctuations are among the most relevant for the investigation of the QCD phase diagram. 
Within the assumption of thermodynamic equilibrium, cumulants of net-particle multiplicity distributions become directly related to thermodynamic susceptibilities, and can be compared to results from lattice QCD  (see Sect.~\ref{sub:susceptibilities}) to extract information on the chemical freeze-out line \citep{Alba:2014eba, Alba:2015iva, Alba:2020jir}. Moreover, large, relatively long-range fluctuations are expected in the neighborhood of the conjectured QCD critical point, making fluctuation observables very promising signatures of criticality \citep{Stephanov:1999zu,Stephanov:2008qz,Athanasiou:2010kw}.  It has also been proposed that the finite-size scaling of critical fluctuations could be employed to constrain the location of the critical point \citep{Palhares:2009tf,Fraga:2010qef,Lacey:2014wqa,Lacey:2016tsw}. 
In \cite{Fraga:2010qef}, finite-size scaling arguments  were applied to mean transverse-momentum fluctuations measured by STAR \citep{STAR:2005vxr,STAR:2006rqs} to exclude a critical point below $\mu_B \lesssim 450$ MeV. 

Fluctuations of the conserved charges $B,S$, and $Q$ are of particular importance. 
As mentioned already in subsection \ref{sub:susceptibilities}, these fluctuations can be used to probe the deconfinement transition, as well as the location of the critical endpoint. Calculated via the susceptibilities expressed in equation \eqref{eq:susceptibilities}, which can be evaluated via lattice QCD simulations or HRG model calculations, they can also be related to the corresponding cumulants of conserved charges $C^{BSQ}_{lmn}$ \footnote{Note that such cumulants are often referred to as $\kappa^{BSQ}_{lmn}$ in the literature; we used a different symbol here to avoid confusion with the $\kappa$ coefficients of the pseudo-transition temperature parametrisation from lattice QCD in subsection \ref{sub:pseudo-trans_line}.}, following the relation
\begin{equation}
    C^{BSQ}_{lmn} = VT^3 \times \chi^{BSQ}_{lmn} \ ,
    \label{eq:susc-cumulants_relation}
\end{equation}
with the volume $V$ and temperature $T$  of the system, and $l,m,n  \in \mathbb{N}$ \citep{Luo:2017faz}. The cumulants are also theoretically related to the correlation length of the system $\xi$, which is expected to diverge in the vicinity of the critical endpoint. In particular, the higher order cumulants are proportional to higher powers of $\xi$, making them more sensitive to critical fluctuations \citep{Stephanov:1999zu,Stephanov:2008qz,Athanasiou:2010kw}.

In heavy-ion collisions, however, it is impossible to measure the fluctuations of conserved charges directly, because one cannot detect all produced particles (e.g., neutral particles are not always possible to measure, so that baryon number fluctuations do not include neutrons). 
Nevertheless, it is common to measure the cumulants of identified particles' net-multiplicity distributions, using some hadronic species as proxies for conserved charges \citep{Koch:2005vg}. Net-proton distributions are used as a proxy for net-baryons \citep{STAR:2010mib,STAR:2019dow,STAR:2021iop}, net-kaons \citep{STAR:2017tfy, Ohlson:2017wxu, STAR:2019ans} or net-lambdas \citep{STAR:2020ddh} are used as a proxy for net-strangeness, and net-pions+protons+kaons \citep{STAR:2019ans} has been recently used as a proxy for net-electric charge, instead of the actual net-charged unidentified hadron distributions. Mixed correlations have also been measured \citep{STAR:2019ans}, although alternative ones have been suggested, that would provide more direct comparisons to lattice QCD susceptibilities \citep{Bellwied:2019pxh}. 

These net-particle cumulants can be used to construct ratios, as they are connected with usual statistic quantities characterizing the net-hadron distributions $N_\alpha = n_\alpha - n_{\overline{\alpha}}$ (with $n_{\alpha/\overline{\alpha}}$ being respectively the number of hadrons or anti-hadrons of hadronic specie $\alpha$).
Hence, relations between such ratios and the mean $\mu_\alpha$, variance $\sigma_\alpha$, skewness $S_\alpha$ or kurtosis $\kappa_\alpha$ can be expressed as follows
\begin{align}
    \frac{\sigma_\alpha^2}{\mu_\alpha} 
    & = \frac{C_2^\alpha}{C_1^\alpha} 
    = \frac{\langle (\delta N_\alpha)^2 \rangle}{\langle N_\alpha \rangle} \ , \\
    S_\alpha \sigma_\alpha 
    & = \frac{C_3^\alpha}{C_2^\alpha} 
    = \frac{\langle (\delta N_\alpha)^3 \rangle}{\langle (\delta N_\alpha)^2 \rangle} \ , \\
    \kappa_\alpha \sigma_\alpha^2 
    & = \frac{C_4^\alpha}{C_2^\alpha} 
    = \frac{\langle (\delta N_\alpha)^4 \rangle}{\langle (\delta N_\alpha)^2 \rangle} - 3{\langle (\delta N_\alpha)^2 \rangle} \ , \\
    \frac{\kappa_\alpha \sigma_\alpha}{S_\alpha} 
    & = \frac{C_4^\alpha}{C_3^\alpha} 
    = \frac{\langle (\delta N_\alpha)^4 \rangle - 3{\langle (\delta N_\alpha)^2 \rangle}^2}{\langle (\delta N_\alpha)^3 \rangle} \ ,
\end{align}
with $\delta N_\alpha = N_\alpha - \langle N_\alpha \rangle$, and $\langle \, \rangle$ denoting an average over the number of events in a fixed centrality class at a specific beam energy \citep{Luo:2017faz}.
These ratios allow for more direct comparisons to theoretical calculations of susceptibilities because the leading order dependence on volume and temperature cancels out (see Eq.~\eqref{eq:susc-cumulants_relation}).  

While direct comparisons between theoretically calculated susceptibilities and multi-particle cumulants have been made, certain caveats exist.  
First of all, these comparisons are only valid if the chosen particle species are good proxies for their respective conserved charge (see e.g. \citealt{Chatterjee:2016mve, Bellwied:2019pxh}).  Additionally, there are fundamental conceptual differences between the assumed in-equilibrium and infinite volume lattice QCD calculations on one side, and the highly dynamical, far-from-equilibrium, short-lived, and finite-size system created in heavy-ion collisions on the other side.
While building cumulant ratios cancels the trivial dependence on volume and temperature, it does not prevent volume fluctuations that can affect the signal, especially for higher order cumulants \citep{Gorenstein:2011vq,Konchakovski:2008cf,Skokov:2012ds,Luo:2017faz}. Calculating the cumulants as a function of the centrality class will generally increase the signal, as the volume of systems  varies within a single centrality class \citep{Luo:2013bmi}. 
Also known as the centrality bin-width effect, this artificial modification of the measured fluctuations can be minimized by using small centrality classes, and some correction methods \citep{Sahoo:2012wn,Gorenstein:2015ria}. 
A second consequence is the fact that the finite size of the system limits the growth of $\xi$. The correlation length must be smaller than the size of the system itself and $\xi$ is even smaller when the system is inhomogeneous \citep{Stephanov:1999zu}. Because the system only approaches the critical point for a finite period of time, the growth of $\xi$ would be consequently limited, restraining even more the size of measured fluctuations amplitude \citep{Berdnikov:1999ph,Hippert:2015rwa,Herold:2016uvv}. Critical lensing effects may somewhat compensate for some of these effects by drawing more of the system towards the critical regime \citep{Dore:2022qyz}.

Finally, the width of the rapidity (angle with respect to the beam line) window in which particle cumulants are measured is important. The signal from critical fluctuations is expected to have a correlation range of $\Delta y \sim 1$, hence cumulants should be measured with particles in a rapidity of this order at least to be sensitive to criticality \citep{Ling:2015yau}. Another alternative is to use factorial cumulants $\hat{C}_n$ (also referred to as correlation functions), which can be expressed as linear combinations of cumulants $C_n$ and are better suited for acceptance dependence studies because of their linear scaling with the rapidity acceptance \citep{Ling:2015yau, Bzdak:2016sxg}. Moreover, acceptance cuts may also affect fluctuation observables and, together with detection efficiency effects, contribute with spurious binomial fluctuations \citep{Pruneau:2002yf,Bzdak:2012ab,Garg:2013ata,Karsch:2015zna,Hippert:2017xoj}. Resonance decays can also lead to spurious contributions, as has been discussed in \cite{Begun:2006jf, Sahoo:2012wn, Garg:2013ata, Nahrgang:2014fza, Bluhm:2016byc, Mishra:2016qyj, Hippert:2017xoj}.

\subsubsection{Net-\texorpdfstring{$p$}{} fluctuations}

Experimental collaborations commonly use the net-proton distribution as a proxy for the baryon number $B$, even though  protons experience isospin randomization during the late stages of the collision \citep{Kitazawa:2011wh,Nahrgang:2014fza}. This process causes the original nucleon isospin distribution to be blurred and is due to the reactions $p + \pi^{0/-} \leftrightarrow \Delta^{+/0} \leftrightarrow n + \pi^{+/0}$ that nucleons undergo several times during the hadronic cascade. These reactions do not affect net-$B$ fluctuations, but do affect net-proton fluctuations. Since protons are the only nucleons measured in the final state, isospin randomization has to be taken into account when comparing both quantities \citep{Kitazawa:2012at}.

\begin{figure}[h!]
    \centering
    \includegraphics[width=0.45\textwidth]{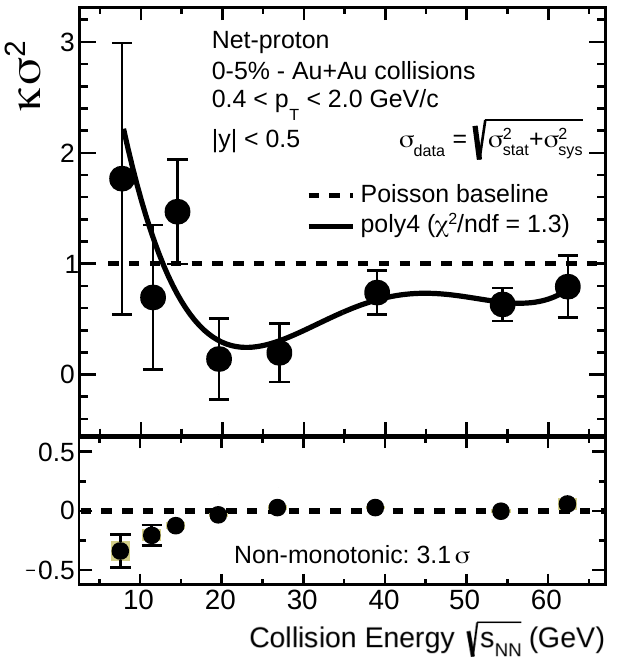}
    \hfill
    \includegraphics[width=0.5\textwidth]{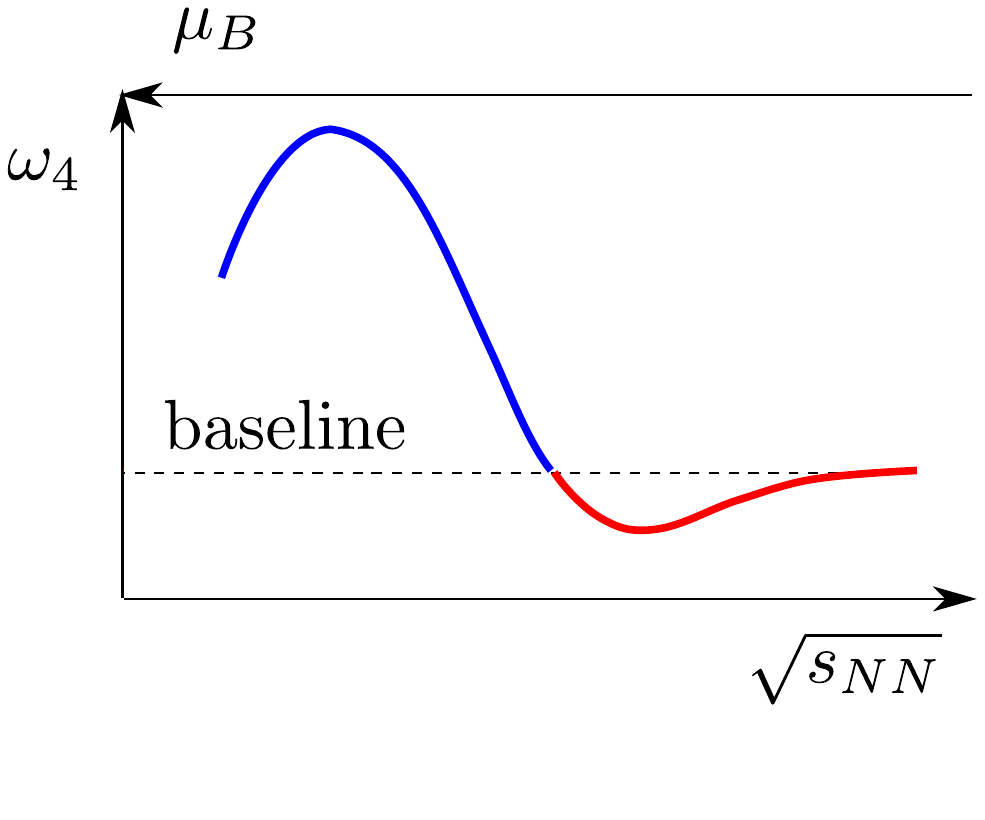}
    \caption{Left: energy dependence of the $C_4/C_2 (=\kappa\sigma^2)$ ratio from net-$p$ distribution for $|y|<0.5$ and $0.4 < p_T < 2.0$ GeV/c, measured in 0--5\% Au-Au collisions by the STAR collaboration. Image adapted from \cite{STAR:2021iop}. 
    Right: expected behavior of the $C_4/C_1(=\omega_4)$ ratio for net-$p$, in the case of a freeze-out line passing through the critical region near the critical endpoint. Image reproduced with permission from \cite{Bzdak:2019pkr}, copyright by Elsevier.}
    \label{fig:C4_over_C2}
\end{figure}

The ALICE collaboration has published measurements of $C_1^p$ and $C_2^p$  net-proton cumulants and their ratios in Pb+Pb collisions at $\sqrt{s_{NN}}=2.76$  TeV \citep{ALICE:2019nbs} and at $\sqrt{s_{NN}}=5.02$ TeV $C_3^p$ was also measured \citep{ALICE:2022xpf}. As the system is created at almost vanishing baryonic chemical potential at such high energies, those cumulants can be compared to lattice QCD susceptibility results like the one discussed in subsection \ref{sub:susceptibilities}, keeping in mind the subtleties of such comparison mentioned in the previous paragraph. However, no critical signal is expected in such collisions, they are mostly used to study correlation dynamics and the effect of global and local charge conservation \citep{ALICE:2022wwr}. Only higher-order cumulants, from $C^p_6$ and beyond, are expected to exhibit $O(4)$ criticality \citep{Friman:2011pf,Almasi:2019yaw}.
At RHIC energies, the STAR experiment has measured net-proton (factorial) cumulants as one of the main objectives of the BES program, in Au+Au collisions from $\sqrt{s_{NN}}=200$ GeV down to $\sqrt{s_{NN}}=7.7$ GeV for cumulants up to $C^p_4$, and even down to $\sqrt{s_{NN}}=3$ GeV for up to $C^p_6$ \citep{STAR:2021iop, STAR:2022vlo}. One of the most interesting results is the energy dependence of the $C^p_4/C^p_2$ ratio, shown in the left panel of Fig. \ref{fig:C4_over_C2}, which exhibits a non-monotonic behavior with a significance of $3.1\sigma$ towards low collision energies \citep{STAR:2021iop}.  The right panel of Fig.~\ref{fig:C4_over_C2} shows a theoretical calculation of one possible critical point from \cite{Stephanov:2011pb} using a 3D Ising model.  Such non-monotonic behavior of the net-proton kurtosis, and more specifically the peak arising after the dip when going to lower energy, had been predicted as an effective sign of the existence a critical region in the phase diagram in \cite{Stephanov:2011pb} -- later work \citep{Mroczek:2020rpm,Mroczek:2022oga} has found exceptions to this using the same framework but incorporating all higher order terms, demonstrating that the peak in kurtosis is the most important signal for the critical point but the bump is not always present. 
Further investigations are already planned to enlighten this special result, by collecting more data in low-energy collisions and especially exploring energies below $\sqrt{s_{NN}} = 7.7$ GeV in the BES-II program \citep{Tlusty:2018rif}. 
They will complete the results of the HADES collaboration, which published a complete analysis for net-$p$ cumulants up to $C^p_4$, in Au+Au collisions at $\sqrt{s_{NN}} = 2.4$ GeV \citep{HADES:2020wpc}.

\subsubsection{Net-charged hadron fluctuations}

\begin{figure}[h!]
    \centering
    \includegraphics[width=0.65\textwidth]{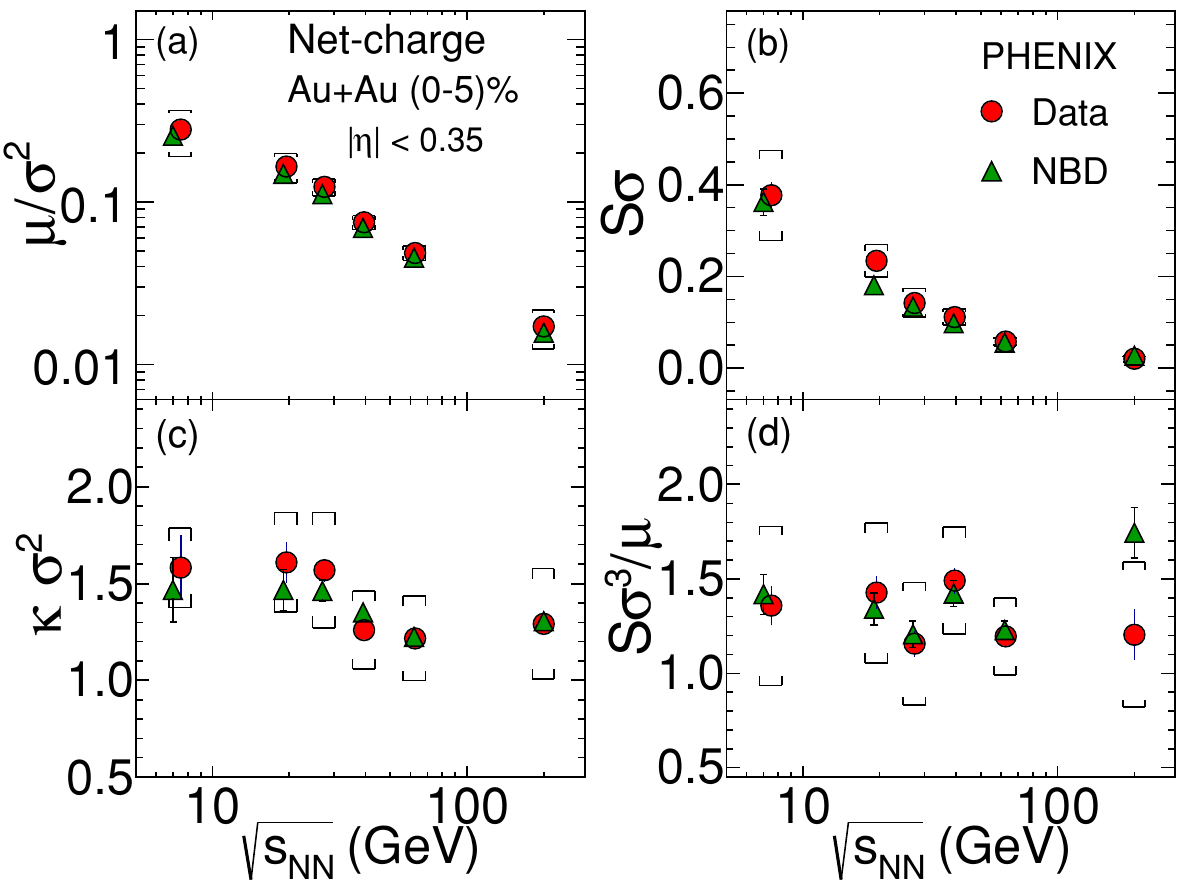}
    \caption{Energy dependence of $\mu/\sigma^2 \sim C_1/C_2$, $S\sigma \sim C_3/C_2$, $\kappa\sigma^2 \sim C_4/C_2$  and $S\sigma^3/\mu \sim C_3/C_1$ for net-$Q$ in central Au+Au collisions for particles with $0.3 < p_T < 2.0$ GeV and within $|\eta| < 0.35$, from the PHENIX collaboration. Data are compared with negative binomial-distribution (NBD). Image reproduced with permission from \cite{PHENIX:2015tkx}, copyright by APS.}
    \label{fig:netQ_cumulants_PHENIX}
\end{figure}

Electric charge fluctuations are the easiest to measure experimentally, as charged particle distributions are accessible even without having to identify the detected particles. Both STAR \citep{STAR:2014egu} and PHENIX \citep{PHENIX:2015tkx} collaborations have published results of net-$Q$ cumulants up to $C_4^Q$ in Au+Au collisions from 7.7 to 200 GeV/A, shown for PHENIX in Fig. \ref{fig:netQ_cumulants_PHENIX}, with no evidence of a peak that could hint at the presence of a critical endpoint.
The same net-$Q$ cumulants have also been measured by the NA61/SHINE experiment in smaller systems (Be+Be and Ar+Sc) for several collision energies within $5.1 \leq \sqrt{s_{NN}} \leq 17.3$ GeV, without any sign of criticality \citep{Marcinek:2022wkm}.  Combining net-p and net-Q fluctuations can be used to extract the  $T,\mu_B$ at freeze-out for a specific $\sqrt{s_{NN}}$ and centrality class (normally central collisions of 0-5\%). This has been done within a hadron resonance gas model where acceptance cuts and isospin randomization can be taken into account \citep{Alba:2014eba,Alba:2015iva,Alba:2020jir} but consistent results have also been found from lattice QCD susceptibilities as well \citep{Borsanyi:2014ewa} that cannot take those effects into account. 

\subsubsection{Net-K, net-\texorpdfstring{$\Lambda$}{} fluctuations}

\begin{figure}[h!]
    \centering
    \includegraphics[width=\textwidth]{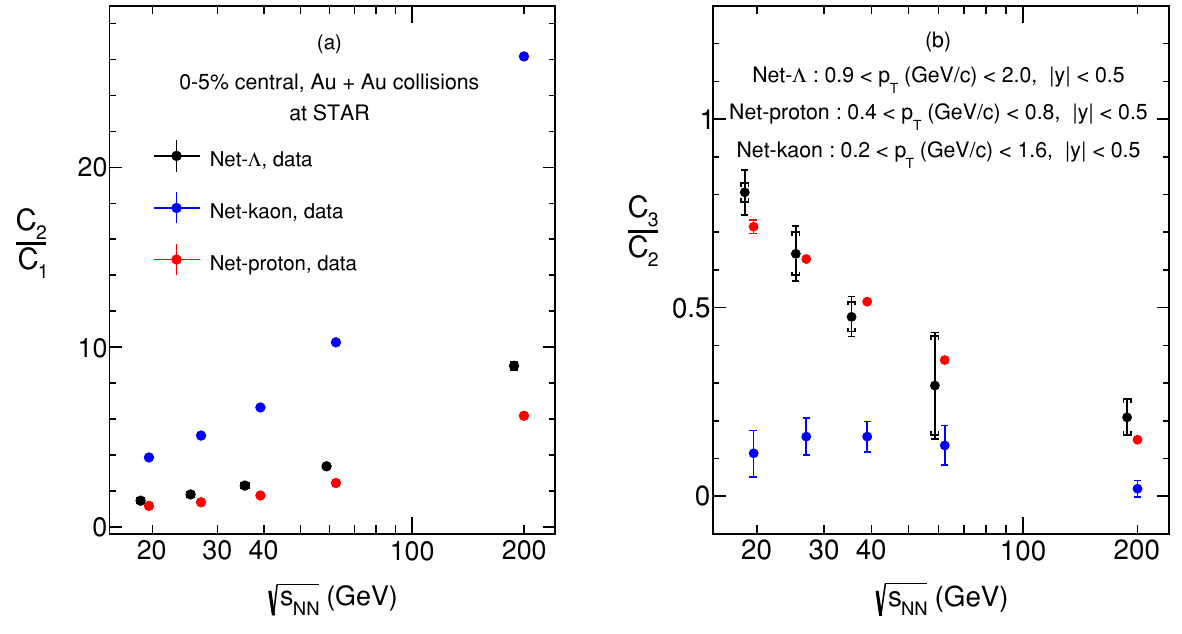}
    \caption{Energy dependence of $C_2/C_1$ and $C_3/C_2$ for net-$\Lambda$, net-$K$ and net-$p$ within $|y| < 0.5$ in central Au+Au collisions, from the STAR collaboration. Image reproduced with permission from \cite{STAR:2020ddh}, copyright by APS.}
    \label{fig:netS_cumulants_STAR}
\end{figure}

In the strangeness sector, net-kaon (specifically $K^\pm$) distributions are used as a proxy, since they are abundantly produced  and easily reconstructed in heavy-ion collisions. Because all other strange particles carry baryon number, the resulting cumulants separated by particle species can provide varying results \citep{Zhou:2017jfk}. These cumulants have been measured extensively by the STAR experiment, again in Au+Au collisions from 7.7 to 200 GeV/A up to $C^K_4$ \citep{STAR:2017tfy}, see Fig.~\ref{fig:netS_cumulants_STAR}. The ALICE collaboration has also published some preliminary results of $C_1$ and $C_2$ for net-$K$  distributions, along with net-$\pi$ and net-$p$ results, in Pb+Pb collisions at 2.76 TeV/A \citep{Ohlson:2017wxu}. Recently, net-$\Lambda$ cumulants up to $C^\Lambda_3$ order and their ratios have also been measured by the STAR experiment, in Au+Au collisions at $\sqrt{s_{NN}}=19.6, 27, 39, 62.4$ and 200 GeV \citep{STAR:2020ddh}. Note that $\Lambda$ results inherently include contamination from $\Sigma^0$ baryons, which decay with a branching ratios of 100\% via the channel $\Sigma^0 \to \Lambda + \gamma$ and cannot be discriminated from primary $\Lambda$ production.
Such results on event-by-event fluctuations of $\Lambda$ baryons are important to investigate the interplay between baryon number and strangeness conservation at hadronization. 

It was proposed in \cite{Bellwied:2013cta} from lattice QCD that there may be a flavor hierarchy wherein strange particles freeze-out at a higher temperature than light particles.  The idea relies on the change in the degrees of freedom in comparisons between various lattice susceptibilities  for light and strange quarks to the hadron resonance gas model.  The original lattice QCD paper suggested strange particles hadronize at $T\sim 10-15$ MeV higher temperatures than light particles.  Using the net-K and net-$\Lambda$ results, it has been shown from both a hadron resonance gas \citep{Bellwied:2018tkc,Bluhm:2018aei} and lattice QCD \citep{Noronha-Hostler:2016rpd} that a preference for a higher freeze-out temperature for strangeness is preferred (although the exact temperature is somewhat model dependent). 

\subsubsection{Mixed conserved charges}

\begin{figure}[h!]
    \centering
    \includegraphics[width=1\textwidth]{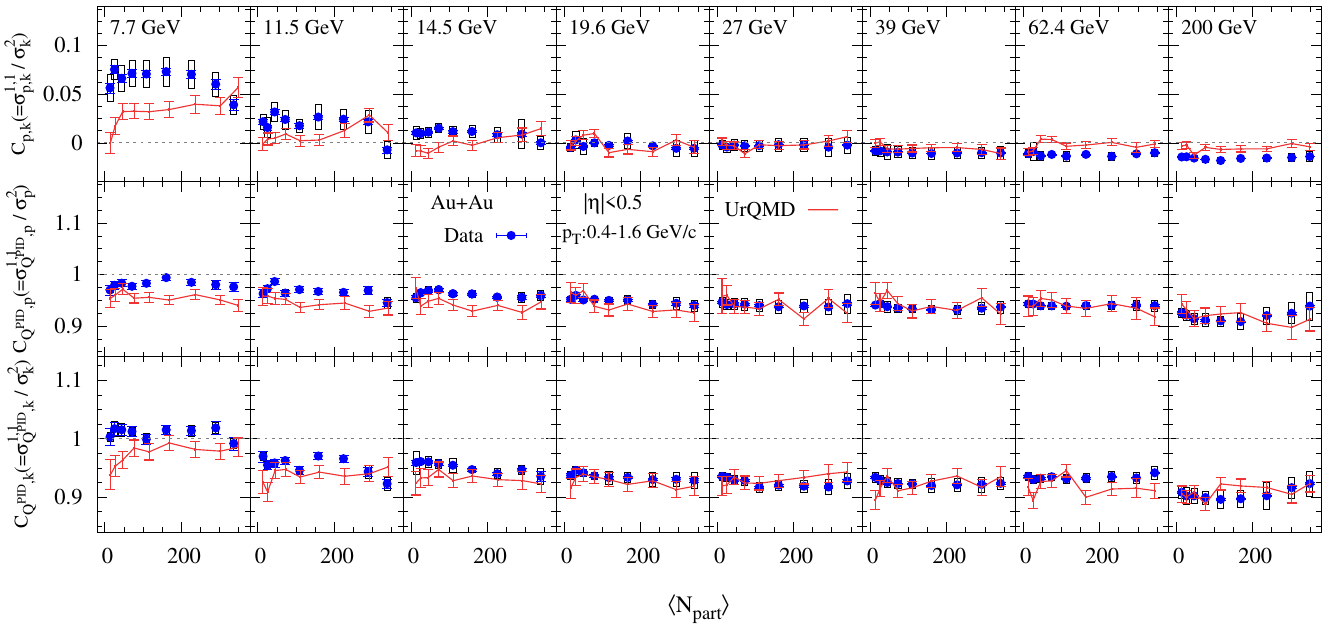}
    \caption{From \cite{STAR:2019ans}. Centrality dependence of covariance to variance ratios for net-$Q$, net-$p$, and net-$K$ distributions in Au+Au collisions at different collision energies from the STAR collaboration, compared with results from ultra-relativistic quantum molecular dynamics (UrQMD) simulations.}
    \label{fig:covariances_vs_Npart}
\end{figure}

In addition to the so-called ``diagonal'' cumulants, i.e., cumulants of net-multiplicity distribution for hadronic species related to a single conserved charge, the STAR collaboration measured off-diagonal cumulants that represent correlations between different conserved charges \citep{Koch:2005vg, Majumder:2006nq}. STAR extracted results for covariances, i.e. $C_2$ mixed-cumulants of net-$Q$, net-$p$, and net-$K$ distributions (proxies for $B$ and $S$ respectively) and their ratios, from Au+Au collisions in the usual BES-I collision energy range $7.7 \leq \sqrt{s_{NN}} \leq 200$ GeV \citep{STAR:2019ans}. The full suite of experimental observables was $\sigma^{1,1}_{Q,p}$, $\sigma^{1,1}_{Q,K}$, $\sigma^{1,1}_{Q,K}$, $\sigma^{1,1}_{Q,p}/\sigma^{2}_{p}$, $\sigma^{1,1}_{Q,K}/\sigma^{2}_{K}$, $\sigma^{1,1}_{p,K}/\sigma^{2}_{K}$.
While both $\sigma^{1,1}_{Q,p}/\sigma^{2}_{p}$ and $\sigma^{1,1}_{Q,K}/\sigma^{2}_{K}$ ratios show only a small collision energy dependence and no peculiar behavior, the $\sigma^{1,1}_{p,K}/\sigma^{2}_{K}$ ratio exhibits a global sign change around $\sqrt{s_{NN}}\sim 20$ GeV, as can be seen in Fig. \ref{fig:covariances_vs_Npart}. Even though not straightforward to interpret, this result might provide important insight into the onset of deconfinement. In \cite{Bellwied:2019pxh} it was argued that the current off-diagonal cumulants were not the best to reproduce lattice QCD due to ``missing" hadrons that could not be measured experimentally (e.g. neutrons).  Thus, it was suggested to instead measure $\sigma^2_\Lambda/\left(\sigma^2_K+\sigma^2_\Lambda\right)$ and $\sigma^2_K/2\left(\sigma^2_\Lambda+\sigma^2_K\right)$ to assess, respectively, strange baryon correlations and strange electric charge correlations. These can be reconstructed using the data from \cite{STAR:2019bjj} and \cite{STAR:2020ddh}. 
% \jj{completed the part about \cite{Bellwied:2019pxh}}

\subsubsection{Summary of cumulant observables for conserved charge proxies}

In terms of theoretical comparisons to these data points, each theoretical approach has its own caveats.  In principle, lattice QCD only provides results for the infinite volume limit, it cannot account for decays, cannot account for limited particle species (i.e. effects like isospin randomization are missed), nor can it account for kinematic cuts.  However, using partial pressures \citep{Noronha-Hostler:2016rpd} it is at least possible to capture fluctuations of certain hadronic species (i.e. kaons) more directly from lattice QCD. A hadron resonance gas approach does have the advantage of fitting lattice QCD very well at temperatures below $T\sim 150-165$ MeV (the exact temperature depends on the observable), can take into account isospin randomization and kinematic cuts, and can calculate quantities for specific particle species.  However, a hadron resonance gas model is dependent on the particle list considered (incomplete particle lists can lead to misleading results) and cannot take into account dynamical effects or out-of-equilibrium effects.  A third option is also often used, which are hadron transport codes like UrQMD \citep{Bass:1998ca, Bleicher:1999xi} or SMASH \citep{Weil:2016zrk, Hammelmann:2022yso} that can take into account all the dynamical and out-of-equilibrium effects. However, transport models have the caveats that they cannot take into account decays and interactions of more than 2 bodies (i.e. $1\rightarrow 3$ body decays are excluded, even though they are known to exist experimentally) and their connection to temperature is more tenuous. 

In Table \ref{table:cumulants} we summarize the latest results from ALICE, STAR, HADES, and NA61/SHINE on all the cumulants for net-proton, net-charge, and net-strangeness.  We note that new results are anticipated later this year from STAR's Beam Energy Scan II program that will significantly reduce the error bars from BES I and also provide new beam energies in the fixed target regime i.e. between $\sqrt{s_{NN}}=3-7.7$ GeV.

\begin{landscape}
\begin{table}[h!]
\caption{Summary of cumulant ratios of different particle species for which measurements have been published by experiments across several collision energies ($\sqrt{s_{NN}}$) and systems.}
\label{table:cumulants}
\centering
\begin{tabular}{cccc}
\hline \hline
   Ratio  & System & Experiment & $\sqrt{s_{NN}}$\\ \hline \hline
   $C^p_2/C^p_1$   & Au+Au  & HADES \citep{HADES:2020wpc}   & 2.4 GeV \\
                   &        & STAR \citep{STAR:2021iop}     & 7.7, 11.5, 14.5, 19.6, 27, 39, 54.4, 62.4, 200 GeV \\
   \hline
   $C^p_2/\langle p+\bar{p}\rangle$ & Pb+Pb  & ALICE \citep{ALICE:2019nbs,ALICE:2022xpf} & 2.76, 5.02 TeV \\
   \hline
   $C^p_3/C^p_1$   & Au+Au  & PHENIX \citep{PHENIX:2015tkx} & 7.7, 19.6, 27, 39, 62.4, 200 GeV \\
   \hline
   $C^p_3/C^p_2$   & Au+Au  & HADES \citep{HADES:2020wpc}   & 2.4 GeV \\
                   &        & STAR \citep{STAR:2021iop}     & 7.7, 11.5, 14.5, 19.6, 27, 39, 54.4, 62.4, 200 GeV \\
   \hline
   $C^p_4/C^p_2$   & Au+Au  & HADES \citep{HADES:2020wpc}   & 2.4 GeV \\
                   &        & STAR \citep{STAR:2022vlo,STAR:2021iop} & 3, 7.7, 11.5, 14.5, 19.6, 27, 39, 54.4, 62.4, 200 GeV \\
   \hline
   $C^p_5/C^p_1$   & Au+Au  & STAR \citep{STAR:2022vlo}    & 3, 7.7, 11.5, 14.5, 19.6, 27, 39, 54.4, 62.4, 200 GeV \\
   \hline
   $C^p_6/C^p_2$   & Au+Au  & STAR \citep{STAR:2022vlo}    & 3, 7.7, 11.5, 14.5, 19.6, 27, 39, 54.4, 62.4, 200 GeV \\
   \hline
   $C^Q_2/C^Q_1$   & Au+Au  & PHENIX \citep{PHENIX:2015tkx} & 7.7, 19.6, 27, 39, 62.4, 200 GeV \\
                   & Au+Au  & STAR \citep{STAR:2014egu} & 7.7, 11.5, 19.6, 27, 39, 62.4, 200 GeV \\
   \hline
   $C^Q_3/C^Q_1$   & Au+Au  & PHENIX \citep{PHENIX:2015tkx} & 7.7, 19.6, 27, 39, 62.4, 200 GeV \\
   \hline
   $C^Q_3/C^Q_2$   & Au+Au  & PHENIX \citep{PHENIX:2015tkx} & 7.7, 19.6, 27, 39, 62.4, 200 GeV \\
                   &        & STAR \citep{STAR:2014egu} & 7.7, 11.5, 19.6, 27, 39, 62.4, 200 GeV \\
   \hline
   $C^Q_4/C^Q_2$   & Au+Au  & PHENIX \citep{PHENIX:2015tkx} & 7.7, 19.6, 27, 39, 62.4, 200 GeV \\
                   &        & STAR \citep{STAR:2014egu} & 7.7, 11.5, 19.6, 27, 39, 62.4, 200 GeV \\
   \hline
   $C^K_2/C^K_1$   &        & STAR \citep{STAR:2017tfy} & 7.7, 11.5, 14.5, 19.6, 27, 39, 62.4, 200 GeV \\
   \hline
   $C^K_3/C^K_2$   &        & STAR \citep{STAR:2017tfy} & 7.7, 11.5, 14.5, 19.6, 27, 39, 62.4, 200 GeV \\
   \hline
   $C^K_4/C^K_2$   &        & STAR \citep{STAR:2017tfy} & 7.7, 11.5, 14.5,  19.6, 27, 39, 62.4, 200 GeV \\
   \hline
   $C^\Lambda_2/C^\Lambda_1$ &        & STAR \citep{STAR:2020ddh} & 19.6, 27, 39, 62.4, 200 GeV \\
   \hline
   $C^\Lambda_3/C^\Lambda_2$ &        & STAR \citep{STAR:2020ddh} & 19.6, 27, 39, 62.4, 200 GeV \\
\hline \hline
\end{tabular}
\end{table}
\end{landscape}

\subsubsection{Other types of fluctuations}

\begin{figure}[h!]
    \centering
    \includegraphics[width=0.65\textwidth]{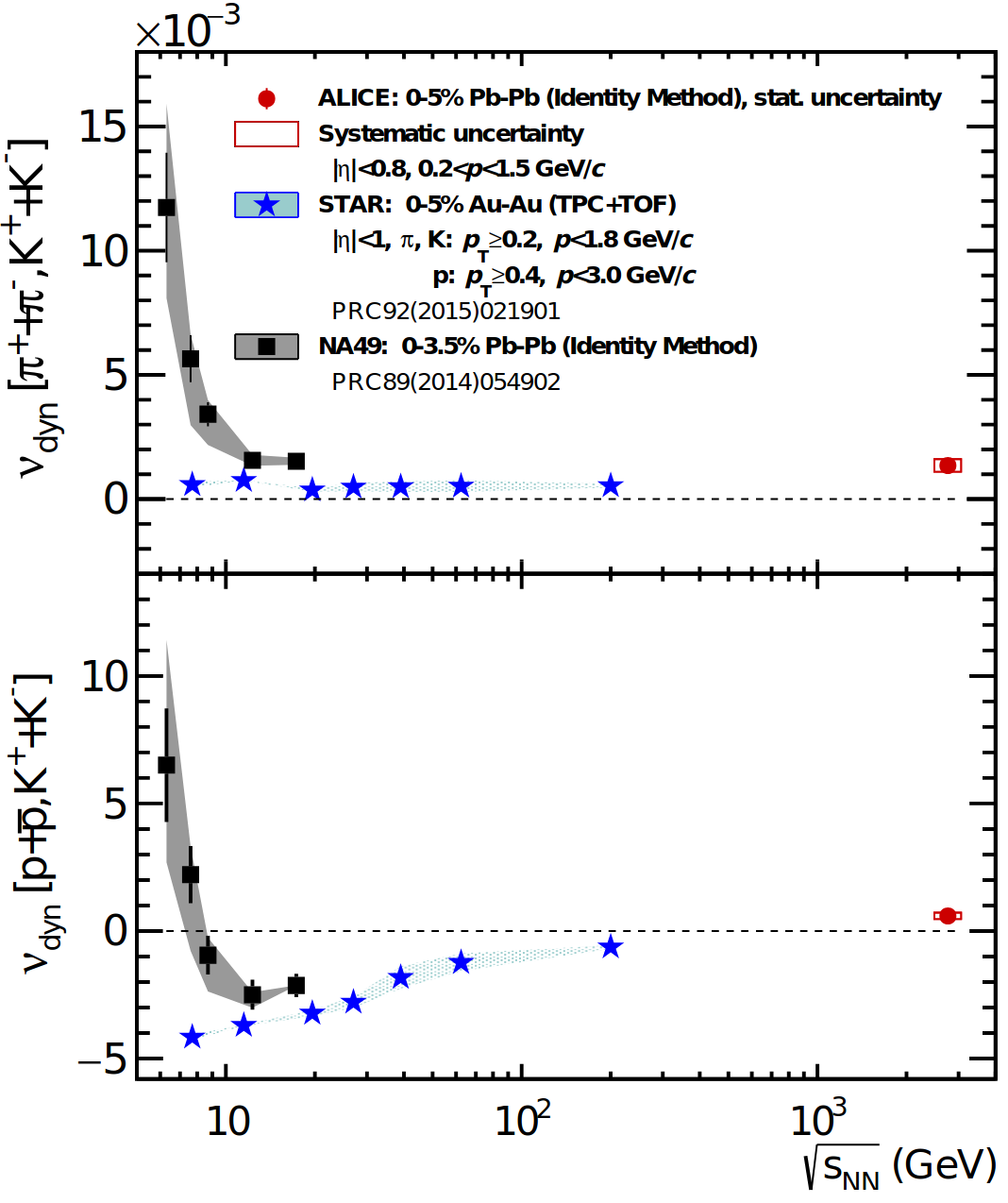}
    \caption{Combined results from STAR measurements in central Au+Au collisions, and ALICE and NA49 measurements in central Pb+Pb collisions on the energy dependence of $\nu_\text{dyn}[\pi,K]$ and $\nu_\text{dyn}[p,K]$ \citep{ALICE:2017jsh}.}
    \label{fig:fluct_correl_piK+pK}
\end{figure}

Other observables used to study two-particle correlations are the dynamical fluctuations, $\nu_\text{dyn}$, which can be used in the case of incomplete particle detection, even though $\nu_\text{dyn}$ intrinsically depends on multiplicity \citep{Gavin:2001uk, Pruneau:2002yf}. 
Two-species correlations between proton, kaon, and pion distributions have been investigated by the NA49, STAR, and ALICE collaborations, in very central Pb+Pb and Au+Au collisions for $6.3 \leq \sqrt{s_{NN}} \leq 17.3$ GeV, $7.7 \leq \sqrt{s_{NN}} \leq 200$ GeV and $\sqrt{s_{NN}} = 2.76$ TeV, respectively \citep{Anticic:2013htn, STAR:2014nuj, ALICE:2017jsh}. The energy dependence of $\nu_\text{dyn}[\pi,K]$ and $\nu_\text{dyn}[p,K]$ from NA49 data, in particular, displays a strong variation below center-of-mass energy $\sqrt{s_{NN}} \sim 10$ GeV, as can be seen in Fig. \ref{fig:fluct_correl_piK+pK}. This could indicate a change in the production mechanism of such particles,  hinting at differences in the phases probed in these collisions \citep{ALICE:2017jsh}.  The ALICE collaboration recently presented preliminary results on the system-size dependence of $\nu_\text{dyn}[+,-]$ normalized by charged particle density to remove its intrinsic multiplicity dependence. The data collected from p+p, p+Pb, and Pb+Pb collisions at $\sqrt{s_{NN}}=5.02$ TeV, and Xe+Xe collisions at $\sqrt{s_{NN}}=5.44$ TeV display a decreasing trend with the size of collided systems that no model has been able to completely reproduce \citep{Sputowska:2022gai}.

Intensive and strongly intensive quantities can be used to investigate hadron number fluctuations, getting rid of the volume dependence and volume fluctuations, as proposed in \cite{Gorenstein:2011vq}. Some of the results introduced earlier, like the scaled cumulants published by ALICE in \cite{ALICE:2022xpf}, STAR in \cite{STAR:2021iop} or NA61/SHINE in \cite{Marcinek:2022wkm} are in fact intensive quantities by construction.

Scaled factorial moments $F_r(M)$ of order $r$ can be used to study the presence of the critical endpoint or phase transition, as they are defined to be sensitive to  large multiplicity fluctuations caused by criticality \citep{Bialas:1985jb, Satz:1989vj, Bialas:1990xd}. 
They are defined as
\begin{equation}
    F_r(M) = \frac{\left\langle \frac{1}{M^D} \sum^{M^D}_{i=1} n_i(n_i-1)...(n_i-q+1) \right\rangle}{\left\langle \frac{1}{M^D} \sum^{M^D}_{i=1} n_i \right\rangle^q}
    \ ,
\end{equation}
with $M^D$ the number of bins in which the $D$-dimensional momentum space is partitioned, $n_i$ the multiplicity of particles of interest in the $i^\text{th}$ bin, and $r$ the order of the moment of interest \citep{Wu:2019mqq}.
The NA61/SHINE collaboration has measured $F_2(M)$ for protons in central Pb+Pb and Ar+Sc collisions, and $F_{2,3,4}(M)$ for negatively charged hadrons at several energies below $\sqrt{s_{NN}}=17$ GeV, showing no signal of any criticality in such systems \citep{Adhikary:2022sdh}. 

Fluctuations of the mean transverse momentum, as those of conserved quantities, would also diverge at the critical point in equilibrium and  are expected to be enlarged in its vicinity \citep{Stephanov:1999zu}. Mean transverse momentum correlations have been measured by the STAR collaboration, for beam energies of $\sqrt{s_{NN}} = 20, 62, 130$ and $200$ GeV \citep{STAR:2005vxr,STAR:2006rqs}. 

\subsection{Flow harmonics}

Another important class of observables giving information about the dynamics of heavy-ion collisions and the EoS of nuclear matter are the flow harmonics $v_n$. They are the coefficients of the Fourier expansion of an $N$-particle triple-differential distribution
\begin{equation}\label{eq:angular_dist}
E\frac{d^3N}{d^3p}  = \frac{1}{2\pi}.\frac{d^2N}{p_T dp_T dy}\left( 1 + 2 \sum_{n=1}^{\infty} v_n. \cos{(n(\phi-\Psi))} \right) \ ,
\end{equation}
where $E$ is the particle energy, $p_{(T)}$ its (transverse) momentum, $y$ its rapidity, $\phi$ its azimuthal angle and $\Psi$ the event plane angle.
The flow coefficients are defined as $v_n = \langle \cos(n(\phi-\Psi)) \rangle$, $\langle \dots \rangle$ denoting an average over many collision events \citep{Voloshin:1994mz, Poskanzer:1998yz}.  These flow coefficients measure the azimuthal anisotropies of particle distributions and are a signature of collective expansion, with $v_2$ measurements playing an important role in the conclusion that QGP formation was observed in Au+Au collisions at RHIC \citep{PHENIX:2004vcz, PHOBOS:2004zne, STAR:2005gfr}.

\subsubsection{Measuring collective flow across $\sqrt{s_{NN}}$: event-plane vs multi-particle cumlant methods}

While collective flow harmonics may initially appear deceptively simple as just a $\cos$ term, there are a number of subtle details that the reader must be aware of before making direct theory-to-experiment comparisons.  There are two primary methods used to calculate $v_n$ for low $p_T$ particles (for the relevance to the EoS we will focus only on low $p_T$ particles but at high $p_T$ other technical details exist both on the theory and experimental side, some of which are discussed in \cite{Betz:2016ayq}). Before explaining the two methods, it is important to first understand that collective flow is not just a scalar quantity but rather it is a vector that contains both the magnitude of the flow $v_n$ that is 0 for a circular event and 1 for the extreme of the corresponding shape (i.e. for elliptical flow, $v_2$, it would appear as a line) and the corresponding event-plane angle $\Psi_n$ that is the direction of the flow vector. Then, the flow harmonic can be defined as the complex vector 
\begin{equation}\label{eq:complex_v_n}
    V_n=v_n e^{in\Psi_n}\ .
\end{equation}
To be clear, $\Psi_n$ is the event plane angle reconstructed by the detector that includes all the usual caveats of having finite number of particles, acceptance cuts, and efficiencies. However, in principle, if all particles were measured to infinite precision one could rigorously define the underlying event plane $\Phi_n$ that is the actual event plane of the given event.  Due to the previously mentioned detector effects, $\Psi_n\neq \Phi_n$ and the dispersion in this relationship can be defined as the resolution, $R,$ for a specific flow harmonic 
\begin{equation}
    R\left(v_n\right)\equiv \langle e^{in\left(\Psi_n-\Phi_n\right)}\rangle\ ,
\end{equation}
wherein the bracket $\langle\dots \rangle$ indicates an averaging over a large ensemble of events. 

At this point, we can discuss the two different methods for calculating flow harmonics. 
The first one is the {\bf ``event plane method'' } \citep{Poskanzer:1998yz}, which was the first technique used to calculate flow harmonics.  At that time it was assumed that dynamical fluctuations would have a negligible effect on the extraction of the event plane, such that $v_n$ would essentially be the same for all events within a fixed centrality class. In this method, $\Psi_n$ is determined from two or more subsets of particles (A and B, known as subevents) within a single event such that
\begin{equation}
    \psi_n^{A,B}\equiv \frac{1}{n}\frac{\sum_{i\in A,B} w_i \sin\left(n\phi_i\right)}{\sum_{i\in A,B} w_i \cos\left(n\phi_i\right)}\ ,
\end{equation}
where $n$ is the number of particles considered and $w_i$ is a relevant weight (such as energy or momentum).  At this point, it is important to discuss the type of particles considered.  The most standard flow measurements are all charged particles.  However, it can be of interest to study the flow of \emph{identified particles} such as protons, pions, or kaons.  In that case, one ``particle of interest", subevent A, is taken (e.g. kaons) and one ``reference particle", subevent B, is taken from all charged particles.  Because particles of interest tend to be rarer, in most cases only one particle of interest is considered rather than 2 (although exceptions exist). Next, the flow harmonic is determined via
\begin{equation}
    v_n\left\{EP\right\}\equiv \frac{\langle \cos\left(\phi_n-\Psi_n^A\right) \rangle_{poi} }{\sqrt{\langle \cos\left(\phi_n-\Psi_n^A\right) \rangle}_{all}}\ ,
\end{equation}
where the average in the numerator is only of particle of interest and the average in the denominator is over all charged particles.

The second method is the ``multi-particle cumulant'' method \citep{Borghini:2001vi,Bilandzic:2010jr,Bilandzic:2013kga} that correlates $m$ number of particles. In this method (here we follow the formalism used in \cite{Luzum:2013yya}) one can correlate $m$ particles such that
\begin{equation}\label{eqn:m_particle_correlation}
    \langle m\rangle_{n_1,n_2,\dots,n_m}\equiv \langle\langle\cos \left(n_1\phi^{a_1}+n_2\phi^{a_2}+\dots+n_m\phi^{a_m}\right) \rangle_m\rangle_{ev}\ ,
\end{equation}
where averages over particles are indicated by the subscript $m$ and averages over events are indicated by a subscript $ev$.  While not shown here, often these averages also include weights (such as by multiplicity) when averaging over multiple events (we will revisit this concept later). At this point we should note that Eq.\ (\ref{eqn:m_particle_correlation}) leads to two distinct contributions that are flow $v_n$ (single particle distribution) and non-flow $\delta_{n,p}$ (genuine $p$-particle correlations that arise from things like a $\rho\rightarrow \pi\pi$ decay).  Because hydrodynamics leads to only flow, experimentalists use various methods to minimize non-flow in their data analysis.  Returning to Eq.\ (\ref{eqn:m_particle_correlation}), a 2-particle correlation of all charged particles (i.e. with no particle of interest) leads to 
\begin{equation}
    v_n\left\{2\right\}\equiv \sqrt{\langle v_n^2\rangle_{ev}+\langle \delta_{2,2}\rangle_{ev} }\ ,
\end{equation}
where one gets a contribution both from genuine flow and 2-particle correlations.  In order to minimize the non-flow contribution, rapidity gaps are taken within the experiments. They remove decay and jet effects that occur close to each other in rapidity. Thus, after these cuts it is reasonable to assume that
\begin{equation}\label{eq:2part-flow}
    v_n\left\{2\right\}\approx \sqrt{\langle v_n^2\rangle_{ev} }\ ,
\end{equation}
such that in theoretical calculations one can directly calculate $v_n$ for a single event and then take the root-mean-squared over many events to  calculate $v_n\left\{2\right\}$. Notice that event-by-event flow fluctuations will also contribute to Eq.~\eqref{eq:2part-flow} and to multi-particle cumulants in general, and should be taken into account when comparing to hydrodynamic simulations. 

Depending if one is in a high resolution limit (i.e. $v_n\gg 1/\sqrt{N}$ and $R\rightarrow 1$) or the low resolution limit (i.e. $v_n\sqrt{N}\ll 1$ and $R\rightarrow v_n\sqrt{N}$) the correct theoretical quantity to calculate varies \citep{Luzum:2012da} such that
\begin{eqnarray}
    v_n\left\{EP\right\}&\xrightarrow[\text{high res}]{}& \langle v_n\rangle\ ,\\ 
    v_n\left\{EP\right\}&\xrightarrow[\text{low res}]{}& \sqrt{\langle v_n^2\rangle}\ . 
\end{eqnarray}
However, most calculations fall between the high and low resolution limits, leading to ambiguous comparisons to experimental data.  Thus, only the multi-particle cumulant method that explicitly defines a two particle correlation as 
\begin{equation}\label{eq:v_n_2}
    v_n\left\{2\right\} \equiv \sqrt{\langle v_n^2\rangle} \ ,
\end{equation}
provides a method for unambiguous comparisons between theory and experiment.

Due to these uncertainties in the comparison between theory and experiment, only the multi-particle cumulant method ensures an apples-to-apples comparison \citep{Luzum:2012da}.  However, currently multi-particle cumulants have not been adopted uniformly across $\sqrt{s_{NN}}$ but rather, at low energies the event-plane method is still used and at high-energies multi-particle cumulants are the standard. As explained in \cite{Luzum:2012da}, this may lead to just a difference of few percentage points in the results, but for precision calculations that can lead to ambiguities.

An additional caveat when comparing high and low beam energies is the choice of event plane angle that is used in the experimental analysis. At high $\sqrt{s_{NN}}$, in the rare occasions in which the event-plane angle is used, the latter is always consistent with the flow harmonic such that $v_2$ is measured with $\Psi_2$, $v_3$ is measured with $\Psi_3$ and so on.  However, at low beam energies all collective flow harmonics are measured with the event-plane method,  relative to
spectator reaction plane, $\Psi_1$ \citep{FOPI:2011aa}.  Thus, the interpretation of the flow harmonics is quite different than those measured at high $\sqrt{s_{NN}}$.  At the time of writing, we are not aware of a systematic study within a theoretical model comparing the differences between these measurements across $\sqrt{s_{NN}}$. However, experiments have compared these methods and found differences between them \citep{FOPI:2005ukb,ALICE:2011ab}. 

Other caveats exist when comparing theory to experiment.  The averaging over events, $\langle\dots \rangle_{ev} $ normally is weighted by a certain factor $W_i$ that is a function of the multiplicity, $M$, of a given event such that 
\begin{equation}\label{eqn:weighting}
    \langle\dots \rangle_{ev} \equiv \frac{\sum_i^{ev}\langle\dots \rangle_i W_i}{\sum_i^{ev} W_i}\ ,
\end{equation}
where for a two-particle correlation $W=M(M-1)$. The averaging in Eq.\ (\ref{eqn:weighting}) biases experimental observables (within a fixed centrality class) to events with high multiplicities because events with large $M$ have a large weight.
 Additionally, experiments often perform calculations in smaller centrality bins (e.g. $0.5\%$) which are then re-assembled into broader centrality bins such as ranges of $5\%$ or $10\%$ depending on the statistics.  While these are minor effects, they do play a role at the few percent level, especially for correlations of 4+ particles \citep{Gardim:2016nrr,Betz:2016ayq}.

For the following sections, we will consider integrated flow harmonics, which include all particles in certain kinematic ranges in transverse momentum, $p_T$, and rapidity, $y$, or pseudorapidity, $\eta$. Later we will discuss differential flow, where one still integrates over (pseudo)rapidity but lets $p_T$ vary.  For the latter approach, the particle of interest is at a fixed $p_T$ and the reference particles are taken across a much wider $p_T$ range. For these calculations, a scalar product must be used
\begin{equation}\label{eq:diff_flow}
    v_n\left\{SP\right\}(p_T)\equiv \frac{\langle v_n v_n(p_T) \cos n\left(\psi_n-\psi_n(p_T)\right) \rangle }{\sqrt{\langle v_n^2\rangle}}\ , 
\end{equation}
where $v_n(p_T)$ and $\psi_n(p_T)$ indicate the magnitude and angle of the flow harmonic at a fixed $p_T$. One can write a very similar equation integrating over $p_T$ and instead varying (pseudo)rapidity.

\subsubsection{Directed flow, $v_1$}

Due to event-by-event fluctuations with rapidity, there are two contributions \citep{Teaney:2010vd,Gardim:2011qn} to directed flow ($v_1$)
\begin{equation}
    v_1(y) e^{i\Psi_1(y)}=v_{1,even}(y) e^{i\Psi_{1,even}(y)}+v_{1,odd}(y) e^{i\Psi_{1,odd}(y)} \ , 
\end{equation}
where $v_{1,even}(y)=v_{1,even}(-y)$, $\Psi_{1,even}(y)=\Psi_{1,even}(-y)$, and  $\Psi_{1,odd}(y)=\Psi_{1,odd}(-y)$ such that they are even across rapidity, whereas $v_{1,odd}(-y)=-v_{1,odd}(y)$ such that it is odd in rapidity.  Before the discovery of event-by-event fluctuations (i.e. if one assumes no initial state fluctuations)  only  $v_{1,odd}(y)$ is relevant.  Thus, $v_{1,even}(y)$ is an entirely fluctuation-driven quantity.  

Even though knowledge of $v_{1,even}(y)$ has existed for over a decade, it has been studied very little because  issues related to momentum conservation make it difficult to perform meaningful comparisons between theory and experiment. On the other hand, $v_{1,odd}(y)$ has been used quite extensively at low $\sqrt{s_{NN}}$ to study the EoS.  In fact, it is more common to consider the slope of directed flow with respect to the rapidity i.e. $dv_1/dy$, which is a notoriously difficult quantity to reproduce in theoretical calculations. Thus, from this point forward we will only consider $v_{1,odd}(y)$ and drop the ``odd" sub-index for convenience.

\begin{figure}[h!]
    \centering
    \includegraphics[width=0.75\textwidth]{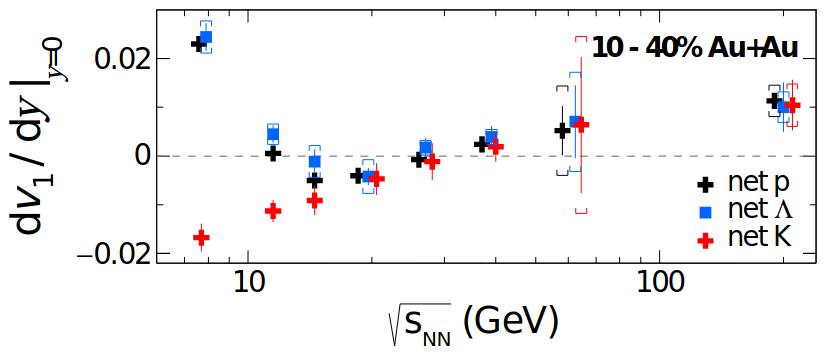}
    \caption{Directed flow of net-$p$, $\Lambda$ and $K$ particle distributions at mid-rapidity as a function of the energy, for Au+Au collisions at intermediate centrality (10-40\%). Image adapted from \cite{STAR:2017okv}.}
    \label{fig:v1_energy_IDhadrons}
\end{figure}

Directed flow is  expected to be sensitive to the EoS, in particular to the presence of a $1^\text{st}$ order phase transition, or the compressibility of nuclear matter \citep{Stoecker:2004qu, Reisdorf:1997fx}. 
For this reason, $v_1$ of charged particles has been measured in Pb+Pb collisions by the NA49 experiment at $\sqrt{s_{NN}}=8.7$ GeV and  $\sqrt{s_{NN}}=17.3$  GeV \citep{NA49:2003njx} and by the ALICE experiment at  $\sqrt{s_{NN}}=2.76$ TeV \citep{ALICE:2013xri}.
The same has been done in Au+Au collisions, from $\sqrt{s_{NN}}=2$ GeV to $\sqrt{s_{NN}}=8$ GeV by the E895 experiment \citep{E895:2000maf}, and at several collision energies within $7.7 \leq \sqrt{s_{NN}} \leq 200$ GeV by the PHOBOS experiment \citep{PHOBOS:2005ylx}, and by the STAR experiment for identified particles \citep{STAR:2014clz, STAR:2017okv}. An interesting observation from these results is the discontinuous slope of the mid-rapidity proton directed flow as a function of the energy, exhibiting a minimum in the $\sqrt{s_{NN}}=10-20$ GeV energy range, as displayed in Fig.~\ref{fig:v1_energy_IDhadrons}.
The interpretation of such results is still unclear though, despite the fact that it has been proposed as a sign for a $1^\text{st}$ order phase transition, as no model can explain this behavior properly \citep{Singha:2016mna}.
At last, $v_1$ measurements have been achieved in asymmetric Au+Cu collisions at $\sqrt{s_{NN}}=200$ GeV too, for charged and identified particles by both PHENIX \citep{PHENIX:2015zbc} and STAR collaborations \citep{STAR:2017ykf}.

\subsubsection{Elliptic flow, $v_2$}

Elliptic flow, another name for the second flow harmonic $v_2$, characterises the ellipticity of the final-state particle momentum distribution, and is proportional to the initial spatial eccentricity of the collision system \citep{Snellings:2011sz}. A nonzero value of $v_2$ is naturally expected to arise due to the initial asymmetry of the colliding part of the system in non-central heavy-ion collisions, especially when going towards more peripheral events, where the almond shape of the overlapping region is accentuated. The pressure gradient, created by the very high initial energy density in the overlapping region, will hence convert the spatial anisotropies into momentum anisotropies while the system cools down \citep{Snellings:2011sz}. 
%Some corrections have to be taken into account when measuring $v_2$ experimentally though, in order to correct from non-flow contribution originating for instance from jets or resonance decays \citep{Borghini:2001vi, Bilandzic:2010jr}.
Using hydrodynamics or transport models, one can then extract the EoS \citep{Pratt:2015zsa} and transport coefficients of the quark-gluon plasma like the shear viscosity over entropy ratio $\eta/s$ \citep{Bernhard:2019bmu}, which are nevertheless highly model-dependent and thus out of the scope of this paper. Because the connection to the EoS is highly dependent on a number of other quantities such as the initial state amongst many others, we refer an interested reader to reviews, e.g. \cite{Heinz:2013th}.

\begin{figure}[ht]
    \centering
    \includegraphics[width=\textwidth]{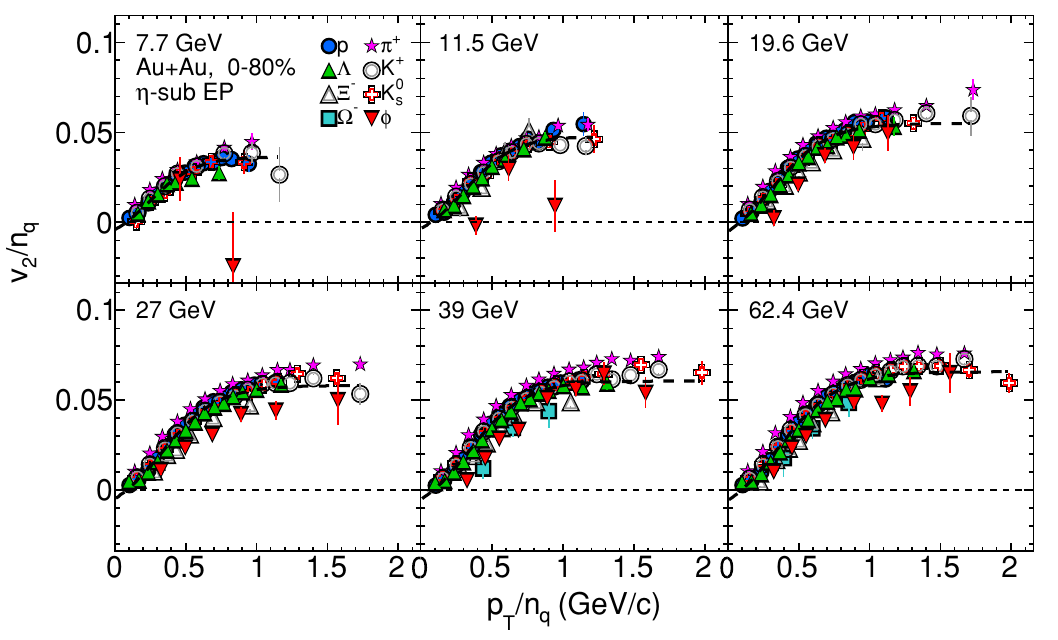}
    \caption{NCQ scaling of the elliptic flow of identified hadrons ($\pi^+$, $K^+$, $K^0_s$, $\phi$, $p$, $\Lambda$, $\Xi^-$ and $\Omega^-$) for 0--80\% Au+Au collisions at $\sqrt{s_{NN}} = 7.7$, 11.5, 19.6, 27, 39 and 62.4 GeV. Image reproduced with permission from \cite{STAR:2013ayu}, copyright by APS.}
    \label{fig:v2_NCQ_scaling}
\end{figure}

Elliptic flow has been measured by several collaborations at the LHC in Pb+Pb collisions at $\sqrt{s_{NN}}=2.76$ TeV and $\sqrt{s_{NN}}=5.02$ TeV for charged particles \citep{ATLAS:2012at, CMS:2012zex, ALICE:2018rtz,CMS:2013bza} and also for identified hadrons \citep{ALICE:2014wao,ALICE:2018yph}, as well as in Xe+Xe collisions at $\sqrt{s_{NN}}=5.44$ TeV \citep{ALICE:2018lao, ATLAS:2019dct,CMS:2019cyz}.
At collision energies of the BESI program, between $\sqrt{s_{NN}}=7.7$ and 200 GeV, $v_2$ has also been extensively studied in Au+Au collision systems for both charged particles \citep{STAR:2002hbo, PHOBOS:2004nvy, PHOBOS:2004vcu, STAR:2012och, PHENIX:2014yml} and identified particles \citep{STAR:2001ksn, STAR:2005npq, STAR:2015rxv, STAR:2022gav} by the STAR, PHENIX and PHOBOS collaborations.
Elliptic flow of smaller systems, namely Cu+Cu collisions at $\sqrt{s_{NN}}=62.4$ and 200 GeV \citep{PHOBOS:2006dbo, STAR:2010ico, PHENIX:2014yml}, Ru+Ru and Zr+Zr collisions at $\sqrt{s_{NN}}=200$ GeV \citep{Sinha:2022cvj} have been measured as well. In addition, $v_2$ data have also been extracted for the large, deformed system of U+U collisions at $\sqrt{s_{NN}}=193$ GeV \citep{STAR:2021twy}.
All of these results show two strong indicators of the common origin of produced particles from a moving fluid, namely the QGP. The first one is the mass ordering of the $p_T$-differential $v_2$ from identified hadrons at low $p_T$ ($<2$ GeV), with lighter species exhibiting a larger $v_2$ than heavier ones \citep{STAR:2010ico, STAR:2013ayu}. It can be understood by assuming that all particles originate from common cells of fluid with a given velocity, thus causing a shift of the $v_2$ signal in momentum due to the mass differences \citep{Huovinen:2001cy,Huovinen:2006jp}. The second one is the so-called Number of Constituent Quarks (NCQ) scaling \citep{Molnar:2003ff} at intermediate $p_T$ ($> 1$ GeV), namely the fact that all $v_2$ signals from different species match when plotted as a function of $p_T$ or the reduced transverse mass, also called the transverse kinetic energy, $K\,E_T=m_T-m_0$, all of them divided by the number of valence quarks $N_q$ ($=2$ for mesons and $=3$ for baryons) \citep{PHENIX:2006dpn,STAR:2010ico, STAR:2013ayu,ALICE:2014wao,ALICE:2018yph,ALICE:2021ibz}. 
Assuming a common origin of the produced hadrons, this scaling shown in Fig.~\ref{fig:v2_NCQ_scaling} illustrates that collectivity arises from quarks as degrees of freedom, before being translated  into flow of the hadrons formed by quark coalescence \citep{Fries:2003vb, Molnar:2003ff}. Such observations constitute, among other probes, a good indicator of the formation of a quark-gluon plasma.
However, $N_q$ scaling starts breaking down in minimum bias Au+Au collisions from $\sqrt{s_{NN}}=11.5$ GeV and below, as can be seen in Fig. \ref{fig:v2_NCQ_scaling} with the $v_2/N_q$ of $\phi$ mesons (although this interpretation is discussed in \cite{Hirano:2007ei}).  

At even lower beam energies of Au+Au collisions at $\sqrt{s_{NN}}= 3$ GeV, the $v_2$ measurements by the STAR experiment \citep{STAR:2021yiu} show no $N_q$ scaling at all. This result tends to indicate the absence of efficient quark coalescence in the system created in such collisions, thus suggesting a dominance of the hadronic phase over deconfined partonic flow. Current hydrodynamic calculations can provide a reasonable match to experimental data down to $\sqrt{s_{NN}}=7.7$ GeV \citep{Shen:2022oyg} and $\sqrt{s_{NN}}=4.3$ GeV \citep{Schafer:2021csj}. However, we point out that other effects may destroy $N_q$ scaling that have not yet been explored at these energies in full (3+1)D relativistic viscous hydrodynamical simulations. For example, the existence of a critical point  leads to critical fluctuations that may have dramatic effects on transport coefficients, which in turn affect collective flow (see \citealt{Lovato:2022vgq} for further discussion on flow at these beam energies).  We also caution that NCQ scaling is only approximate and expected to hold only at intermediate $p_T$ \citep{Molnar:2003ff}. Precise NCQ scaling for $v_2\{2\}(p_T/N_q)/N_q$ is not observed even  at LHC energies \citep{ALICE:2014wao,ALICE:2018yph,ALICE:2021ibz}, so one must be careful when drawing conclusions from its absence. Overall, a much clearer scaling can be seen when plotting against $K\,E_T/N_q$ \citep{PHENIX:2006dpn,ALICE:2014wao}.

\subsubsection{Triangular flow $v_3$ and beyond}

Let us first discuss triangular flow before moving onto higher-order flow harmonics.  The existence of triangular flow was not understood in the field of heavy-ion collisions for many years. In part, this occurred because experimentalists measured triangular flow with the wrong event plane angle (e.g. using $\Psi_2$ instead of $\Psi_3$, leading to a result of $v_3(\Psi_2)=0$).  However, the key issue was that  the importance of event-by-event fluctuations and how significantly they could affect flow harmonics was not understood.  The calculations from \cite{Takahashi:2009na} of 2 particle correlations, were performed in the first theoretical framework to use event-by-event fluctuating initial conditions and was the only model able to reproduce the experimental results at the time. This made it possible for experimentalists to understand the significance of triangular flow and discover the correct method for calculating it \citep{Alver:2010gr}. This discovery of nonzero $v_3$ that can only arise from event-by-event fluctuating initial conditions led to a radical change in the field of heavy-ion collisions because many assumptions had to be changed, but it allowed for new ways to probe the QGP that did not exist previous to the discovery.

\begin{figure}[h!]
    \centering
    \includegraphics[width=0.7\textwidth]{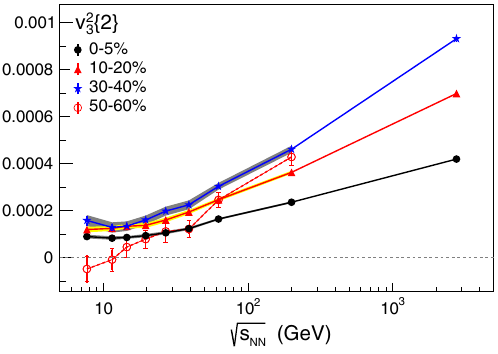}
    \caption{All charged particles $v_3^2\left\{2\right\}$ from the STAR collaboration for various centralities in Au+Au collisions at $\sqrt{s_{NN}} = 7.7$, 11.5, 19.6, 27, 39 and 62.4 GeV. Image reproduced with permission from \cite{STAR:2016vqt}, copyright by APS.}
    \label{fig:v3_2part}
\end{figure}

\begin{figure}[ht!]
    \centering
    \includegraphics[width=\textwidth]{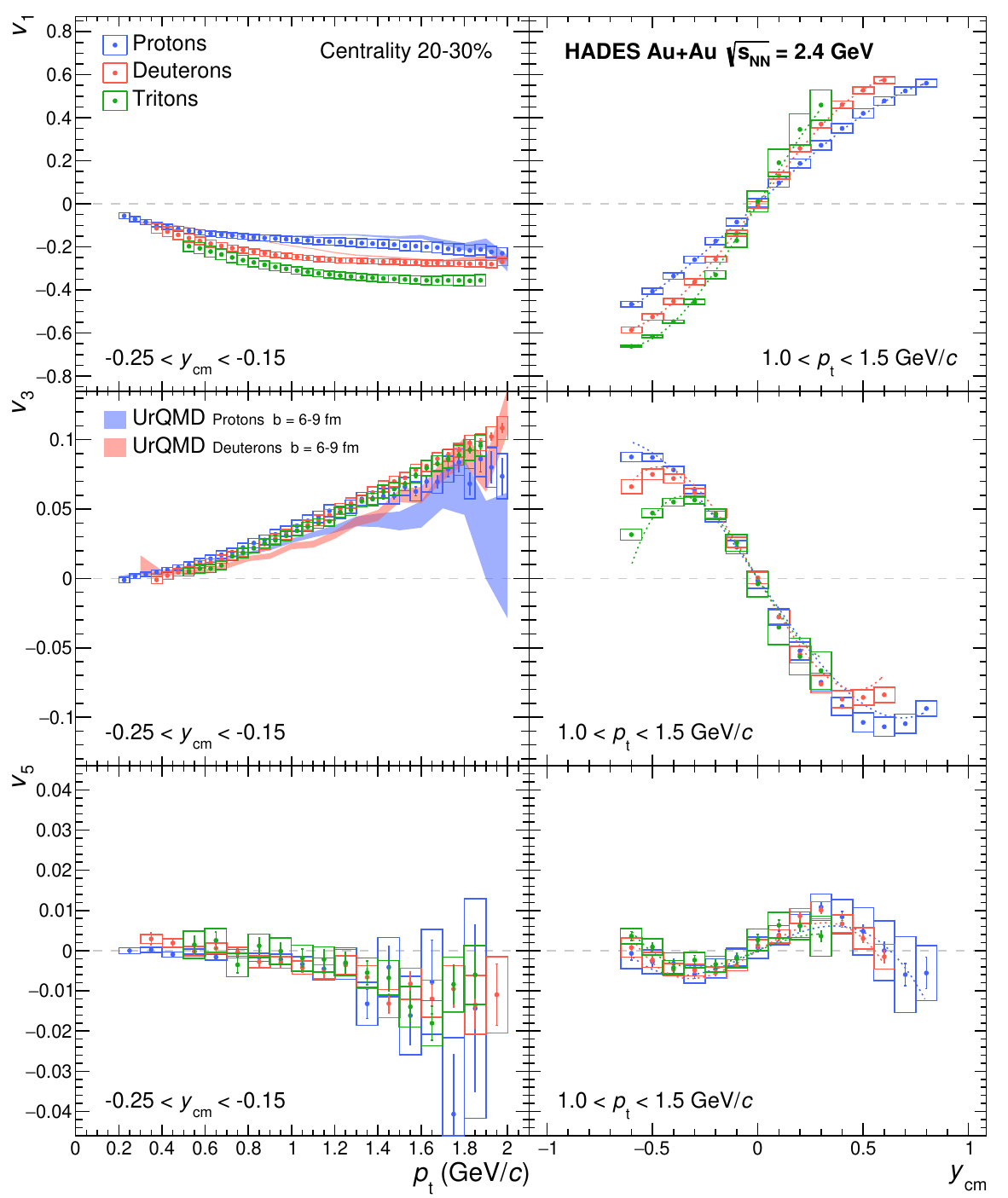}
    \caption{All charged particles collective flow from HADES  in Au+Au collisions at $\sqrt{s_{NN}} = 2.4$ GeV. Image reproduced with permission from \cite{HADES:2020lob}, copyright by APS. }
    \label{fig:HADESflow}
\end{figure}

Higher-order flow harmonics, from $v_3$ and up to $v_6$, have also been measured in order to get complementary information concerning what can be already determined from $v_{1,2}$ results. 
As they require a rather high number of events to be measured with reasonable uncertainties, they have been measured almost exclusively at top energy of the BES program in Au+Au collisions, i.e., $\sqrt{s_{NN}} = 200$ GeV \citep{PHENIX:2014uik, STAR:2020gcl, STAR:2022ncy}, and also in Pb+Pb collisions at $\sqrt{s_{NN}}=2.76$ TeV \citep{ALICE:2011ab,ATLAS:2012at,ALICE:2016cti,CMS:2013bza,CMS:2013wjq} and $\sqrt{s_{NN}}=5.02$ TeV \citep{ALICE:2016ccg,ATLAS:2018ezv,ALICE:2020sup}, as well as in Xe+Xe collisions at $\sqrt{s_{NN}}=5.44$ TeV \citep{ALICE:2018lao,ATLAS:2019dct}. STAR (AuAu $\sqrt{s_{NN}} = 7.7 - 200$ GeV) has now measured the suppression of $v_3\left\{2\right\}$ at low beam energies \citep{STAR:2016vqt} where a minimum is reached (but $v_3\left\{2\right\}$ does not precisely reach 0) whose results are still open for interpretation, see Fig.~\ref{fig:v3_2part}.   Additionally, HADES has also calculated higher-order flow harmonics at $\sqrt{s_{NN}}=2.4 $ GeV \citep{HADES:2020lob}, albeit with the event-plane method with respect to the $\Psi_1$ event plane angle, see Fig.~\ref{fig:HADESflow}. 

In the context of BESI, the STAR collaboration has performed a comprehensive analysis of 3-particle azimuthal anisotropy correlations $C_{m,n,m+n}$, between different harmonics $m$, $n$ and $m+n$, over a wide range of multiplicities, transverse momenta, and energies from $\sqrt{s_{NN}}=7.7\, \mathrm{GeV}$ to $200\, \mathrm{GeV}$ \citep{STAR:2017idk}. The dependence displayed by results on the  pseudo-rapidity differences between particles suggests the breaking of longitudinal boost invariance or dominance of unconventional non-flow contributions. 

\subsubsection{4+ particle flow cumulant observables}

The multi-particle cumulant method used to extract $v_n\{2\}$ in Eq.~\eqref{eq:v_n_2} can be extended to cumulants of higher order, denoted $v_n\{2k\}$ \citep{Borghini:2000sa,Borghini:2001vi,Borghini:2001zr,Bilandzic:2010jr,Bilandzic:2013kga,DiFrancesco:2016srj,Moravcova:2020wnf}. In general, measurements of $v_n\{2k\}$ depend on irreducible $2k$-particle correlations, so that non-collective contributions to $v_n$ are suppressed by a combinatorial factor, which decreases with $k$.  
The four-particle $v_n\{4\}$, for instance, is given by
\begin{equation}
    - v_n\{4\}^4 = \langle v_n^4 \rangle - 2\,\langle v_n^2 \rangle^2\ ,
\end{equation}
where 2-particle contributions are subtracted by the second term (see, for instance \citep{Giacalone:2017uqx}). 

Non-flow contributions set aside, differences between multi-particle cumulants are also sensitive to how the coefficients $v_n$ fluctuate event-by-event \citep{ATLAS:2013xzf,ATLAS:2014qxy}. In the regime in which these coefficients scale linearly with anisotropies of the initial state, i.e. $v_n \propto \varepsilon_n$ \citep{Teaney:2010vd,Gardim:2011xv,Niemi:2012aj,Teaney:2012ke,Qiu:2011iv,Gardim:2014tya,Betz:2016ayq}, where $\varepsilon_n$ quantifies the initial-state anisotropy, ratios between different cumulants depend only on fluctuations of the initial anisotropies, characterized by cumulants $\varepsilon_n\{2k\}$ \citep{Ma:2016hkg,Giacalone:2017uqx}. For instance, 
\begin{equation}
    \frac{v_n\{2 k\}}{v_n\{2\}} \approx \frac{\varepsilon_n\{2k\}}{\varepsilon_n\{2\}}\ ,
\end{equation}
where   the proportionality factor relating initial-state and flow anisotropy cancels out \citep{Yan:2013laa}, so that multi-particle flow measurements can be employed to isolate information on initial-state fluctuations from information on the response of the medium \citep{Giacalone:2017uqx,Bhalerao:2019fzp,Bhalerao:2019uzw}. However, in peripheral collisions \citep{Noronha-Hostler:2015dbi}, small systems \citep{Sievert:2019zjr}, differential flow \citep{Hippert:2020kde}, and to a much lesser extent lower beam energies \citep{Rao:2019vgy} this linear scaling begins to break down, and it requires non-linear response as well. Multiparticle cumulants and fluctuations of flow coefficients have been  measured at $\sqrt{s_{NN}}= 130$ GeV by the STAR collaboration \citep{STAR:2002hbo}, across BESI $\sqrt{s_{NN}} = 7.7 - 200$ GeV from the STAR collaboration \citep{STAR:2022gki}, and at $\sqrt{s_{NN}} = 2.76 - 5.02$ TeV by ATLAS, CMS and ALICE in Pb+Pb collisions \citep{ATLAS:2013xzf,ATLAS:2014qxy,CMS:2017glf,CMS:2017xnj,ALICE:2018rtz}, by ALICE, CMS, and ATLAS in p-Pb collisions \citep{ALICE:2014dwt,CMS:2015yux,CMS:2016fnw,CMS:2017xnj,ATLAS:2017hap,ALICE:2019zfl}, and by ALICE in p+p and Xe+Xe collisions \citep{ALICE:2019zfl}.

While multi-particle cumulants of a single flow coefficient can be used as proxies of its fluctuations, correlations between different coefficients can be studied with so-called symmetric cumulants \citep{Bilandzic:2013kga}
\begin{equation}
     \textrm{SC}(n,m) = \langle v_n^2\, v_m^2 \rangle - \langle v_n^2 \rangle \langle v_m^2 \rangle \ ,
\end{equation}
or normalized symmetric cumulants $\textrm{sc}(m,n)=\textrm{SC}(n,m)/ (\langle v_n^2 \rangle \langle v_m^2 \rangle)$ \citep{Bhalerao:2011yg,Bhalerao:2019fzp,Moravcova:2020wnf}, which can also be generalized to higher orders \citep{Mordasini:2019hut}.  For symmetric cumulants, it is important to ensure that multiplicity weighing and centrality binning are taken into account in order to reproduce experimental data \citep{Gardim:2016nrr}. 

Normalized symmetric cumulants are related to event-plane correlations \citep{Bhalerao:2014xra,Giacalone:2016afq}, 
which can be quantified by examining how combinations of event planes for different harmonics,  $\Psi_{n_1,n_2,\ldots,n_k} \equiv a_1\Psi_{n_1} + a_2\Psi_{n_2} + \ldots+a_m\Psi_{n_k}$ fluctuate \citep{Jia:2012ma,Jia:2012ju,Bhalerao:2011yg,Qin:2011uw}. Taking the Fourier series of the distribution  $dN/d\Psi_{n_1,n_2,\ldots,n_k}$, one finds coefficients
\begin{equation}
    V^j_{n_1,n_2,\ldots,n_k} \equiv \langle \cos\left( j\, \Psi_{n_1,n_2,\ldots,n_k} \right)\rangle\ ,
\end{equation}
where $n$-fold symmetry of the event-plane angle $\Psi_n$ requires that $a_i=n_i\,c_i$, $c_i\in \mathcal{Z}$, and invariance of correlations under global rotations imposes that $\sum_i n_i\,c_i = 0$. The event-plane correlations have been suggested as a method to possibly constrain viscous effects \citep{Niemi:2015qia}.  

The PHENIX collaboration has investigated event-plane correlations in Au+Au collisions at $\sqrt{s_{NN}}= 200$ GeV at RHIC \citep{PHENIX:2011yyh}. 
Results for symmetric cumulants and normalized symmetric cumulants in Au+Au collisions at center-of-mass energies of $\sqrt{s_{NN}}= 27$ GeV, $39$ GeV, $54.4$ GeV and $200$ GeV were also published by STAR in \cite{STAR:2018fpo,STAR:2023akz}, where they are compared to LHC results.  
ATLAS and ALICE have performed measurements of 
event-plane correlations in Pb+Pb collisions at $\sqrt{s_{NN}}= 2.76$ TeV \citep{ATLAS:2014ndd,ALICE:2017fcd}, 
 while ALICE, CMS, and ATLAS  have  published measurements of symmetric cumulants \citep{ALICE:2016kpq,CMS:2017kcs,ATLAS:2018ngv,ALICE:2021adw}.

\subsubsection{Differential flow and scalar products}

Another important tool for constraining  QGP properties is the study of flow anisotropies \emph{differentially} in transverse momentum $p_T$ or (pseudo-rapidity $\eta$) \citep{Luzum:2008cw,Schenke:2011bn,Nijs:2020roc}.    
In theory this amounts to considering the $p_T$   in Eq.~\eqref{eq:angular_dist}, so that the flow coefficients become $v_n = v_n(p_T)$. In practice, this entails binning particles of interest according to their momenta. However, a good determination of the reference event plane $\psi_n$ demands that particles of reference be defined over a wider and suitably chosen momentum range. The differential flow coefficients can thus be found from the scalar product of Eq.~\eqref{eq:diff_flow} (see e.g. \citealt{Bilandzic:2010jr} for details). 

Flow harmonics in different $p_T$ regions reveal different physics. 
In the low $p_T$ region, differential flow observables allow for the investigation of the hydrodynamic response of the QGP as a function of transverse momentum. 
 As discussed above, the mass ordering of elliptic flow at low $p_T$, with lighter particles presenting stronger flow, has been regarded as a signature of hydrodynamic expansion, while the NCQ scaling of flow harmonics at intermediate $p_T$  has been viewed as a signature of quark coalescence \citep{Huovinen:2001cy,Molnar:2003ff,Gyulassy:2004vg,Gyulassy:2004zy,Huovinen:2006jp,STAR:2005npq,STAR:2008ftz,STAR:2008ftz}. At higher $p_T$, flow harmonics encode information on energy loss \citep{Gyulassy:2000gk,Wang:2000fq}.   

Differential flow harmonics have been measured for a variety of systems and beam energies.  For Pb+Pb collisions, at  the LHC differential flow measurements have been made at 
$\sqrt{s_{NN}}=2.76$ TeV  from ALICE \citep{ALICE:2011ab}, CMS \citep{CMS:2012tqw} and ATLAS \citep{ATLAS:2011ah},
 and at $\sqrt{s_{NN}}=5.02$ GeV  from ALICE \citep{ALICE:2016ccg,ALICE:2018yph,ALICE:2022zks} and CMS \citep{CMS:2017xgk,CMS:2019nct}. Measurements have also been performed at RHIC at high beam energies for  Au+Au \citep{STAR:2004jwm,PHENIX:2003qra,PHENIX:2009cjr,PHENIX:2003ccl,PHOBOS:2004vcu}, Cu+Cu \citep{PHENIX:2014yml} and Cu+Au \citep{PHENIX:2015zbc} collisions at $\sqrt{s_{NN}}=200$ GeV and $62.4$ GeV \citep{PHENIX:2014yml},   and Au+Au collisions at $\sqrt{s_{NN}}=130$ GeV \citep{STAR:2000ekf}. 
A comprehensive analysis of p+Au, d+Au, Cu+Cu and Cu+Au collisions at   $\sqrt{s_{NN}}=200$ GeV, as well as U+U collisions at  $\sqrt{s_{NN}}=193$ GeV, from the STAR collaboration can be found in \cite{STAR:2019zaf}. 
 In the context of the BESI there have been measurements from $\sqrt{s_{NN}}=7.7-200$ GeV \citep{STAR:2013cow,STAR:2015rxv,STAR:2008ftz}. 
 
At very low beam energies, sometimes differential flow measurements are reported in terms of the transverse mass $m_T$ such that this is related to $p_T$ through
\begin{equation}
     m_T=\sqrt{m^2+p_T^2}\ ,
\end{equation}
where $m$ is the mass of the particle(s) considered. If only one particle species is considered in the flow, it is possible to translate between $p_T$-dependent flow and $m_T$-dependent flow. However, if multiple identified particles are considered, it is not possible to make this translation without more information from the experiment. 
Fixed target measurements have been made at the STAR fixed target program for $\sqrt{s_{NN}}=3$ GeV \citep{STAR:2021yiu} (in this work it was also normalized by the number of quarks such that $N_q=2$ for mesons and $N_q=3$ for baryons) for $\sqrt{s_{NN}}=4.5$ GeV \citep{STAR:2020dav}. At HADES, using Au-Au collisions at $\sqrt{s_{NN}}=2.4$ GeV \citep{HADES:2020lob}, differential flow was measured in terms of $m_T$. 
Differential flow provides a more stringent test for models of strongly interacting matter, with transport models, represented by SMASH \citep{Mohs:2020awg},  currently unable to reproduce results from HADES \citep{Kardan:2018hna} -- although this may be fixed with further improvements \citep{Sorensen:2023zkk}. 

An interesting consequence of the $p_T$ dependence of flow coefficients is that one may investigate how anisotropies over different $p_T$ ranges correlate with one another \citep{Kikola:2011tu,Gardim:2012im,Heinz:2013bua}. In particular, 2-particle correlations are given by the covariance matrix
\begin{equation}
    V_{n\Delta}^{ab} \equiv V_{n\Delta}(p_T^a,p_T^b) = \langle  V_n^*(p_T^a)\,V_n(p_T^b) \rangle\ ,
\end{equation}
where $p_T^{(a,b)}$ denote the transverse momentum in two $p_T$ bins $a$ and $b$, and we generalize the complex $V_n = v_n \,e^{i\Psi_n}$ of Eq.~\eqref{eq:complex_v_n} for differential flow. In the absence of flow fluctuations or non-flow effects, independent particle emission from the fluid implies that the covariance matrix ``factorizes'' as  
$V_{n\Delta}^{ab} = v_n(p_T^a)\, v_n(p_T^b)$. The breaking of this factorization can signal non-flow contributions \citep{Kikola:2011tu,Jia:2011tse,Adare:2011hd,ATLAS:2012at} or, more interestingly, event-by-event flow fluctuations \citep{Gardim:2012im}, and can be quantified by the factorization-breaking coefficient
\begin{equation}\label{eq:factorization_breaking}
    r_n(p_T^a,p_T^b) \equiv \frac{ V_{n\Delta}(p_T^a,p_T^b)}{\sqrt{V_{n\Delta}(p_T^a,p_T^a)\,V_{n\Delta}(p_T^b,p_T^b)}}
    \leq 1\ ,
\end{equation}
with the inequality being saturated for the case of perfect factorization. The same coefficient can be defined for pseudo-rapidity instead of transverse-momentum bins.   Equation~\eqref{eq:factorization_breaking} is such that, in the regime of linear response to initial anisotropies, the response of the medium cancels out and $r_n$ is determined mainly by initial-state fluctuations. 
This coefficient, and its counterpart for $\eta$-differential flow, has been measured by CMS for both Pb+Pb collisions at $\sqrt{s_{NN}}=2.76$ TeV and p-Pb collisions at $\sqrt{s_{NN}}=5.02$ TeV \citep{CMS:2015xmx,CMS:2017mzx}, and was found to be generally close to, yet below 1. 

Factorization breaking can also be understood in terms of the spectral decomposition and principal component analysis of $V_{n\Delta}^{ab}$, with the case of perfect factorization corresponding to a single nonzero eigenvalue. The principal component analysis of flow fluctuations was proposed in \cite{Bhalerao:2014mua} as a more transparent  alternative to $r_n$, and measured by CMS, in Pb+Pb  collisions,  at $\sqrt{s_{NN}}=2.76$ TeV  \citep{CMS:2017mzx}. However, it has been suggested that the original prescription for carrying out this measurement led to the contamination of subleading anisotropic flow modes by fluctuations of radial flow \citep{Hippert:2019swu}.
Results from hydrodynamic simulations indicate that the factorization-breaking coefficient and subleading components of the PCA of $V_{n\Delta}$ are capable of probing initial-state fluctuations at smaller length-scales, compared to the initial transverse size of the system \citep{Kozlov:2014fqa,Gardim:2017ruc,Hippert:2020kde}. 

\subsection{HBT}

Hanbury Brown--Twiss (HBT) interferometry \citep{Heinz:1999rw, Wiedemann:1999qn, Lisa:2005dd} provides a way to probe the space-time evolution of heavy-ion collisions using correlations between pairs of identical particles emitted from these collisions, such as pions or kaons.\footnote{The more general term \emph{femtoscopy} typically may include, in addition to HBT, non-identical particle correlations and coalescence analyses as well, which may yield additional insights into the space-time dynamics of heavy-ion collisions \citep{Lisa:2005dd}.}  By analyzing the resulting two-particle correlation functions at small relative momentum, one finds that their inverse widths (the `HBT radii' $R^2_{ij}$) correspond to the average separation between the emission points of any two identical particles, and, therefore, provide a way of quantifying the space-time structure of the particle emission process.  Insofar as the bulk of particle production occurs towards the end of the system's lifetime \citep{Plumberg:2015eia, Muller:2022htn}, one expects that HBT measurements should be sensitive to the entire history of the fireball, and, therefore, should depend on the EoS as well.

This expectation turns out to be correct.  Previous studies have established at least five different connections between HBT and the nuclear EoS.  First, the space-time information extracted from HBT can be used to estimate the volume of the system at freeze out.  The freeze-out volume can then be used to estimate the peak energy density and, thereby, obtain evidence for the onset of deconfinement \citep{Heinz:1999rw, Lisa:2005dd}.  Second, the specific observables $R^2_o - R^2_s$ and $R_o/R_s$ are sensitive to the emission duration and rate of expansion of the collision system \citep{Heinz:1999rw, Pratt:2008qv, HADES:2018gop}; measuring these quantities as functions of the pair momentum $K_T$ can constrain the existence or absence of a first-order phase transition at different beam energies and chemical potentials \citep{Lisa:2005dd, Li:2022iil}.  For the same reason, these observables have also been identified as potential signatures of the QCD critical point, by analyzing their scaling with beam energy $\sqrt{s_{NN}}$ to signal a softening in the EoS \citep{Lacey:2014wqa}.  A third connection between HBT and the nuclear equation state uses the moments of collision-by-collision fluctuations in the HBT radii to isolate the geometric effects of critical fluctuations on the system's evolution \citep{Plumberg:2017tvu}, and thereby provides an independent way of probing the QCD critical point.  Fourth, HBT is sensitive to the speed of sound as a function of temperature.  HBT measurements in different collision systems and beam energies (Au+Au at RHIC vs. Pb+Pb at the LHC) have been shown in a multi-observable Bayesian analysis \citep{Pratt:2015zsa, Sangaline:2015isa} to directly constrain a parameterized form of $c_s^2(T)$.  Fifth, the multiplicity dependence of the HBT radii in different size collision systems -- specifically, the slope of $R_{ij}$ vs. $(dN_{ch}/d\eta)^{1/3}$ in p+p, p+Pb, and Pb+Pb -- may also reflect the influence of $c^2_s$ on the rate of the system's expansion \citep{Plumberg:2020jod}.

In addition to these examples, the global shape of the fireball at freezeout is sensitive to the EoS in at least two ways.  First, the HBT radii can be analyzed experimentally as functions of the azimuthal pair emission angle $\Phi_K$, which indicates the direction of a given particle pair's average momentum in the transverse plane.  By expanding this azimuthal dependence in a Fourier series and studying the extracted coefficients as functions of $\sqrt{s_{NN}}$, one acquires sensitivity to the EoS via the freeze-out eccentricity $\varepsilon_F$ (roughly, the normalized second-order Fourier coefficients of the $R^2_{ij}$), which measures the extent to which the initial elliptic geometry of the collision system is inverted by the subsequent dynamical expansion \citep{STAR:2014shf, HADES:2019lek}.  Second, one can also observe sensitivity to the EoS in the transition from prolate to oblate freezeout configurations, which is done by correlating transverse and longitudinal HBT radii as a function of collision energy.  The resulting non-monotonic trend observed in the data \citep{STAR:2020dav} is attributed to a transition from dynamical evolution dominated by nucleon stopping at low energies to boost-invariant evolution at higher energies.

HBT measurements have been carried out in a vast number of different collision systems and over a wide range of collision energies and particle species.  We focus here on HBT with Bose-Einstein correlations, primarily using charged pion pairs.  Important recent measurements include: azimuthally sensitive analyses in Pb+Pb at 2.76 TeV \citep{ALICE:2017gxt, ALICE:2018fdu}, p+Pb at 5.02 TeV \citep{ATLAS:2017jts}, Cu+Au and Au+Au \citep{E895:2000opr, STAR:2003ytv, STAR:2004qya, PHENIX:2014pnh, Khyzhniak:2020xpf}, and Pb+Au at 40, 80, and 158 GeV \citep{CERES:2002tur, CERES:2008oot}.  The majority of studies present results for azimuthally averaged analyses.  At the LHC, HBT has been measured: by ALICE, in Pb+Pb at 2.76 TeV \citep{ALICE:2011dyt}, p+p at 0.9 and 7 TeV \citep{ALICE:2011kmy, ALICE:2019bdw}; by CMS, in p+p, p+Pb, and Pb+Pb at various energies 0.9, 2.76, 5.02, and 7 TeV; and by ATLAS in p+Pb at 5.02 TeV \citep{ATLAS:2017shk}.  At RHIC, STAR has further measured HBT: in U+U collisions at 193 GeV \citep{Campbell:2018qgv}; in Au+Au at collision energies of $\sqrt{s_{NN}} = 4.5$, 7.7, 9.2, 11.5, 19.6, 27, 39, 62.4, 130, and 200 GeV \citep{STAR:2001gzb, STAR:2009sxc, Anson:2011ik, STAR:2014shf, Zbroszczyk:2022zsl}; in p+Au and d+Au at 200 GeV \citep{Khyzhniak:2020ogf}; and in p+p collisions at 200 GeV \citep{STAR:2010yvd}.  PHENIX has conducted measurements in d+Au and Au+Au at 200 GeV \citep{PHENIX:2014dmi} and in Au+Au at 7.9, 19.6, 27, 39, 62, and 200 GeV \citep{Soltz:2014dja}.  Au+Au collisions have been further studied by the HADES collaboration at 2.4 GeV \citep{HADES:2018gop, HADES:2019lek, Greifenhagen:2020wsz}.  The majority of analyses study two-pion correlations for statistical reasons, but HBT has also been studied for kaon pairs by various collaborations at both RHIC and the LHC \citep{STAR:2013che, Nigmatkulov:2016nbe, ALICE:2017iga, ALICE:2019kno} and for multi-pion correlations in various systems \citep{PHENIX:2015jaj, ALICE:2015ryj}.

%%%%%%%%%%%%%%%%%%%%%%%%%%%%%%%%%%%%%%%%%%%%%%%%%%%%%%%%%%%%%%%%%%%%%%%%%%%%%%%%%%%%%%%%%%%%%%%%%%%%%%%%%%%%%%%%%%%%%%%%%%%%%%%%%%%%%%%%%%%%%%%%%%%%%%%%%%%%%%%%%%%%%%%%%%%%%%%%%%%%%%%%%%%%%%%%%%%

\section{Experimental constraints: low-energy nuclear physics}
\label{sec:nuclear}

In the following section, we will review various empirical observations of constraints relevant to low-energy nuclear physics for isospin-symmetric and asymmetric matter.

\subsection{Isospin symmetric matter  at  saturation density}

As neutron stars are in many ways like giant nuclei with $\sim10^{57}$ nucleons \citep{Glendenning:1997wn}, at densities which are typical of nuclei they should reproduce the same properties. Therefore, astrophysicists use laboratory measurements of nuclear experiments to calibrate neutron stars (or dense matter in general) properties. Low-energy laboratory conditions can be considered at $T\sim 0$ when describing fermions, as their order of chemical potential and mass are 10$^5$ times larger than the temperature of fully evolved neutron stars (10$^{-2}$ MeV $\sim 10^{8}$ K) and even more when compared to laboratories. At effectively zero temperature, antifermions do not contribute and the relevant degrees of freedom at $n_{\sat}$  are nuclei (composed of neutrons and protons), which are approximately isospin symmetric, with charge fraction $Y_Q\sim0.5$. The minimal condition on any dense matter theory is to reproduce the experimental results of zero temperature, isospin symmetric nuclear matter at saturation: saturation density $n_{\sat}$, binding energy per nucleon $B/A$, and compressibility $K$. Additionally, for models that contain exotic degrees of freedom, hyperon potentials $U_H$ and $\Delta$-baryon potential $U_\Delta$ can be used. 

These empirical values are known within uncertainties and are extracted using extrapolation within phenomenological models, since infinite nuclear matter does not exist. Here, we  enlist various empirical investigations of the above-mentioned properties. The values for the observables related to isospin-symmetric nuclear matter are summarized in Table \ref{table:iso_cons}, those related to exotic matter in Table \ref{table:exo_cons}.

\subsubsection{Saturation density}

 The saturation density (actually the number density, as it defines the number of baryons per volume) indicates that there is a size saturation in atomic nuclei, preventing them from  expanding or collapsing as a result of the nuclear force's powerful attraction and repulsion. This is the most important constraint, as often other constraints are determined at this density.

The simplest way to calculate a given number density is by dividing the number of baryons in a nucleus over the volume of the nucleus, as \citep{Haensel:1981p} 
\beq
n_{\sat}=\frac{A}{4\pi R^3/3}= \frac{3}{4\pi (R/A^{1/3})^3}=\frac{3}{4\pi R_0^3}=0.17\pm0.03 ~~ \mathrm{fm}^{-3}\ ,
\label{den_sat_1}
\eeq
where it was assumed that the value of the nuclear radius constant is   {$R=R_0 A^{1/3}=1.04-1.17$  fm}.   The error bar reflects the uncertainty on $R_0$, obtained from electron scattering and  $\mu$-mesic atom experiments \citep{Myers1977}.

\begin{figure}[h!]
\includegraphics[scale=0.4]{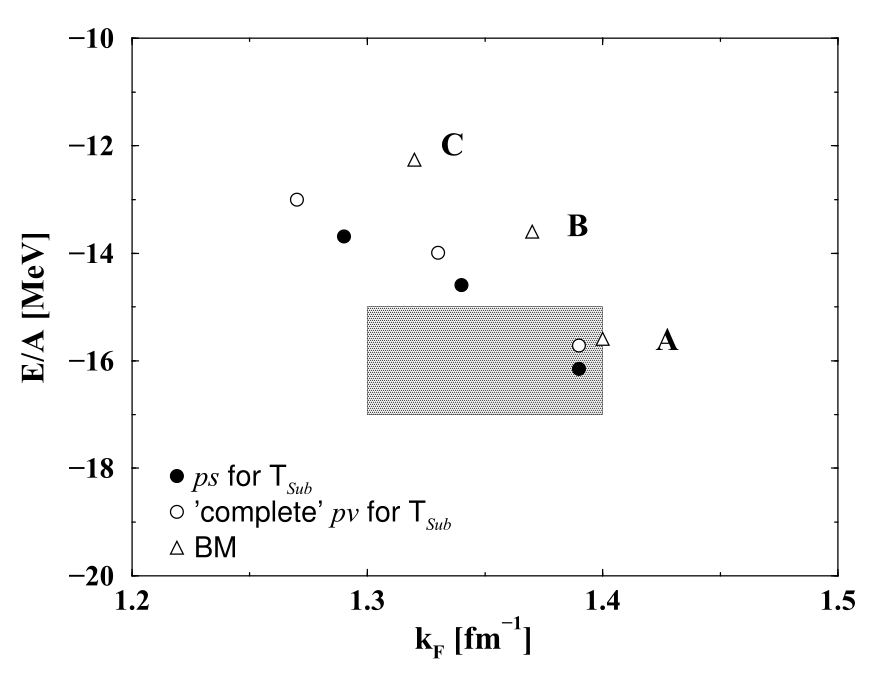}
\centering
\caption{  Binding energy per nucleon ( in this review, referred to as $B/A$) vs. Fermi momentum $k_f$, with the shaded box representing saturation. The  Bonn potentials shown via letters A, B, and C are used for bare nucleon-nucleon interaction. In addition, the three legends describe the T-matrix calculations using a subtraction scheme with the $ps$ representation (top), the $pv$ representation (middle), and Brockmann and Machleidt, $BM$ (bottom) \citep{Brockmann:1990cn}. Image reproduced with permission from \cite{Gross-Boelting:1998qhi}, copyright by Elsevier.}
\label{fig:Binding_energy}
\end{figure}

Recently, in the elastic scattering of
longitudinally polarized electrons extracted from $^{208}$Pb,  the PREX collaboration reported a precise measurement of the parity-violating asymmetry A$_\mathrm{PV}$ term \citep{PREX:2021umo}, where the  interior baryonic density at saturation was derived from the measured interior weak density $n^0_W$ i.e.
\beq
  n_{\sat}=0.1480\pm0.0038 ~~\mathrm{fm}^{-3}\ .
\label{den_sat_3}
\eeq
Here, the uncertainty contains both theoretical and experimental contributions.

\subsubsection{Binding energy per nucleon at saturation}

Binding energy ($B$)  is the energy  required to separate a nucleus, and is given by the semi-empirical mass formula

{\begin{equation}
    B=a_v A-a_s A^{2 / 3}-a_c \frac{Z(Z-1)}{A^{1 / 3}}-a_a \frac{(N-Z)^2}{A}+\delta(A, Z)\ ,
    \label{eq:SEMF}
\end{equation}
where
\begin{equation}
\delta(A, Z) = \begin{cases}
+a_p A^{-1/2} & \text{for even } Z, \text{ even } N \\
0 & \text{for odd } Z, \text{ even } N \text{ or  even } Z, \text{ odd } N \\
-a_p A^{-1/2} & \text{for odd } Z, \text{ odd } N
\end{cases}.
\end{equation}}
In Eq.~\eqref{eq:SEMF}, the first term representing the volume effects ($B_{\mathrm{vol}}$), the second term denoting the surface effects ($B_\mathrm{surf}$), the third term describing the Coulomb interactions ($B_\mathrm{Coul}$), the fourth term describing the effect of isospin asymmetry ($B_\mathrm{asym}$), and the last term covering the effects of the pairing ($B_\mathrm{p}$) \citep{Martin:2019}. The exponent of $A$ in the pairing term is derived from experimental binding-energy data. While it was previously commonly assumed to be 
$-3/4$, more recent experimental data suggest that a value closer to $-1/2$ is more accurate \citep{Myers1977}.  Furthermore, the coefficients for various terms are provided as  $
a_{\mathrm{v}}=15.56$ MeV,  $a_{\mathrm{s}}=17.23$ MeV,  $a_{\mathrm{c}}=0.697$ MeV,  $a_{\mathrm{a}}=93.14$ MeV, and $a_{\mathrm{p}}=12$ MeV. These values constitute one of the datasets employed to adjust the binding energy curve for $A>20$ \citep{Martin:2019}. Since there are no surface effects and Coulomb effects are also ignored in infinite isospin-symmetric nuclear matter, the net binding energy is approximated as the binding energy's volume term ($B \approx B_{vol}$). In \cite{Myers:1966zz}, the volume term of the binding energy per nucleon $B/A$=-15.677 MeV at $n_{\sat}$=0.16146 fm$^{-3}$ was obtained from a non-relativistic semi-empirical four-parameter mass formula with coefficients from the liquid-drop model using the experimental masses of 49 heavy nuclei. In addition,  using experimental data from 1654 ground state masses of nuclei with N, Z $
\ge$ 8,   $B/A$=-16.24 MeV at $n_{\sat}$=0.16114 fm$^{-3}$ was obtained from non-relativistic semi-empirical four-parameter mass formula with coefficients from shell-corrected Thomas--Fermi model \citep{Myers:1995wx}. 

\subsubsection{Compressibility at saturation}

The incompressibility, usually referred to as compressibility, or compression modulus 
$K_{\infty}=\left. 9 \frac{d P}{d n_B}\right|_{n_{\sat}}$ of infinite nuclear matter at saturation is considered one of the vital constraints for dense matter, as it determines the stiffness of the EoS\footnote{ (R3 \#14) At saturation density one can also write $K_{\infty}=9\left[n_B^2 \frac{d^2 E / A}{d n_B^2}\right]_{n_{s a t}}$.}. In finite nuclear systems,  the isoscalar giant monopole resonance (ISGMR) establishes a direct empirical link between the finite nucleus compressibility, $K_A$ and  the centroid energy $E_{\mathrm{ISGMR}}$ \citep{Colo:2013yta},
\bea
E_{\mathrm{ISGMR}}=\sqrt{\frac{\hbar^{2} K_{A}}{m\left\langle r^{2}\right\rangle_{m}}}\ ,
\label{E_ISGMR}
\eea
where $\left\langle r^{2}\right\rangle_{m}$ is the mean square mass-radius in the ground state and $m$ is the nucleon mass.

Furthermore, the finite compressibility can be expressed in terms of the liquid-drop mass formula \citep{Blaizot:1980tw}
\bea
K_{A}=K_{\mathrm{vol}}+K_{\mathrm{surf}} A^{-1 / 3}+K_{\mathrm{asym}} \Big(\frac{N-Z}{A}\Big)^{2}+K_{\mathrm{Coul}} \frac{Z^{2}}{A^{4 / 3}}\ ,
\label{K_A}
\eea
where $K_{\mathrm{vol}}$, $K_{\mathrm{surf}}$, $K_{\mathrm{asym}}$ and $K_{\mathrm{Coul}}$ define the volume, surface, asymmetry and Coulomb terms, respectively. The  $K_{\mathrm{vol}}$  term is related to the nuclear matter properties such that $K_{\mathrm{vol}} \approx K_{\infty}$  \citep{Blaizot:1980tw}.   Inelastic scattering of isoscalar probes can be used to determine the ISGMR strength distribution experimentally. The $\alpha$ particle  has been the most widely used and emerges as effective probe for such observations. The  discussion of ISGMR collective nucleon excitations from $^{90}$Zr and $^{208}$Pb nuclei suggests $K_{\infty}=240\pm20$ MeV \citep{Colo:2013yta,Todd-Rutel:2005yzo,Colo:2004mj,Agrawal:2003xb}, but there might be a softening in this value caused by pairing \citep{Cao:2012dt,Vesely:2012dw}.

\cite{Khan:2012ps} argue that properties of nuclei do not set constraints on the EoS at saturation density, but rather at an average density of $\sim 0.11$ fm$^{-3}$, that they design as crossing density. Considering that the  giant monopole resonance of a chain of nuclei constrains  the third derivative of the energy per unit of volume at this density, they arrive at  $K_\infty=230\pm40$~MeV with 17\% uncertainty.

In another work, the authors reviewed in detail the methods of analysis of giant monopole resonance (GMR) data, as well as values obtained using different techniques and theories between 1961 and 2016 covering a range $K_{\infty}=100-380$ MeV (see references within Table~1 of \cite{Stone:2014wza}), with a trend to higher values in relativistic than in non-relativistic mean-field models.  Without any microscopic model assumptions, except (marginally) the Coulomb effect,  $250 < K_{\infty} < 315$ MeV was obtained and it was shown that surface characteristics have a crucial influence in vibrating nuclei \citep{Stone:2014wza}.

\begin{table}[t!]
\caption{Experimental constraints related to isospin-symmetric nuclear matter at saturation.}
\label{table:iso_cons}
\centering
\begin{tabular}{cc m{4cm}}
\hline \hline
Constraints & Value & Reference   \\ \hline \hline
Saturation density, $n_{\sat}$ (fm$^{-3}$)& 0.17 $\pm$ 0.03  &  \cite{Haensel:1981p}   \\ 
& 0.148 -- 0.185  & \cite{Gross-Boelting:1998qhi}  \\ 
 &  0.148 $\pm$ 0.0038&  \cite{PREX:2021umo}   \\ 
\hline
Binding energy per nucleon, $B/A$ (MeV) & -15.677  &   \cite{Myers:1966zz}   \\ 
     & -16.24  &    \cite{Myers:1995wx}    \\ 
\hline
Compressibility, $K_\infty$ (MeV) &  240 $\pm$ 20 &   \cite{Colo:2013yta,Todd-Rutel:2005yzo,Colo:2004mj,Agrawal:2003xb}   \\ 
  & 210 -- 270 & \cite{Khan:2012ps}\\
   & 251 -- 315 &    \cite{Stone:2014wza}  \\ 
\hline
\hline
\end{tabular}
\end{table}

\subsubsection{Hyperon potentials}
	
In this section, we list various  experimental results relevant to hyperon potentials. In relativistic models, the hyperon potential is defined as $U_H$=vector interaction+scalar interaction, while in non-relativistic models it is defined as $U_H=E_H-T_H-m_H$, the difference between single-particle energy and kinetic energy (plus mass), both usually discussed at saturation. Although more constraining data exists for the potential of the $\Lambda$ hyperon, data also exists for the $\Sigma$ and $\Xi$ potentials.

The analysis of data from the level spectra of $\Lambda$ hypernuclei from $\pi^+K^+$ and $K^-\pi^-$ reactions produced at emulsion and bubble chambers showed that the single-particle energies of $\Lambda$-hypernuclei  vary smoothly with number of nucleons $A$  and are well reproduced considering the potential at nuclear saturation density $U(\Lambda-N)\equiv U_\Lambda=-28$ MeV \citep{Millener:1988hp}.
Considering a slightly different renormalization of the data obtained from the KEK 12-GeV PS and  superconducting kaon spectrometer (SKS, \citealt{Hasegawa:1996fj}), the value $U_\Lambda=-30$ MeV was obtained in \cite{Gal:2016boi}.
In \cite{Shen:2006nv},  using a relativistic mean-field approach, in particular the TM1 model \citep{Sugahara:1993wz}, the data of \cite{Hasegawa:1996fj} have been used to fit the $\sigma$-$\Lambda$ coupling, while considering that the  $\Lambda$-vector mesons  couplings are determined from the SU(6) quark model.  A $\Lambda$-hyperon potential in nuclear matter  at saturation density  equal to  $U_\Lambda=-30$ MeV was obtained. Similar values, $U_\Lambda=-30$ to $-32$ MeV,  have been obtained  with other relativistic mean-field models using the same constraints \citep{Fortin:2017dsj}.

As discussed in \cite{Gal:2016boi}, measurements at KEK (the High Energy Accelerator Research Organization in Japan)  of the $\Sigma^-$ spectrum \citep{Noumi:2001tx,Saha:2004ha}  have indicated that the $\Sigma$-nucleon potential is strongly repulsive. A $\Sigma$ potential in symmetric nuclear matter considered reasonable  is of the order $U_\Sigma=30\pm20$ MeV \citep{Gal:2016boi}. Note that the $\Sigma$-hyperon is predicted by some models, such as the QMC model, not to appear in dense matter for the regime relevant for neutron stars \citep{Stone:2019blq}.

Concerning the $\Xi$-nucleon potential, the measurement of $^{12}_\Xi$Be \citep{AGSE885:1999erv} was described with a Wood--Saxon potential of the order of $U_\Xi=-14$ MeV \citep{Gal:2016boi}. The more recent Kiso event \citep{Nakazawa:2015joa} for $^{15}_\Xi$Ca, if interpreted as a 1$p$ state, seems to indicate a deeper potential.
Relativistic mean field models have been constrained by this measurement and a depth of the order of $U_\Xi=-15$ to $-19$ MeV was obtained \citep{Fortin:2020qin}.  Recently,
considering five two-body $\Xi$ capture events to two single $\Lambda$-hypernuclei obtained by KEK \citep{KEKE176:2009jzw,Nakazawa:2015joa} and J-PARC (the Japan Proton Accelerator Research Complex, \citealt{J-PARCE07:2020xbm}), \cite{Friedman:2021rhu} have calculated an attractive $\Xi$-nucleon interaction with a depth $U_\Xi\gtrsim 20$ MeV ($U_\Xi = 21.9\pm 0.7$ MeV).

Note, however, that a much  less attractive $\Xi$ potential has been calculated  by the HAL-QCD collaboration \citep{Inoue:2019jme}.
Using hyperon interactions extracted from a (2+1) lattice QCD in Brueckner--Hartree--Fock (BHF) calculation with a statistical error of approximately $\pm$2 MeV related with the QCD Monte Carlo simulation, the HAL QCD collaboration anticipated $U_\Lambda=-28$ MeV, $U_\Sigma=+15$ MeV, and $U_\Xi=-4$ MeV \citep{Inoue:2019jme} at $n_{\sat}$, which agrees with $p-\Xi^-$ correlation functions from the ALICE collaboration using 3-momenta measured at $s = \sqrt{13}$ TeV \citep{Fabbietti:2020bfg,ALICE:2020mfd}. 

In a recent study, \cite{Friedman:2022bpw} conducted a direct optical potential analysis, examining the binding energies of both $1s_{\Lambda}$ and $1p_{\Lambda}$ states across the periodic table, covering nuclei from $A = 12$ to $A = 208$. This analytical approach relied on nuclear densities constrained by the charge root-mean-square (r.m.s.) radii. They investigated  the  three-body $\Lambda NN$ (repulsive) interactions and found  $U^{(3)}_\Lambda=13.9\pm1.4$ MeV. The combined effect on $U_{\Lambda}$, computed as the sum of $U^{(2)}_{\Lambda}$ and $U^{(3)}_{\Lambda}$, was calculated to be $-26.7\pm1.7$ MeV at the saturation density. 

\subsubsection{$\Delta$-baryon potential}

Here, we list the empirical observations of spin 3/2 $\Delta$ baryon-nucleon potential at saturation, which is a helpful quantity when constraining its values in dense matter models. As already discussed, $\Delta$'s can replace hyperons without softening the EoS as much, which can be a solution to the hyperon puzzle \citep{Bednarek:2011gd}.  

The introduction of a phenomenological $\Delta$-nucleus spin-orbit interaction was used to improve the fit to the experimental $\pi$-nucleus angular distributions for  $\pi-^{16}$O  at 114 and 240 MeV, $\pi-^{4}$He  at 220 and 260 MeV and $\pi-^{12}$C  at 180 and 200 MeV \citep{Horikawa:1980cv}. It was anticipated that the strength of the $\Delta$-nucleus spin-orbit interaction term is similar (attractive) to the nucleon-nucleon one, i.e. $U_{\Delta} \approx U_N$ \citep{Horikawa:1980cv}.  In another study, the cross-section measurement of electron-nucleus scattering observed in 2.5 GeV Synchrotron at Bonn gives a density-dependent average binding potential $U_{\Delta}(n_B) \simeq$ $-75$ $n_B(r)/n_{\sat}$ MeV \citep{Koch:1985qz}. Furthermore, a range of uncertainty is estimated from the electron-nucleus, pion nucleus, scattering and photoabsorption experiments i.e. $-30$ MeV $+U_{N}<U_{\Delta}<U_{N}$, with $U_{N}\simeq-(50-60)$ MeV, which leads to the constraint $-90$ MeV $<U_{\Delta}<-50$ MeV \citep{Drago:2014oja}.

\begin{table}[h!]
\caption{Empirical constraints related to isospin-symmetric exotic matter at saturation.}
\label{table:exo_cons}
\centering
\begin{tabular}{ccc}
\hline \hline
 Constraints & Value  & Reference   \\ \hline \hline
  Hyperon Potentials (MeV)& $U_\Lambda$ = -28  &  \cite{Inoue:2019jme,Millener:1988hp}   \\ 
  & $U_\Lambda = -30$  &  \cite{Gal:2016boi}\\ 
    & $U_\Lambda = -26.7\pm1.7$  & \cite{Friedman:2022bpw} \\ 
  \hline
& $U_\Sigma = 15$  & \cite{Inoue:2019jme}  \\ 
& $U_\Sigma = 10-50$   & \cite{Gal:2016boi}  \\ \hline
& $U_\Xi$ = $-14$ &  \cite{Gal:2016boi}   \\ 
& $U_\Xi = -21.9 \pm 0.7$ &  \cite{Friedman:2021rhu}   \\ 
& $U_\Xi = -4$ &  \cite{Inoue:2019jme}   \\ \hline
Delta Baryon Potential (MeV) & $U_\Delta \sim U_N$   &   \cite{Horikawa:1980cv}   \\ 
   & $U_\Delta(n_B) = -75 n_B(r)/n_{\sat}$ &    \cite{Koch:1985qz}  \\
   & $-90 < U_\Delta< -50$ &   \cite{Drago:2014oja}   \\  \hline \hline
\end{tabular}
\end{table}

\subsection{Symmetry energy $E_{\sym}$ and derivative $L$}

To translate between isospin symmetric matter  and matter in neutron stars, with very low charge fraction, we make use of the symmetry energy $E_{\sym}$.
The determination of the saturation density $E_{\rm{sym}}$, $L$, $K_{\sym}$, and $J$ parameters is very challenging, and involves large experimental and theoretical efforts \citep{SRIT:2021gcy}. At low densities, we have some fundamental understanding of the symmetry energy but, at large densities, the uncertainty becomes extremely large. At subsaturation densities, nuclear structure probes can generally confine the symmetry energy most effectively, whereas astronomical observations and heavy-ion collisions are two significant methods for constraining the symmetry energy from subsaturation to suprasaturation densities \citep{Zhang:2015ava}. The main source of uncertainty in the $E_{\rm{sym}}$ parameters are the poorly understood many-nucleon interactions.

Within nuclei, the distributions of neutrons and protons differ, and  change with Z and A. As a result, nuclear property measurements, particularly for neutron-rich nuclei, hold promise for restricting nuclear symmetry energy parameters. It is shown that, in addition to the Fermi momentum and the isospin asymmetry parameter, the neutron skin thickness  ($R_{\skin}=R_n-R_p$) of asymmetric semi-infinite nuclear matter is a function of the Coulomb energy,  $E_{\sym,\sat}$, $L_{\sat}$, and $K_{\sym,\sat}$ (see \citealt{Suzuki:2022mow} and references therein). The neutron skin thickness of heavy nuclei such as $^{208}$Pb has been shown to correspond linearly to the slope parameter $L$, which can also be written as $L=3 n_{B} \frac{d E_{\sym}}{d n_B}$, regulating the density dependence of $E_{\sym}$ around the saturation density $n_{\sat}$. As a result, the high-accuracy measurement of $R_{\skin}$ is a significant limit on the density dependence of $E_{\sym}$ at subnormal densities.

\subsubsection{$E_{\sym}$ and $L$ at saturation}
\label{sec:sym_E}

\begin{landscape}
\begin{figure}[ht]
\includegraphics[trim={0 1.9cm 0 1.9cm},clip,scale=0.326]{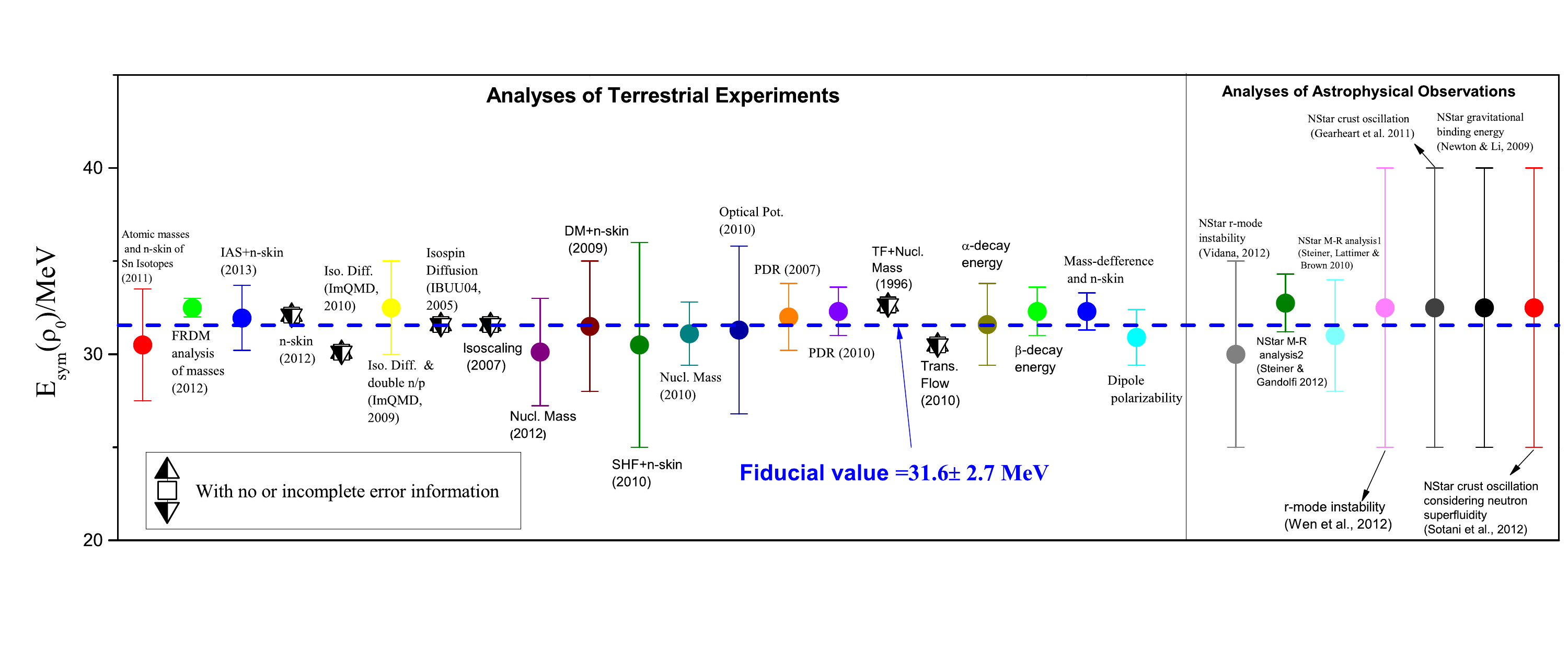}
\includegraphics[trim={0 1.9cm 0.45cm 1.9cm},clip,scale=0.33]{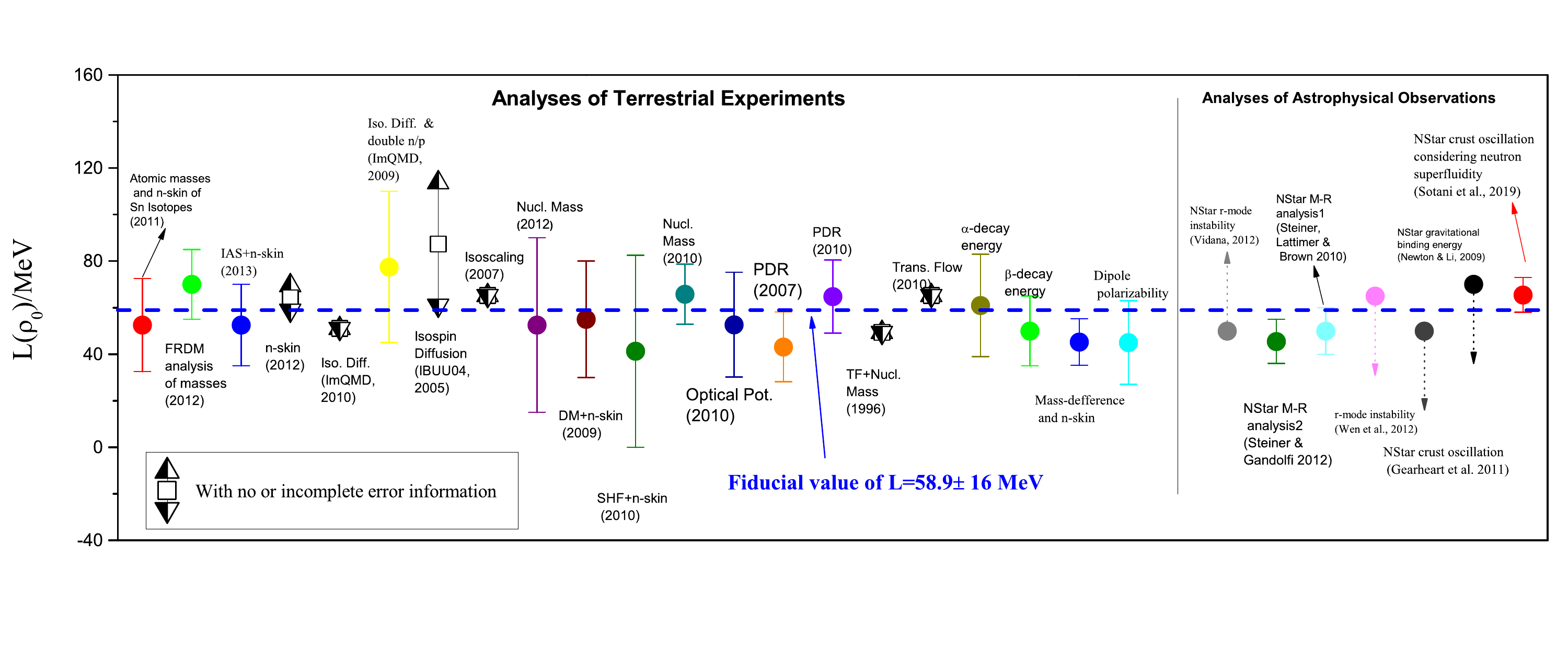}
\centering
\caption{Constraints on saturation values of the symmetry energy (upper panel) and slope (lower panel) at saturation from  terrestrial and astrophysical experiments. Image reproduced with permission from \cite{Li:2019xxz}, copyright by SIF/Springer.}
\label{fig:Esym_L_fiducial}
\end{figure}
\end{landscape}

In this section, we review various experiments to determine the symmetry energy parameters at saturation density. \cite{Li:2019xxz} illustrated a comprehensive list of $E_{\sym}$ and $L$ at saturation density from 28 model assessments of terrestrial nuclear tests and astrophysical data  with the fiducial values of (31.6$\pm$2.7) MeV and (58.9$\pm$16) MeV, respectively (see Fig.~\ref{fig:Esym_L_fiducial}). The S$\pi$RIT collaboration from the Radioactive Isotope Beam Factory (RIBF) at RIKEN measured the  spectra of charged pions produced by colliding rare isotope tin (Sn) beams with isotopically enriched Sn targets  to restrict their contributions at suprasaturation densities. The calculated slope of the symmetry energy is ($42 < L < 117$) MeV using ratios of charged pion spectra observed at high transverse momentum \citep{SRIT:2021gcy}. Furthermore, using the available experimental nuclear masses of heavy nuclei, $E_{\sym}$ was determined, which was employed further to extract $L$=(50.0 $\pm$ 15.5) MeV at $n_{\sat}$ = 0.16 fm$^{-3}$ \citep{Fan:2014rha}. As discussed earlier, the thickness of the neutron skin of $^{208}$Pb offers a strict laboratory restriction on the symmetry energy. Using the strong correlation between $R_{\skin}$ and $L$, the updated Lead Radius Experiment (PREX-II) reported a large $L$=(106 $\pm$ 37) MeV based on both theoretical and actual data, consistently overestimating present limits \citep{Reed:2021nqk}. In another PREX-II study, for $^{208}$Pb the  parity-violating asymmetry $A_{PV}$ was studied using non-relativistic  and  relativistic energy density functionals \citep{Reinhard:2021utv}. A neutron skin thickness of $R^{208}_{\skin}$=(0.19$\pm$0.02) fm, and a small value of derivative $L$=(54$\pm$8) MeV were obtained. 
Recent studies have considered both CREX and PREXI+II measurements  in their analysis. In \cite{Lattimer:2023rpe}, the symmetry energy slope was determined   to be $ L=53\pm13$ MeV. However, \cite{Reinhard:2022inh} have shown the PREX and CREX measurements cannot be simultaneously described by  nuclear models, defined in the framework of an energy density functional theory, which are also consistent with a pool of global nuclear properties.
 The two panels of Fig.~\ref{fig:Esym_L_fiducial}  are in line with earlier astrophysical estimates and show
considerably lower  values for the symmetry energy and its slope at saturation than those recently published using a specific set of relativistic
energy density functionals to describe PREXII as in \cite{Reed:2021nqk}.

\subsubsection{$E_{\sym}$ and $L$ below saturation}

\begin{figure}[h!]
\includegraphics[scale=0.9]{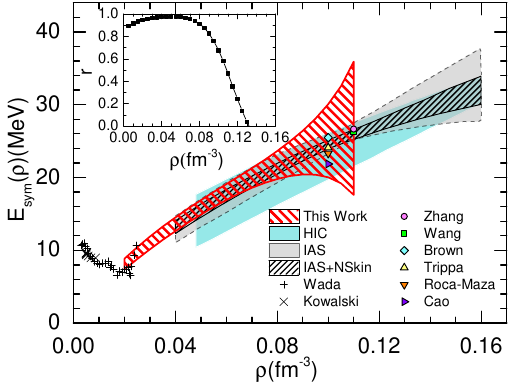}
\centering
\caption{Empirical constraints on symmetry energy  vs baryon density. Image reproduced with permission from \cite{Zhang:2015ava}, copyright by APS.}
		\label{Esym_vs_rho}
\end{figure}

\begin{figure}[h!]
\includegraphics[scale=0.42]{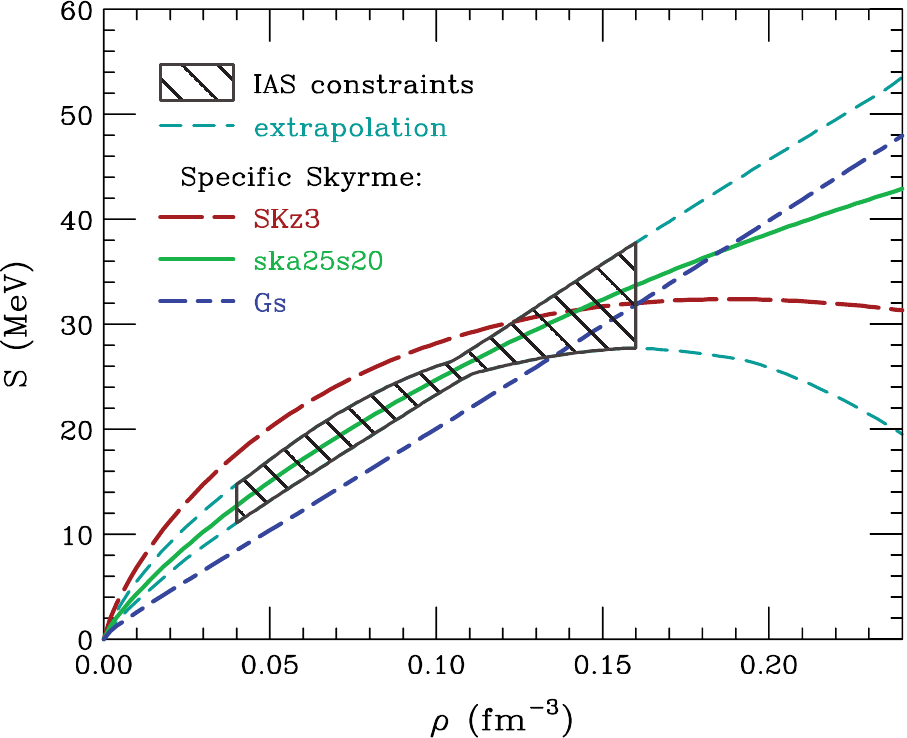}
\includegraphics[scale=0.42]{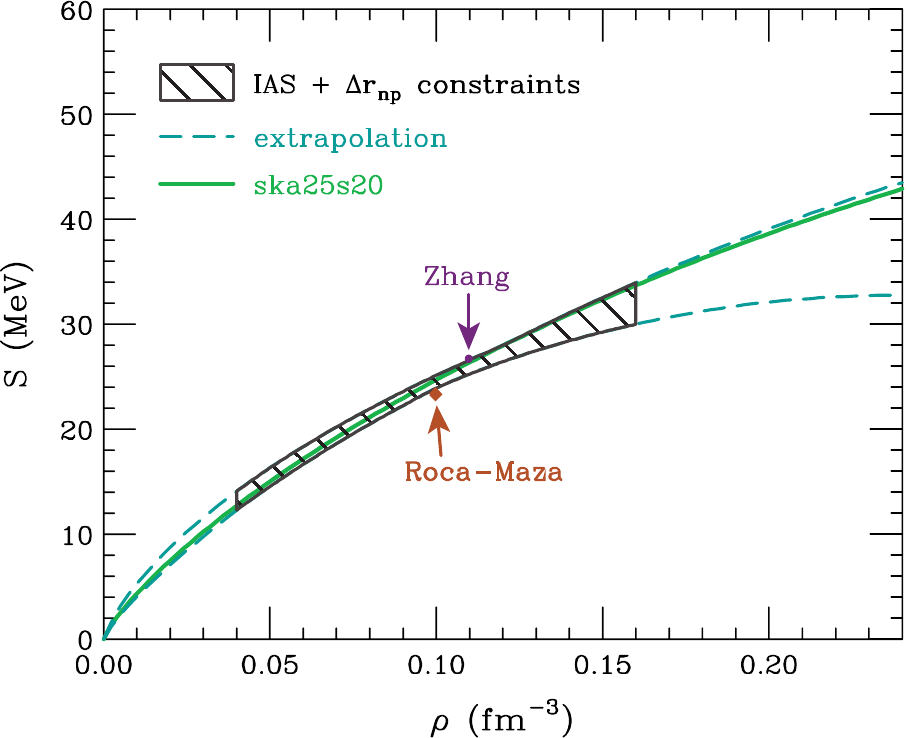}
\centering
\caption{Symmetry energy  vs baryon density without (Left:) and with (Right:) neutron-skin constraints. Image reproduced with permission from \cite{Danielewicz:2013upa}, copyright by Elsevier.}
\label{fig:daniel}
\end{figure}

Constraining the symmetry energy parameters at subsaturation density also helps to restrict the EoS of the neutron-star matter. Here, we discuss experimental measurements for $E_{\sym}$ and $L$ below saturation density. The moderate-temperature nuclear gases formed in the  collisions of  $^{64} \mathrm{Zn}$ projectiles with $^{92} \mathrm{Mo}$ and $^{197} \mathrm{Au}$ target nuclei showed a large degree of $\alpha$ particle clustering at low densities. From isoscaling analyses of the yields of nuclei with $A \leq 4$, the temperature and density-dependent symmetry energy coefficients of these gases at densities $n_B=0.01~n_{\sat}-0.05~n_{\sat}$  were evaluated as $E_{\sym}=9.03-13.6$ MeV (see Table~1 of \citealt{Kowalski:2006ju}).  From the measured excitation energies of the isovector giant quadrupole resonance (IVGQR) in $^{208}$Pb, the symmetry-energy value $E_{\sym}=(23.3 \pm 0.6)$ MeV at $n_B$ = 0.1 fm$^{-3}$ was estimated \citep{Roca-Maza:2012uor}. Furthermore, the magnitude of the symmetry energy $E_{\text {sym }}= (26.65 \pm 0.20)$ MeV at a subsaturation cross density $n_B \approx 0.11$ fm$^{-3}$ was determined by the  binding energy difference $\Delta E$ between a heavy isotope pair \citep{Zhang:2013wna}. Using the available experimental nuclear masses of heavy nuclei $E_{\sym,\sat}$ was determined, which was employed further to extract the density dependent symmetry-energy value, $E_{\sym}=(25.98 \pm 0.01)$ MeV  and  the derivative,  $L=(49.6 \pm 6.2)$ MeV at $n_B$ = 0.11 fm$^{-3}$ \citep{Fan:2014rha}. The magnitude of the symmetry energy $E_{\text {sym }}$ at densities around $n_{\sat}/3$ was determined uniquely by the electric dipole polarizability $\alpha_{\mathrm{D}}$ in ${ }^{208}\mathrm{Pb}$. A stringent constraint of $E_{\mathrm{sym}}=(15.91\pm0.99)$ MeV was observed  at $n_B=0.05$ fm$^{-3}$ and for the range $n_B=0.02-0.11$ fm$^{-3}$, results are illustrated by the red band in Fig.~\ref{Esym_vs_rho} \citep{Zhang:2015ava}.  
In addition, the nuclear symmetry coefficients were extracted using excitation energies to isobaric analog states (IAS) and charge invariance. A narrow  constraint of $E_{\sym}=$32.2 $\pm 2.4$ MeV  was obtained at saturation and for the range of density $(0.04 \lesssim n_B \lesssim 0.13)$ fm$^{-3}$, the behavior is illustrated in  the left panel of Fig. \ref{fig:daniel}. In addition, inclusion of the skin constraints narrows down the  constraints for the same density range (see right panel of Fig.~\ref{fig:daniel}) and at saturation $E_{\sym}$ becomes (32.2$\pm1.1)$ MeV \citep{Danielewicz:2013upa}. Table~\ref{table:asym_cons} summarizes the empirical constraints on $E_{\sym}$ and $L$.
  
\begin{table}[h!]
\caption{Empirical constraints related to isospin asymmetric matter for the symmetry energy and  slope parameter. }
\label{table:asym_cons}
\centering
\begin{tabular}{cccc}
\hline \hline
Constraints \ \ \ \ \ \ & Value\ \ \ \ \ \ \ \ \ \ \ \ \ & $n_B$ & Reference   \\ \hline \hline
$E_{\sym}$ (MeV)& 31.6 $\pm$ 2.7  & $n_{\sat}$& \cite{Li:2019xxz}   \\ 
&9.03 - 13.6  &0.01-0.05 fm$^{-3}$ &\cite{Kowalski:2006ju}  \\    
&15.91  $\pm$ 0.99   &0.05 fm$^{-3}$   & \cite{Zhang:2015ava}   \\ 
&23.3   $\pm$ 0.6    & 0.1 fm$^{-3}$ &\cite{Roca-Maza:2012uor}   \\ 
&26.65  $\pm$ 0.20   & 0.11  fm$^{-3}$ &\cite{Zhang:2013wna}   \\ 
&25.98  $\pm$ 0.01   & 0.11 fm$^{-3}$ & \cite{Fan:2014rha}  \\
&32.20  $\pm$ 2.4    &0.04 $\leq$  $n_B$ (fm$^{-3}$) $\leq$  0.13& \cite{Danielewicz:2013upa}  \\\hline 
$L$ (MeV)& 58.16 $\pm$ 16  & $n_{\sat}$& \cite{Li:2019xxz}   \\ 
&50     $\pm$ 15.5  & $n_{\sat}$& \cite{Fan:2014rha}  \\ 
&54     $\pm$ 8     & $n_{\sat}$& \cite{Reinhard:2021utv}   \\ 
&106    $\pm$ 37    & $n_{\sat}$& \cite{Reed:2021nqk}   \\ 
&49.6   $\pm$ 6.2   & 0.11 fm$^{-3}$ &  \cite{Fan:2014rha}  \\  \hline
\hline
\end{tabular}
\end{table}
  
\subsubsection{Correlation of symmetry energy and slope parameter}

\begin{figure}[h!]
\includegraphics[scale=0.4]{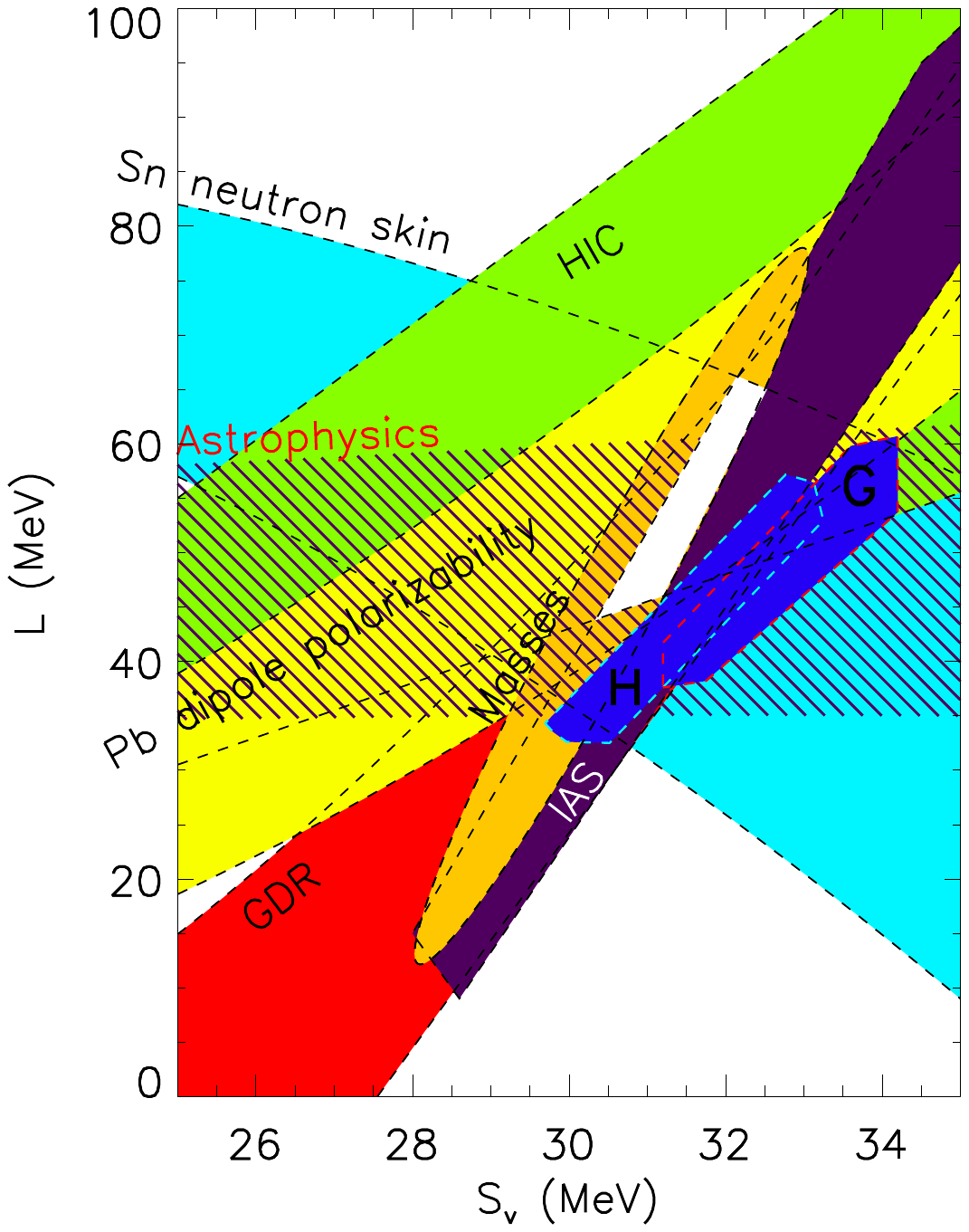}
\centering
\caption{Correlation between symmetry energy and slope at saturation. Image reproduced from \cite{Lattimer:2014sga}, copyright by SIF/Springer.}
		\label{fig:correlation}
\end{figure}

As discussed earlier, the symmetry energy and its slope play an important role in constraining the EoS of neutron-star matter. Thus, the correlation between both parameters can be useful to understand their interdependence. \cite{Lattimer:2014sga} illustrate the experimentally determined values of $E_{\sym}$ and $L$ parameters. Figure~\ref{fig:correlation} contains  different regimes coming from the observations of low-energy heavy-ion collisions, astrophysics, neutron skin, giant dipole resonance (GDR), isobaric analog states (IAS), and dipole polarizability.

\subsubsection{Heavy-ion collision measurements of neutron skin of $^{197}$Au and $^{238}$U}

Recently, from heavy-ion collisions it was discovered that it is possible to measure the nucleus charge radius by using the energy dependence of the Breit--Wheeler process \citep{Wang:2022ihj}.  Additionally, the STAR experiment has also found the total matter radius determined using the diffractive photoproduction in ultra-peripheral collisions to observe a unique spin interference pattern in the angular distribution of $\rho\rightarrow \pi^++\pi^-$ decays \citep{STAR:2022wfe}.  Combining these two measurements, they found that for $^{197}$Au there was a neutron skin of 0.17 $\pm$ 0.03 (stat.) $\pm$ 0.08 (syst.) fm and a for $^{238}$U there was a neutron skin of 0.44 $\pm$ 0.05 (stat.) $\pm$ 0.08 (syst.) fm \citep{STAR:2022wfe}. To date these results have not yet been used in theoretical calculations but they do appear to be consistent with PREXII results.

\subsection{Isospin-symmetric nuclear matter liquid-gas  critical point}
\label{subsecLG}

The similarity between nucleon-nucleon  and Van der Waals interactions suggested the  possibility of also having a liquid-gas phase transition in nuclear matter if the effects of Coulomb and limited size are ignored. If that is the case, the existence of a coexistence line associated with a first order transition stopping at a critical point is expected. Indeed in the liquid-gas phase diagram, the surface tension of a nucleus diminishes as its temperature rises, eventually disappearing at a point known as the critical point \citep{Landaustatbook}. The liquid-gas phase transition line divides two separate nuclear phases, i.e., nuclei (towards low density) and bulk matter (towards high density). In many models, if just nuclear matter is taken into account and Coulomb interactions are disregarded, this phase transition is anticipated as first-order \citep{Soma:2009pf,Fiorilla:2011sr} whereas in astrophysical environments, the production of nuclei involves the Coulomb and other interactions that occur in finite nuclei, where some characteristics of the first order phase transition may be retained \citep{Buyukcizmeci:2012it}.

\begin{figure}[h!]
\includegraphics[scale=0.7]{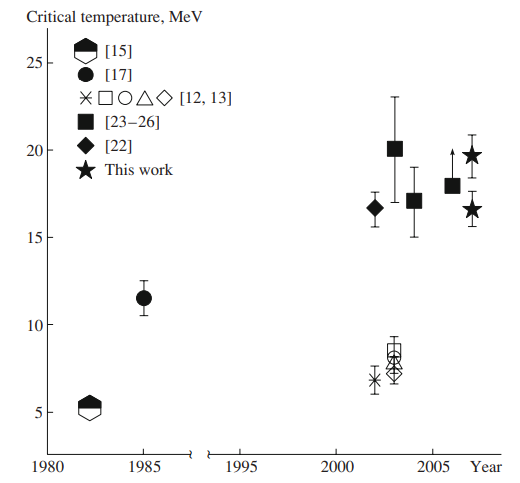}
\centering
\caption{Constraints on the critical temperature of the liquid-gas phase transition. Image reproduced with permission from \cite{Karnaukhov:2008be}, copyright by Pleiades.}
		\label{fig:critical_temperature}
\end{figure}

In this section, we review the different  nuclear reaction experiments relevant to the liquid-gas phase diagram of  symmetric nuclear matter. The critical temperature found in the studies on finite nuclei \citep{Karnaukhov:2003vp,Elliott:2013pna,Natowitz:2002nw,Karnaukhov:2008be} is in the range $T_c  \approx 17-20$ MeV. In a study from empirical observations of limiting temperatures, the critical temperature of infinite nuclear matter  was found to be $T_c = 16.6 \pm 0.86\,$ MeV \citep{Natowitz:2002nw}. Also, in p $+$ Au collisions at 8.1 GeV, the charge distribution of the intermediate mass fragments was analyzed in the framework of the statistical multifragmentation model, and the critical temperature  was found to be $T_c = 20 \pm 3\,$ MeV \citep{Karnaukhov:2003vp}.  Furthermore, in another study, a critical temperature $T_c > 15$ MeV was anticipated for the nuclear liquid-gas phase transition from fission and multifragmentation data (see Fig.~\ref{fig:critical_temperature} for a list of $T_c$ obtained from different studies \citep{Karnaukhov:2008be}).  In the latest study, six different sets of experimental data obtained from  the Lawrence Berkeley National Laboratory
 88-inch cyclotron, Indiana Silicon Sphere Collaboration and Equation of State Collaboration indicated that infinite nuclear matter has critical rms values for temperature $(T_c = 17.9 \pm 0.4)\,$ MeV, density ${n_B}_c =( 0.06 \pm 0.01)\,$ fm$^{-3}$, and  pressure $P_c = (0.31 \pm 0.07)$, MeV fm$^{-3}$ \citep{Elliott:2013pna}. The critical temperature is essentially a parameter that affects how quickly the surface tension falls when the nuclei heat up. The values mentioned in this work are useful to constrain nuclear matter models at finite temperature. They are summarized in Table~\ref{table:LG_cons}.

\begin{table}[h!]
\caption{Empirical constraints for the critical temperature, pressure, and baryon density related to liquid-gas critical point. }
\label{table:LG_cons}
\centering
\begin{tabular}{ccc}
\hline \hline
 Constraints & Value & Reference   \\ \hline \hline
  $T_c$ (MeV)& 16.6 $\pm$ 0.86  &  \cite{Natowitz:2002nw}   \\ 
& 20 $\pm$ 3& \cite{Karnaukhov:2003vp}  \\ 
& $>$15& \cite{Karnaukhov:2008be}  \\ 
 & 17.9 $\pm$ 0.4&  \cite{Elliott:2013pna}   \\ \hline
  $P_c$ (MeV fm$^{-3}$) \ \ \ & 0.31 $\pm$ 0.07   &  \cite{Elliott:2013pna}  \\  \hline
  ${n_B}_c$ (fm$^{-3}$) & 0.06 $\pm$ 0.01  &   \cite{Elliott:2013pna}    \\   \hline \hline
\end{tabular}
\end{table}

%%%%%%%%%%%%%%%%%%%%%%%%%%%%%%%%%%%%%%%%%%%%%%%%%%%%%%%%%%%%%%%%%%%%%%%%%%%%%%%%%%%%%%%%%%%%%%%%%%%%%%%%%%%%%%%%%%%%%%%%%%%%%%%%%%%%%%%%%%%%%%%%%%%%%%%%%%%%%%%%%%%%%%%%%%%%%%%%%%%%%%%%%%%%%%%%%%%%%%%%%%%%%%%%%%%%

\section{Observational constraints: astrophysics}
\label{sec:astro}

Measurements of stellar masses, radii and tidal deformabilities play a central role in establishing a link between neutron stars' microscopic and macroscopic properties.  These observables depend upon the internal structure and composition of the matter that makes up the neutron star. This microphysical information is relayed through the equations of stellar structure by the EoS, which cannot be directly measured in laboratory experiments at the densities, temperatures, and isospin asymmetries relevant for neutron star cores. In the astrophysical context, the EoS is probed by a variety of astronomical observations of neutron stars, particularly electromagnetic observations in the radio and X-ray bands, as well as gravitational waves from the coalescence of compact binaries. These probes assume general relativity is the correct description of nature, as otherwise, observables would also depend on the theory of gravity that is considered (see e.g. \citealt{Yagi:2013bca,Yagi:2013awa}). 

Neutron stars, whether in isolation or in binaries systems, are described by the Einstein equations, which require, as input, the EoS to connect the pressure to the energy density and temperature inside the stars. When in isolation and when considering non-rotating stars, the Einstein equations reduce to the Tolman--Oppenheimer--Volkoff (TOV) equations \citep{Tolman:1939jz,Oppenheimer:1939ne}, whose solution determines the mass and radius of the non-rotating star for a given central energy density. Choosing a sequence of central densities yields a sequence of pairs of masses and radii, which together form a ``mass-radius curve.'' For slowly-rotating stars, one can expand the Einstein equations in small rotation through the Hartle--Thorne approximation\citep{Hartle:1967he,Paschalidis:2016vmz}, and the solution to the expanded equations yields the moment of inertia at first order in rotation, the quadrupole moment and a mass correction at second order. When a neutron star is not in isolation, but instead is perturbed by some external object (like a binary companion), the perturbations can also be studied by solving the linearized Einstein equations about a neutron star background, which yields the tidal deformabilities of the star. In this way, one can construct various curves, including the mass-radius, moment of inertia-mass, quadrupole moment-mass, tidal deformability-mass, and so on, that represent neutron stars of various central densities. These curves can change shape depending on the EoS prescribed. Since one usually expects neutron stars to all be described by a unique EoS \footnote{There is also the possibility of a two-family scenario, where more than one distinct EoS is used to describe neutron stars \citep{Drago:2015cea}}, there is a single set of such curves that correctly describes Nature. Astrophysical observations can determine which set is the correct one, and thus, which is the correct EoS to describe neutron stars. 

In the next subsections, we discuss one by one the empirical macroscopic constraints obtained from various astronomical observations.
We provide summaries of current constraints with only brief explanations of the underlying measurements.
See, e.g., \cite{Chatziioannou:2020pqz} for a more thorough review.

It is important to remember that the densities relevant for the cores of neutron stars can certainly reach several times nuclear saturation density (see, e.g., Fig.~5 of \cite{Legred:2021hdx}, Fig.~6 of \cite{Mroczek:2023zxo}, Fig.~7 of \cite{Pang:2023dqj}, Fig.~5 of \cite{Koehn:2024set} for recent estimations), and connecting constraints from theoretical calculations and terrestrial experiments at lower densities (Sects.~\ref{sec:eft} and~\ref{sec:nuclear}) to neutron star observations will almost certainly involve some extrapolation.
The amount of freedom allowed in extrapolation schemes may be, but are not always, informed by physical models.
Some care is, therefore, warranted when interpreting results combining low-density and high-density constraints as the conclusions often depend on the particular extrapolation scheme chosen.
See, e.g., \cite{Essick:2021kjb, Essick:2021ezp, Legred:2022pyp} for more discussion.

%%%%%%%%%%%%%%%%%%%%%%%%%%%%%%%%%%%%%%%%%%%%%%%%%%%%%%%%%%%%%%%%%%%%%%%%%%%%%%%%%%%%%%%%%%%%%%%%%%%%%%%%%%%%%%%%%%%%%%%%%%%%%%%%%%%%%%%%%%%%%%%%%%%%%%%%%%%%%%%%%%%%%%%%%%%%%%%%%%%%%%%%%%%%%%%%%%%%%%%%%%%%%%%%%%%%

\subsection{Neutron-star maximum mass}

The measurement of a neutron star's mass sets a lower bound on the mass above which the neutron star must undergo gravitational collapse to a black hole.
Most observed neutron stars spin slowly compared with their Keplerian break-up rotation rate.
Hence, mass observations are often used to bound the maximum mass achievable by cold, non-rotating neutron stars even though spin can support up to approximately $\sim 20\%$ additional mass \citep{Breu:2016ufb}.

The radio pulsar PSR J1614-2230 was the first massive pulsar  to be measured with a mass  close to two solar masses \citep{Demorest:2010bx}. Recently,
its mass was updated to 1.908$^{+0.016}_{-0.016}\, M_\odot$ \citep{NANOGrav:2017wvv}.  The pulsar PSR J1614-2230 has a low mass companion and the determination of its mass was possible through the measurement of the Shapiro delay, a retardation effect of the pulse signal that originates on  the curvature of the space-time close to the companion, and is given by \citep{vanStraten:2001zk}
$$\Delta_S=2r \ln[1-s\cos(\phi-\phi_0)], $$
where $s=sin\, i$ with $i$ the inclination angle, $r=G\,m_2/c^3$ with $m_2$ the companion mass, $\phi$  is the orbital phase,
and $\phi_0$ is the phase
where the pulsar is on the opposite side of the companion from Earth. The measurement of $\Delta_S$ allows the determination of the companion mass.

The record for the highest precisely and reliably measured neutron star mass currently belongs to the 2.8-ms radio pulsar (PSR) J0740+6620, which is likely orbiting an ultracool white-dwarf companion \citep{Fonseca:2021wxt}.
The data sets of this study are integrated pulse arrival-time measurements, obtained using the Canadian Hydrogen Intensity Mapping Experiment (CHIME) and 100-meter Green Bank Telescope (GBT).
Timing solutions for PSR J0740+6620 were produced by the GBT and CHIME/Pulsar collaborations using narrow-band and wide-band times of arrival.
While comparing the credible intervals of the Shapiro delay parameters across different dispersion measures (DM) evolution models, all calculated solutions were found to be statistically compatible with different DMX models at 68.3\% credibility, with a 1\% fluctuation in the credible intervals of the Shapiro delay.
The Shapiro delay, also known as the gravitational time delay effect, is an increase in travel time of a signal when it passes near a massive object.
The estimated mass of PSR J0740+6620 is $m_{p}=2.08_{-0.07}^{+0.07}$ M$_{\odot}$ (in solar masses) and of its companion is $m_{c}=0.253_{-0.005}^{+0.006}\, M_{\odot}$ \citep{Fonseca:2021wxt} calculated at $1\sigma$ (68.3\% credibility), which is consistent with the earlier observations of \cite{NANOGrav:2019jur} (posterior in \citealt{POS208}).

The second-highest precisely and reliably measured pulsar mass belongs to PSR~J0348+0432.  For this pulsar, the mass is estimated to be $2.01\pm 0.04\,M_\odot$ at 68.3\% credibility based on a combination of radio timing of the pulsar and precise spectroscopy of the white dwarf companion, which has a mass of $m_c=0.173\pm 0.003\,M_\odot$ and is in a 2.46-hour orbit around the pulsar \citep{Antoniadis:2013pzd}.

An \emph{upper} limit on the maximum mass of nonrotating neutron stars has been placed, albeit with more substantial astrophysical uncertainties, using the properties of short gamma-ray bursts.  These have long been assumed to be produced by the merger of two neutron stars, with the remnant either forming a black hole immediately or collapsing quickly to a black hole (e.g., \citealt{Murguia-Berthier:2014pta}).  Under this assumption, and working off of the mass distribution of double neutron star systems in our Galaxy, several authors proposed that the maximum mass $M_{\max}$ of a nonrotating neutron star was $\sim 2.3\, M_\odot$ \citep{Bauswein:2013jpa,Lawrence:2015oka,Fryer:2015uia}. 
The gravitational wave and electromagnetic data from the binary neutron-star merger GW170817, for which it was possible to make a good estimate of the total mass, also led to an estimate of $M_{\max}\sim 2.2-2.3\,M_\odot$ 
\cite{Margalit:2017dij}.  The argument for rapid collapse to a black hole is that otherwise the rapidly spinning remnant will spin down and inject energy into the remnant, which has not been seen.  A key but unproven assumption in this argument is that the merger process will generate strong and ultimately poloidal magnetic fields that will spin down the remnant.

Detailed modeling of the electromagnetic afterglow of GW170817 has led to other estimates of the maximum mass, e.g., $M_{\max} \simeq 2.16^{+0.17}_{-0.15}\, M_{\odot}$ \citep{Rezzolla:2017aly}, $M_{\max} \simeq 2.16-2.28\, M_{\odot}$ 
\cite{Ruiz:2017due}, $M_{\max} \simeq 2.13^{+0.17}_{-0.11}\, M_{\odot}$ \citep{Shao:2019ioq}, $M_{\max} \lesssim 2.3\, M_{\odot}$ \citep{Shibata:2019ctb}, $M_{\max} \simeq 2.21^{+0.12}_{-0.12}\, M_{\odot}$ \citep{Nathanail:2021tay}, and if the assumption that the GW170817 event resulted in a black hole is relaxed, $M_{\max} \simeq 2.43^{+0.16}_{-0.12}\, M_{\odot}$ 
\cite{Ai:2019rre}.  All of these estimates are subject to numerical and modeling uncertainties, but they are promising for the future.  The empirical constraints on the maximum neutron-star mass are summarized in Table~\ref{table:mmass_cons}

There have also been several claims of heavier neutron stars.
The discovery of black-widow pulsars with masses estimated as $2.13 \pm 0.04\,M_\odot$ \citep{Romani:2021xmb} for PSR J1810+1744 and $2.35 \pm 0.17\,M_\odot$ \citep{Romani:2022jhd} (68\% confidence) for PSR J0952-0607 have been reported; however, these mass measurements are less secure (due to possible systematics) than the Shapiro delay-based measurements for PSR J0740+6620 and PSR J0348+0432.
Moreover, the $2.59^{+0.08}_{-0.09}\,M_\odot$ secondary component of the compact binary merger GW190814 has been touted as a potential heavy neutron star because it is less massive than known black holes \citep{LIGOScientific:2020zkf}.
However, its proximity in mass to the ostensible $2.7\,M_\odot$ black-hole remnant of GW170817, and the lack of any observed tidal effects or electromagnetic counterpart, have made people think that it may be more likely to be a black hole. However, this conclusion comes with some caveats. Tidal effects for such massive neutron stars are expected to be very small \citep{Tan:2020ics}, and in fact, not observable with the sensitivity of ground interferometers when this event was detected \citep{LIGOScientific:2020zkf}.
Moreover, the lack of an electromagnetic counterpart in the gamma range could be due to the short-gamma ray burst not being pointed toward Earth. An electromagnetic counterpart at other frequencies could have been missed due to the poor localization of the event in the sky through only the use of gravitational wave information. See \cite{deSa:2022qny} for a discussion on many candidates whose masses that lay inside the so called mass gap (between confirmed masses of neutron stars and black holes). 

In another work, the authors used binary inclination $i$ from lightcurve modeling and observed that J1810 offers a lower limit on the NS maximum mass of $2\, M_\odot$ at 99.7\% credibility \citep{Romani:2021xmb}.
A flat, but asymmetric, light-curve maximum and a deep, narrow minimum were observed in the spectrophotometry of the companion of PSR J1810+1744 measured by the Keck telescope.
A hot pole,  surface winds, and severe gravity darkening (GD) around the $L_1$ point are all indicated by the maximum, whereas the minimum denotes significant limb darkening and a low underlying temperature.
Having the radial-velocity amplitude $K_\mathrm{c}=462.3 \pm 2.2$ km s$^{-1}$ recorded by the Keck telescope  provides a precise neutron star mass of $M_{\mathrm{NS}}=2.13 \pm 0.04\, M_\odot$, a value which is very relevant to the understanding of the dense-matter EoS \citep{Romani:2021xmb}.

\begin{table}[t!]
\caption{Pulsar  empirical constraints on the maximum mass of neutron stars. Pulsar timing constraints are less susceptible to modeling assumptions than current gravitational wave constraints. 
}
\label{table:mmass_cons}
\centering
\begin{tabular}{ccc}
\hline \hline
Neutron Star   & $M_{\max}$ ($M_{\odot}$) & Reference   \\ \hline \hline
PSR J0740+6620 & $\geq$ 2.08 $\pm$ 0.07 &  \cite{Fonseca:2021wxt}   \\ 
PSR J0348+0432 & $\geq$ 2.01 $\pm$ 0.04 & \cite{Antoniadis:2013pzd}  \\ 
   \hline \hline
\end{tabular}
\end{table}

%%%%%%%%%%%%%%%%%%%%%%%%%%%%%%%%%%%%%%%%%%%%%%%%%%%%%%%%%%%%%%%%%%%%%%%%%%%%%%%%%%%%%%%%%%%%%%%%%%%%%%%%%%%%%%%%%%%%%%%%%%%%%%%%%%%%%%%%%%%%%%%%%%%%%%%%%%%%%%%%%%%%%%%%%%%%%%%%%%%%%%%%%%%%%%%%%%%%%%%%%%%%%%%%%%%%
 
\subsection{Neutron-star mass-radius regions from NICER}

NICER is a soft X-ray telescope  mounted on the International Space Station (ISS) in 2017.
The main aim of NICER is to determine masses and radii of neutron stars using pulse-profile modeling of neighboring rotation-powered millisecond pulsars.
The NICER observations have played an important role in lowering the ambiguity in the EoS of  high-density matter (1.5--5 $n_{\sat}$) at zero temperature.  See Table~\ref{table:mr_cons} for a summary of the results reported from NICER, which we now discuss in detail. A word of warning is due at this stage: the results presented in this table (and in other tables in this section) are a summary, and in particular, the reported masses and radius with error bars are not a square in probability space. Rather, these quantities correspond to the maximum likelihood points and the 90\% confidence regions of the marginalized posterior after a careful Bayesian parameter estimation study. Two-dimensional posteriors on mass and radius are not squares, but rather complicated shapes due to the correlation between parameters. 

\begin{table}[h!]
\caption{Empirical constraints related to mass-radii from NICER.}
\label{table:mr_cons}
\centering
\begin{tabular}{cccc}
\hline \hline
Neutron Star & M (M$_{\odot}$) & Radius (km) & Reference   \\ \hline \hline
PSR J0030+0451
 & $1.34^{+0.15}_{-0.16}$&  $12.71^{+1.14}_{-1.19}$ & \cite{Riley:2019yda}   \\ 
PSR J0740+6620
 & $2.072^{+0.067}_{-0.066}$&  $12.39^{+1.30}_{-1.98}$  &\cite{Riley:2021pdl}  \\ 
PSR J0030+0451
 & $1.44^{+0.15}_{-0.14}$&  $13.02^{+1.24}_{-1.06}$ & \cite{Miller:2019cac}  \\ 
PSR J0740+6620 & $2.08^{+0.07}_{-0.07}$&  $13.7^{+2.6}_{-1.5}$ & \cite{Miller:2021qha}   \\    \hline \hline
\end{tabular}
\end{table}

Using Bayesian parameter estimation based on pulse-profile modeling of the NICER XTI event data for the isolated pulsar PSR~J0030+0451, \cite{Miller:2019cac} (posterior \citealt{pos144}) reported a gravitational mass $M=1.44^{+0.15}_{-0.14}\,M_\odot$ and a circumferential radius $R=13.02^{+1.24}_{-1.06}$ km at 68\% credibility.  For the same data but using slightly different models and a different statistical sampler, \cite{Riley:2019yda} found $M=1.34^{+0.15}_{-0.16}\,M_\odot$ and $R=12.71^{+1.14}_{-1.19}\,\mathrm{km}$ (posterior \citealt{pos134} and raw data \citealt{RAW134}).

The heavy binary pulsar PSR~J0740+6620 has a NICER count rate only $\sim$5\% that of PSR~J0030+0451 and thus NICER data alone are insufficient for precise measurements of the mass and radius.  As a result, for this pulsar, radio data and X-ray data from the X-ray Multi-Mirror (XMM-Newton) satellite were analyzed jointly with the NICER data. \cite{Miller:2021qha} (posterior \citealt{POS208NICER}) found $M=1.97-2.15\,M_\odot$ and $R=12.21-16.33\,\mathrm{km}$, both at 68\% credibility, whereas \cite{Riley:2021pdl} (posterior \cite{pos207}) reported $M=2.072^{+0.067}_{-0.066}~M_\odot$ and $R=12.39^{+1.30}_{-0.98}\,\mathrm{km}$.

The differences between the two results for PSR J0740+6620 (e.g., Miller et al. report a $-1\sigma$ radius of 12.21~km, compared with Riley et al.'s estimate of 11.41~km) led the two groups to explore the reasons for the difference.  As discussed in \cite{Miller:2021qha}, the primary differences are as follows.  First, \cite{Miller:2021qha} use a relative calibration between NICER and XMM-Newton that is consistent with the results of cross-calibration tests, whereas \cite{Riley:2021pdl} assume a much broader range of possible cross-calibration factors.  When \cite{Riley:2021pdl} apply the same cross-calibration as \cite{Miller:2021qha}, they find a $-1\sigma$ radius of 11.75~km rather than 11.41~km, i.e., this accounts for almost half the difference from the 12.21~km $-1\sigma$ radius of \cite{Miller:2021qha}.  Second, \cite{Riley:2021pdl} have a hard prior upper bound on the radius of 16~km, whereas \cite{Miller:2021qha} allow the radius to be anything that fits the data.  When \cite{Miller:2021qha} eliminate all solutions with $R>16$~km, their $-1\sigma$ radius drops to 12.06~km.  The remaining difference, of 0.31~km (compared with the original 0.8~km), is likely to be primarily due to different choices of statistical samplers: \cite{Miller:2021qha} use the parallel-tempered Markov chain Monte Carlo code PT-emcee whereas \cite{Riley:2021pdl} use the nested sampler MultiNest; \cite{Miller:2021qha} argue that at least in this specific case, MultiNest may underestimate the widths of the posteriors.
More generally, there are many moving parts within such analyses and sometimes subtle choices on how data is analyzed or which data is analyzed can affect the resulting constraints; see \cite{Essick:2021ezv} for an example related to data selection.

\subsection{Other observational constraints on neutron star masses and radii}

Quiescent low-mass X-ray binaries (QLMXBs) in globular clusters are promising objects for mass and radius constraints because (i) previous accretion events heat up the surface increasing their luminosity, (ii) their magnetic fields are expected (but not observed) to be relatively small, the magnetic field having been buried by the accretion, and (iii) the globular cluster permits a determination of the distance. There are several QLMXBs which have been used to obtain mass and radius constraints, see \cite{Steiner:2017vmg} for a recent analysis leading to neutron star radii between 10 and 14 km. However, there are still a significant number of potential systematics which may be important: (i) the magnetic field is not observed so the assumption of a small magnetic field is untested, (ii) the determination of the mass and radius requires a model of the neutron star atmosphere and the associated emergent X-ray spectrum, (iii) the composition of the accreted material is not always known, (iv) the temperature is often presumed to be uniform over the entire surface, and (v) scattering by the interstellar medium leads to a reddening of the spectrum which is not fully known. A recent attempt \citep{Al-Mamun:2020vzu} to search for systematic effects in QLMXB models compared QLMXBs to other mass and radius constraints (including those from LIGO/Virgo and NICER) and found no evidence for systematics that poison the QLMXB results. However, this result may change as the LIGO/Virgo constraints improve.

Within the supernova remnant HESS J1731-347, a star that is the centre compact object has been studied, according to \cite{Doroshenko2022}, the mass and radius of this star is estimated to be $M=0.77_{-0.17}^{+0.20}\, M_{\odot}$ and $R=10.4_{-0.78}^{+0.86}\, \mathrm{km}$, respectively, which is based on the modeling of Gaia observations and X-ray spectrum. This result depends critically on the assumption that the entire surface of the star emits at the same temperature, based on the lack of clear modulation with the stellar rotation.  With that assumption, and the assumption that the atmospheric effects of the stellar magnetic field can be neglected, a carbon atmosphere is favored over a hydrogen or a helium atmosphere.  However, as shown by \cite{Alford:2023waw}, nonuniform emission is consistent with the data on several similar sources (and preferred for some).  This means that hydrogen and helium atmospheres are possible, and these could imply standard masses well above one solar mass. According to their estimation, this star can be either the lightest neutron star ever discovered or a strange star with exotic EoS \citep{Doroshenko2022}. The examination of several SN explosions indicates that it is not feasible to form a neutron star (NS) with a mass less  than approximately $1.17 M_\odot$ \citep{Suwa:2018uni}, which begs the issue of what astronomical activity might result in such a small mass. \cite{DiClemente:2022wqp} suggests that in the case of strange quark stars, masses of the order or less than one solar mass can be found and it is conceivable to construct a cogent astrophysical hypothesis that accounts for the object's mass, radius, and gradual cooling.

%%%%%%%%%%%%%%%%%%%%%%%%%%%%%%%%%%%%%%%%%%%%%%%%%%%%%%%%%%%%%%%%%%%%%%%%%%%%%%%%%%%%%%%%%%%%%%%%%%%%%%%%%%%%%%%%%%%%%%%%%%%%%%%%%%%%%%%%%%%%%%%%%%%%%%%%%%%%%%%%%%%%%%%%%%%%%%%%%%%%%%%%%%%%%%%%%%%%%%%%%%%%%%%%%%%%

\subsection{Neutron-star tidal deformability from gravitational waves}

The phasing of the gravitational waves emitted during the inspiral of a compact binary system is sensitive to the tidal deformation experienced by each component as a result of its companion's non-uniform gravitational field.
The size of the deformation is measured by the tidal deformability parameter \citep{Flanagan:2007ix,Hinderer:2007mb}, which depends on the neutron-star mass and the EoS.
The tidal deformabilities of the individual neutron stars appear at leading order in the gravitational waveform as a mass-weighted average known as the chirp or binary tidal deformability, $\tilde{\Lambda}$ \citep{Favata:2013rwa,Wade:2014vqa}, namely
\begin{equation} \label{Ltilde}
    \tilde{\Lambda} = \frac{16}{13} \left[\frac{(m_1 + 12 m_2) m_1^4 \Lambda_1 + (m_2 + 12 m_1) m_2^4 \Lambda_2}{(m_1+m_2)^5}\right]\ , 
\end{equation}
where $m_1$ and $m_2$ are the masses of the binary components and $\Lambda_1$ and $\Lambda_2$ are their (dimensionless) individual tidal deformabilities. Because an EoS predicts a unique $m$--$\Lambda$ relation, a measurement of $\tilde{\Lambda}$ and the binary masses helps to determine the EoS in a manner analogous to constraints on the $m$--$R$ relation.

\subsubsection{Extraction of $\tilde{\Lambda}$ from GW170817}

The first measurement of $\tilde{\Lambda}$ was enabled by the detection of gravitational waves from the binary neutron-star merger GW170817 \citep{LIGOScientific:2017vwq} by LIGO and Virgo. The binary tidal deformability was constrained simultaneously with the system's masses, spins and other source properties via Bayesian parameter estimation, in which a waveform model is compared against the data to produce a likelihood function over the waveform parameters (see e.g. \citealt{Thrane:2018qnx}).
Because the tidal deformability is mass-dependent, and gravitational-wave measurements of the binary mass ratio and spins are correlated, the constraints on $\tilde{\Lambda}$ are sensitive to prior assumptions about spin.
Assuming that both neutron stars in GW170817 had low spins, in keeping with the Galactic double neutron-star population \citep{Burgay:2003jj,Stovall:2018ouw}, LIGO and Virgo measured $\tilde{\Lambda} = 300^{+500}_{-190}$ at 90\% confidence; without this assumption, the constraint is $\tilde{\Lambda} \leq 630 $ \citep{LIGOScientific:2018hze}.
The joint constraints on $\tilde{\Lambda}$ and the binary masses are expressible in terms of a multi-dimensional posterior probability distribution, samples from which are available in the data release accompanying \citep{LIGOScientific:2018hze}.

As before, a word of caution is due at this stage. The  measurements of the tidal deformabilities discussed above and below in this section result from a Bayesian parameter estimation study, and as such, they correspond to the maximum likelihood points and the 90\% confidence region of the marginalized posterior. The posterior is multi-dimensional and covariances exist between the various parameters that enter the waveform model. Therefore, 2-dimensional confidence regions, like that for the chirp tidal deformability and the chirp mass, are not squares, but rather complicated shapes, which can be found in the papers we have referenced here.     

The analysis of the signal \citep{LIGOScientific:2018hze}, and therefore, the tidal deformabilities reported above, do not require both of GW170817's compact objects to share the same EoS, implicitly leaving open the possibility that one of them is a black hole.
Because of the kilonova and short gamma-ray burst counterparts to GW170817, however, it is reasonable to assume that the merger consisted of two neutron stars \citep{De:2018uhw}.
The assumption that all neutron stars share the same EoS implicitly relates $\Lambda_1$ and $\Lambda_2$, since the EoS predicts a unique $m$--$\Lambda$ relation.
\cite{De:2018uhw} approximated this relation as $\Lambda_1 = \Lambda_2{m_2}^6/{m_1}^6$, and used it to further constrain $\tilde{\Lambda} = 222^{+420}_{-138}$ at 90\% confidence, assuming low neutron-star spins. 
An alternative to making such an approximation is to average over many different candidate EoSs drawn from a prior distribution: each EoS sample imposes an exact relation between $\Lambda_1$ and $\Lambda_2$, and the EoS uncertainty encoded in the prior distribution is propagated to $\tilde{\Lambda}$ by the averaging process. Of course, this approach requires modeling the EoS. Typically, this is done phenomenologically, either with a parameterization of the pressure-density relation -- e.g., a spectral decomposition \citep{Lindblom:2010bb} -- or with a \emph{Gaussian process} \citep{Landry:2018prl} -- i.e., a distribution over functions described by a mean function and a  covariance kernel. Caveats exist here, however, because the choice of the functional form of the EoS may not include the entire phase space of possible equations of state that can be conceived from nuclear physics.

The EoS-averaging approach was taken by \cite{Essick:2019ldf}, which modeled the EoS nonparametrically as a Gaussian process and found $\tilde\Lambda = 245^{+361}_{-160}$ (posterior median and 90\% highest-probability-density credible region). The Gaussian process was constructed to explore the entire functional space of EoSs that obey causality (sound speed less than the speed of light) and thermodynamic stability (positive semidenfinite sound speed), subject to the requirement that the EoS support neutron star masses of at least $1.93\,M_{\odot}$ to account for the existence of heavy pulsars. 
For comparison, when using a spectral parameterization for the EoS, one find $\tilde\Lambda = 412^{+313}_{-262}$ for the posterior \emph{median} and 90\% credible regions. If one instead uses the global maximum of the posterior, one finds $\tilde{\Lambda} = 225^{+500}_{-75}$, which is similar to that found using Gaussian processes.

Electromagnetic observations of GW170817's kilonova counterpart have also been used to constrain the binary tidal deformability.
However, these bounds depend on the assumed kilonova lightcurve model and are thus subject to sizeable systematic uncertainty: for example,  $\tilde{\Lambda} \geq 197$ (90\% confidence) \citep{Coughlin:2018miv}, $120 < \tilde{\Lambda} < 1110$ (90\% confidence) \citep{Breschi:2021tbm}, and $109 \leq \tilde{\Lambda} \leq 137$ (68\% confidence) \citep{Nicholl:2021rcr} was found.
Unlike the gravitational-wave measurements of binary tidal deformability, $\tilde{\Lambda}$ does not appear directly as a parameter in the light curve model and it must therefore be constrained via correlations with other observables.
As discussed in \cite{Radice:2017lry}, these constraints come with the possibility of large errors associated with uncertainties in the mass ratio of the system \citep{Kiuchi:2019lls}.

\subsubsection{Extraction of individual tidal deformabilities $\Lambda_1$ and $\Lambda_2$}

The binary tidal deformability $\tilde{\Lambda}$ is the tidal parameter that is best constrained by gravitational-wave observations. However, it is the individual tidal deformability $\Lambda$ that is directly determined by the EoS together with the equations of stellar structure. The relation between the individual tidal deformabilities $\Lambda_1$ and $\Lambda_2$ of each member of a compact binary and the system's binary tidal deformability is expressed in Eq.~\eqref{Ltilde}. Generally, the component masses $m_1$ and $m_2$ are constrained simultaneously with $\tilde{\Lambda}$ through the Bayesian parameter estimation of the gravitational-wave signal. Thus, we have one equation relating three tidal parameters: a measurement of $\tilde{\Lambda}$ only determines $\Lambda_1$ and $\Lambda_2$ up to a residual degeneracy. This degeneracy is broken by a tidal parameter $\delta\tilde{\Lambda}$ that appears at higher order in the gravitational waveform, but it is unfortunately not measurable with current detectors \citep{Wade:2014vqa}.

Nonetheless, this degeneracy is not an obstacle to translating measurements of $\tilde{\Lambda}$ into constraints on $\Lambda_1$ and $\Lambda_2$: it is simply the case that every combination of $\Lambda_1$ and $\Lambda_2$ that produces the same $\tilde{\Lambda}$ is equally likely. Thus, one can build up a posterior for $\Lambda_1$ and $\Lambda_2$ by sampling in the individual tidal deformabilities, and assigning to each sample the likelihood of the $\tilde{\Lambda}$ it predicts (the likelihood is the ratio of the posterior probability to the prior probability). The result of this process for GW170817 is the posterior on the component tidal deformabilities illustrated in Fig.~10 of \cite{LIGOScientific:2018hze}; notice that there is a direction in $\Lambda_1$--$\Lambda_2$ plane along which there is essentially no constraint due to the degeneracy.

One can break the degeneracy by introducing an additional layer of modeling that imposes a common EoS for both components of GW170817: this augments Eq.~\eqref{Ltilde} with a second equation relating the individual tidal deformabilities.  As in the case of $\tilde{\Lambda}$, this can be done either with an approximate relation connecting $\Lambda_1$ and $\Lambda_2$, or by averaging over an EoS prior distribution.

Using the Gaussian process-based EoS representation described above, \cite{Essick:2019ldf} mapped the binary tidal deformability measurement into constraints of $\Lambda_1 = 148^{+274}_{-125}$ and $\Lambda_2 = 430^{+519}_{-301}$ on the component tidal deformabilities, under the assumption of low neutron-star spins. Moreover, since each EoS sample from the Gaussian process prescribes an exact $m$--$\Lambda$ relation, this approach allows the tidal deformability to be inferred at any mass scale. \cite{Essick:2019ldf} also constrained $\Lambda_{1.4} = 211^{+312}_{-137}$ (posterior median and 90\% highest-probability-density credible region). 

Alternatively, one can break the degeneracy with an approximate EoS-insensitive relation connecting the individual tidal deformabilities. The so-called binary Love relations \citep{Yagi:2015pkc} fit the relation between two independent linear combinations of $\Lambda_1$ and $\Lambda_2$ to many candidate EoSs from nuclear theory. With the neutron stars' individual tidal deformabilities related by way of this EoS-insensitive relation \citep{Chatziioannou:2018vzf}, augmented with Gaussian white noise to account for the scatter in the fit, and $\Lambda$'s mass dependence expanded in a series about $1.4\,M_\odot$, \cite{LIGOScientific:2018cki} reduced the joint constraint on binary masses and $\tilde{\Lambda}$ from GW170817 to an estimate of the tidal deformability of a $1.4\,M_\odot$ neutron star: $\Lambda_{1.4} = 190^{+390}_{-120}$ at the $90\%$ credible level, in the low neutron-star spin scenario. The empirical constraints on the tidal deformability from GW170817 are summarized in Table \ref{table:def_cons}.

\begin{table}[h!]
\caption{Empirical constraints on the  tidal deformability for GW170817
 event from  LIGO and VIRGO. }
\label{table:def_cons}
\centering
\begin{tabular}{ccc}
\hline \hline
 Tidal deformability  & Reference & Confidence level  \\ \hline \hline
 $\tilde{\Lambda} = 300^{+500}_{-190}$ & low spins \citep{LIGOScientific:2018hze}  & $90\%$ \\ 
 $\tilde{\Lambda} \leq 630 $& minimal assumptions \citep{LIGOScientific:2018hze}  & $90\%$  \\ 
 $\tilde{\Lambda} = 222^{+420}_{-138}$ & common EoS via analytic approximation \citep{De:2018uhw}   & $90\%$  \\ 
 $\Lambda_{1.4} = 190^{+390}_{-120}$ & common EoS via binary Love relation \citep{LIGOScientific:2018cki}  & $90\%$  \\ 
 $\tilde{\Lambda} = 245^{+361}_{-160}$ & common EoS via Gaussian process \citep{Essick:2019ldf}  & $90\%$  \\ 
 $\Lambda_{1.4} = 211^{+312}_{-137}$ & common EoS via Gaussian process \citep{Essick:2019ldf}  & $90\%$  \\ 
  \hline \hline
\end{tabular}
\end{table}

\begin{figure}[h!]
\includegraphics[scale=0.7]{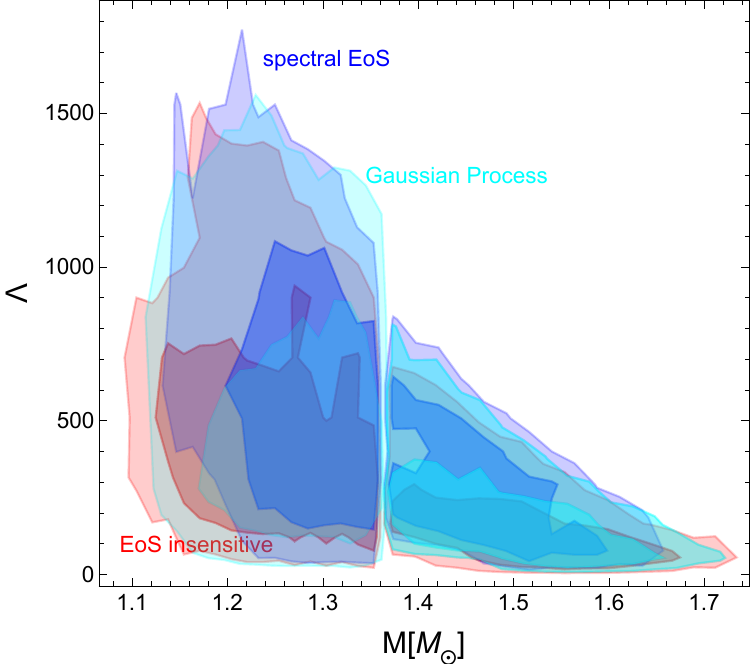}
\centering
\caption{Constraints on the tidal deformability using the universal relations (EOS insensitive), the spectral EOS and an EoS constructed from Gaussian Processes. The universal-relation constraint and spectral-EoS constraint are obtained using GW170817 data. The Gaussian Process constraint is obtained by combining GW170817,  PSR 1614-2230, PSR 0348+0432, and PSR 0740+6620 (mass only) data. Shaded regions are shown at the $68\%$ and $95\%$ confidence regions. }
%\hung{I have changed the caption a bit. Nico, please check. Reed and Phil, please check whether I got the pulsar name right.} \reed{I corrected the list of PSRs based to match Table 1 from https://arxiv.org/pdf/2003.04880.pdf}
		\label{fig:lambda}
\end{figure}

 The results for the extracted tidal deformability posteriors from GW170817 are shown in Fig.~\ref{fig:lambda}.  While there is a lot of overlap for the posteriors of the EoS insensitive (universal relations), spectral EoS, and Gaussian Process EoS, they are not exactly the same and this may lead to incorrect conclusions when extracting the EoS unless one is careful. For instance, most nuclear physics EoS tend towards higher values of $\Lambda$ for a given M.  Thus, the EoS insensitive posterior may seem more restricting than the spectral EoS. However, since these are 90\% confidence regions, all posteriors are actually statistically consistent with each other, and one cannot simply take the edge of one posterior to draw strong conclusions without a careful Bayesian parameter estimation study.  

\subsubsection{Connecting tidal deformabilities to the mass-radius sequence}

Because the neutron-star tidal deformability scales strongly with the stellar radius ($\Lambda \sim R^5/m^5$), and because $R$ is an easily interpretable parameter, gravitational-wave measurements of $\Lambda$ have often been translated into radius constraints in the literature. Nonetheless, we stress that gravitational waves from compact binaries do not directly measure the neutron star radius---the mapping from $\Lambda$ to $R$ necessarily involves additional modeling. This modeling can either be done at the level of the EoS (e.g.~with a spectral or nonparametric EoS representation), or at the level of the mapping itself (e.g., with an EoS-insensitive relation between $\Lambda$ and the stellar compactness $m/R$).

In \cite{LIGOScientific:2018cki}, LIGO and Virgo mapped the joint posterior on component tidal deformabilities and masses from GW170817 (assuming a binary Love relation) into a posterior on $m_1$, $m_2$, $R_1$ and $R_2$ by adopting a spectral parameterization for the EoS. The original posterior was used to compute the likelihood of each spectral EoS realization, and then for each component mass sample from the original posterior, the radius predicted by the EoS realization's mass-radius relation was assigned that likelihood. This procedure led to inferred radii of $R_1 = 11.9^{+1.4}_{-1.4}$ km and $R_2 = 11.9^{+1.4}_{-1.4}$ km.

LIGO and Virgo also implented an alternative approach that used an EoS-insensitive $\Lambda$--$m/R$ relation. For each component mass and component tidal deformability sample from the original posterior, $\Lambda$ was mapped to $m/R$ using the relation from \cite{Yagi:2016bkt}, and the radius was extracted using the sampled mass. The uncertainty in the EoS-insensitive relation fit was modeled as Gaussian white noise in the relation. This procedure led to inferred radii of $R_1 = 10.8^{2.0}_{-1.7}$ km and $R_2 = 10.7^{2.1}_{-1.5}$ km. The small differences vs the radius constraints with the spectral EoSs illustrate the systematic uncertainty that arises from the extra modeling required.

In \cite{De:2018uhw}, a different EoS-insensitive relation was used to map from the tidal deformability measurement to a radius constraint. Imposing the common EoS assumption $\Lambda_1 = \Lambda_2{m_2}^6/{m_1}^6$ described above and assuming that both neutron stars involved in GW170817 had the same radius $R$,  the definition of the binary tidal deformability reduces to $\tilde\Lambda \propto (R/\mathcal{M})^6$, where $\mathcal{M} \equiv \eta^{3/5} (m_1+m_2)$  is the chirp mass with $\eta \equiv m_1 m_2/(m_1+m_2)^2$, in an EoS-insensitive relation that can be fit to a sample of various EoSs. The joint posterior on $\tilde{\Lambda}$ and $\mathcal{M}$ from GW170817 thus determines the common neutron-star radius, which was reported as $R = 10.7^{+2.1}_{-1.6}$ km at 90\% confidence.

\begin{figure}[h!]
\includegraphics[scale=0.7]{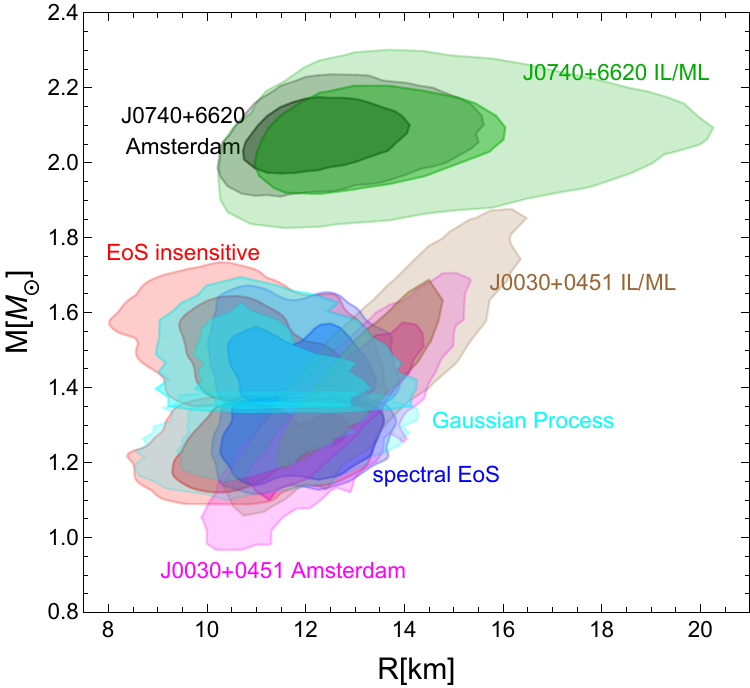}
\centering
\caption{Constraints on the mass-radius using the universal relations (EOS insensitive), the spectral EOS and an EoS sampled through Gaussian Processes. The universal-relation constraint and spectral-EoS constraint are obtained using GW170817 data. The Gaussian Process constraint is obtained by combining GW170817,  PSR 1614-2230, PSR 0348+0432, and PSR 0740+6620 (mass only) data. NICER constraints from both the Amsterdam and Illinois/Maryland groups are show for both J0030+0451 and J0740+6620. Shaded regions show the $68\%$ and $95\%$ confidence region. }
		\label{fig:MR}
\end{figure}

A summary of the resulting mass-radius constraints from both NICER and GW170817 can be see in Fig.~\ref{fig:MR}. The heavier NICER pulsar (J0740+6620) has the highest posterior distribution that is centered nearly directly above the lighter NICER pulsar (J0030+0451) and the GW170817 extracted mass-radius posteriors. However, there is some probability that heavier $M \sim 2 M_\odot$ may bend to the right. The lighter constraints from J0030+0451 and GW170817 overlap in their posteriors as well, although GW170817 generally prefers a slightly smaller radius whereas  J0030+0451's posterior prefers larger radii.

Different methods (EoS insensitive, spectral EoS and Gaussian processes) to obtain the posterior on the mass and radius with GW170817 may look different, but they all are statistically consistent with each other. In particular, we stress that one cannot use these posteriors, over-impose mass-radius curves computed with a given EoS and then determine whether the EoS is allowed or disallowed based on whether it overlaps with the contours or not. This is because each point on this plane, including those outside the 68\% or the 95\% confidence regions, actually has a posterior weight assigned to it (not shown in the figure). Therefore, comparison of the data to a given EoS needs to be done carefully in a Bayesian way. Overall, all posteriors are statistically consistent with each other at this time, and therefore, there is no tension between observations. Eventually, if future observations place better constraints on the mass-radius plane, one expects that a single EoS will be preferred by all data. 

We emphasize that these radius constraints from GW170817 are merely approximate encapsulations of the information contained within the tidal deformability measurements reported above, which are directly constrained by the gravitational waves. When considering gravitational wave data, it is thus preferable to use the tidal deformability measurements to constrain the EoS. 

\subsubsection{GW190425, GW200105, and GW200115}

Gravitational waves also yielded a constraint on tidal deformability in the binary neutron-star merger GW190425, but because this system was heavier (such that its $\tilde{\Lambda}$ is intrinsically smaller) and more distant (such that its signal-to-noise ratio was lower), the constraint is not competitive with GW170817.
Similarly, GW200105 and GW200115 \citep{LIGOScientific:2021qlt} both likely contained neutron stars, but tidal effects were unmeasureably small in these systems due to the large masses of their companions and subsequent rapid orbital evolution through GW emission.

%%%%%%%%%%%%%%%%%%%%%%%%%%%%%%%%%%%%%%%%%%%%%%%%%%%%%%%%%%%%%%%%%%%%%%%%%%%%%%%%%%%%%%%%%%%%%%%%%%%%%%%%%%%%%%%%%%%%%%%%%%%%%%%%%%%%%%%%%%%%%%%%%%%%%%%%%%%%%%%%%%%%%%%%%%%%%%%%%%%%%%%%%%%%%%%%%%%%%%%%%%%%%%%%%%%%

\section{Outlook}
\label{sec:outlook}

In this work, we  compiled various constraints coming from  nuclear and astrophysics related to dense matter and  neutron stars. We encapsulated up to date results from  first principle theories and  experimental  observations. The usage of anticipated next-generation facilities and theoretical advancements are the most promising path for substantial improvement above current limitations. With  enhanced sensitivity and expanded receiver bandwidths, several upcoming astrophysical radio observatories, including the dish Deep Synoptic Array, DSA-2000 \citep{Hallinan:2019qyo}, Canadian Hydrogen Observatory and Radio-transient Detector, CHORD \citep{Vanderlinde:2019tjt}, and the next-generation Very Large Array, ngVLA  \citep{Chatterjee:2018hrp}, will present significant opportunities in pulsar science. Also, we anticipate seeing more multi-messenger binary neutron-star merger detections in the coming years, the INTEGRAL, Fermi, Swift,  and SVOM will observe merging binary neutron stars in conjunction with GW and electromagnetic data during the fourth LIGO-Virgo-KAGRA observing run  which started on May, 2023 \citep{Patricelli:2022hhr,FermiGamma-rayBurstMonitorTeam:2023gjw}. Future detections of neutrinos from supernova explosions in our neighborhood will provide information about larger $Y_Q$ at large densities and intermediate temperatures, while with future LIGO and NICER runs the post-merger signal of gravitation waves from neutron-star mergers will provide information about low $Y_Q$ at large densities and large temperatures \citep{Lovato:2022vgq}.

Several improvements are anticipated in the next phase of BES, thanks to the performance enhancements brought on by collider and detector modifications which will help to dig deeper into the high density regime of the QCD phase diagram. Constraining the EoS is not a simple task and significant work from different communities in nuclear and astrophysics is needed to achieve this.  Within the nuclear physics community, the diverse models and theories, working in a different regime of QCD phase diagram are to be updated with the modern constraints so that we can have a unified EoS for nuclei, bulk baryonic and quark matter.

\bmhead{Acknowledgments}
The authors would like to thank Niseem Abdelrahman, Wei Li, Tetyana Galatyuk, and You Zhou for providing comments and assistance in finding the experimental references in this work.

The MUSES collaboration is supported by NSF under OAC-2103680.
Additional support for the collaboration members includes:
DM is supported NSF Graduate Research Fellowship Program under Grant No. DGE – 1746047.
JSSM is supported by Consejo Nacional de Ciencia y Tecnologia (CONACYT) under SNI Fellowship I1200/16/2020.
CR is supported by NSF under grants no. PHY-1654219, PHY-2208724, PHY-2116686.
MGA is partly supported by DOE, Office of Science, Office of Nuclear Physics under Award No. DE-FG02-05ER41375.
VD is supported by NSF under grants PHY1748621 and NP3M PHY-2116686.
MS is supported by DOE, Office of Science No DE-SC0013470.
MCM is supported in part by NASA ADAP grant 80NSSC21K0649.
ML thanks the São Paulo Research Foundation (FAPESP) for support under grants 2021/08465-9, 2018/24720-6, and 2017/05685-2, as well as the support of the Brazilian National Council for Scientific and Technological Development (CNPq).
ERM acknowledges support as the John Archibald Wheeler Fellow at Princeton University (PCTS), and from the Institute for Advanced Study.
SPH is supported by DOE grant DE-FG02-00ER41132.
PL is supported by the Natural Sciences \& Engineering Research Council of Canada (NSERC).
CP is supported by FCT (Funda\c c\~ao para a Ci\^encia e a Tecnologia, I.P, Portugal) under Projects No. UIDP/-04564/-2020, No. UIDB/-04564/-2020 and 2022.06460.PTDC.
SLS is supported by NSF Grant PHY-2006066 and NASA Grant 80NSSC17K0070, both to UIUC.



% \bibliography{sn-bibliography}% common bib file

\phantomsection
\addcontentsline{toc}{section}{References}
\bibliography{inspire,NOTinspire}

\end{document}